\documentclass{pasj01}
\draft
\usepackage{lscape}
\usepackage{subfigure}
\usepackage{multirow}
\usepackage{url}
\usepackage{threeparttable}
\usepackage{color}
\usepackage{natbib} 

\begin{document}
\SetRunningHead{M.~Mizumoto \etal}{The Iron Feature in GRS 1915$+$105}
\title{Origin of the X-ray Broad Iron Spectral Feature in GRS 1915$+$105}
\author{Misaki \textsc{Mizumoto}\altaffilmark{1,2},
Ken \textsc{Ebisawa}\altaffilmark{1,2},
Masahiro \textsc{Tsujimoto}\altaffilmark{1} \&
Hajime \textsc{Inoue}\altaffilmark{3}
}

\email{mizumoto@astro.isas.jaxa.jp}

\altaffiltext{1}{Institute of Space and Astronautical Science (ISAS), Japan Aerospace Exploration Agency (JAXA), 3-1-1 Yoshinodai, Chuo-ku, Sagamihara, Kanagawa, 252-5210, Japan}
\altaffiltext{2}{Department of Astronomy, Graduate School of Science, The University of Tokyo, 7-3-1 Hongo, Bunkyo-ku, Tokyo, 113-0033, Japan}
\altaffiltext{3}{Meisei University, 2-1-1 Hodokubo, Hino, Tokyo, 191-8506, Japan}

\KeyWords{accretion, accretion disks --- black hole physics --- stars: individual (GRS 1915$+$105) --- X-rays: binaries}

\maketitle

\begin{abstract}
The X-ray spectrum of GRS 1915$+$105 is known to have a ``broad iron spectral feature'' in the spectral hard state.
Similar spectral features are often observed in Active Galactic Nuclei (AGNs) and other black-hole binaries (BHBs), 
and several models have been proposed for explaining it.
In order to distinguish spectral models, time variation provides an important key.
In AGNs, variation amplitude has been found to drop significantly at the iron K-energy band at timescales of $\sim10$~ks.
If spectral variations of black-holes are normalized by their masses,
the spectral variations of BHBs at timescales of sub-seconds should exhibit similar characteristics to those of AGNs.
In this paper, we investigated spectral variations of GRS 1915$+$105 at timescales down to $\sim10$~ms.
This was made possible for the first time with the Suzaku XIS Parallel-sum clocking (P-sum) mode, 
which has the CCD energy-resolution as well as a time-resolution of 7.8~ms.  
Consequently, we found that the variation amplitude of GRS 1915$+$105 does not drop at the iron K-energy band at any timescales from 0.06~s to 63000~s,
and that the entire X-ray flux and the iron feature are independently variable at timescales of hours.
These are naturally understood in the framework of the ``partial covering'' model, 
in which variation timescales of the continuum flux and partial absorbers are independent.
The difference of energy dependence of the variation amplitude between AGN and BHB is presumably due to different mechanisms of the outflow winds, i.e.,
the partial absorbers are due to UV-line driven winds (AGN) or thermally-driven winds (BHB).
\end{abstract}

\section{Introduction}\label{s1}

The ``broad iron spectral features'' have been frequently observed in the X-ray energy spectra of black-hole (BH) objects, 
the nature of which has been under debate for over 20 years.
This feature was detected in the X-ray spectrum of Active Galactic Nucleus (AGN) MCG--6--30--15 with the first CCD detector in orbit (SIS on ASCA; \citealt{tan95}),
and has been subsequently investigated in many AGNs and black-hole binaries (BHB) with CCD detectors (\citealt{mil07} for a review).
There are some competing models to explain this spectral feature.
One is the ``relativistic disk reflection'' model, 
in which X-ray photons emitted from a compact ``lamp-post'' corona just above the central BH are reflected at around the inner radius of the accretion disk and the fluorescent iron line due to the disk reflection is relativistically distorted \citep{fab89,fab02b}.
If this model is valid, we can measure the BH spin and demonstrate effects of the general theory of relativity.
The other is the ``partial covering'' model, in which X-ray absorbers partially cover the X-ray source, producing the iron edge feature that mimics the broad iron emission line (e.g.~\citealt{tan03}; \citealt{miy12}).
If this model is valid, we can observe geometry and kinematics of absorbers or outflowing gas around the central BH, and constrain mechanism of the BH feedback \citep{fab12}.
Unfortunately these models can explain exactly the same time-averaged spectra equally well.
In fact, \citet{mak10} pointed out that the broad iron features couple strongly with the spectral continuum, and 
\citet{nod13} showed that the spectral continuum of AGNs are composed of multi-components and are more complex than previously considered.
Thus, It is very difficult to determine the spectral continuum and the broad iron features uniquely.

Thus, spectral variations have been investigated in order to constrain spectral models.
\citet{mat03} investigated spectral variations of MCG--6--30--15, which has a BH mass of $\sim10^7\,M_\odot$, and found that the variation amplitude significantly drops at around the iron K-feature at $\sim10^{4-5}$~s (see Figure 9 in \citealt{mat03}).
On one hand, this spectral variation can be explained by the ``partial covering'' model,
assuming that the luminosity of the X-ray source is rather constant and that the covering fraction by X-ray absorbers is significantly variable \citep{miy12}.
On the other hand, in the ``relativistic disk reflection'' model, 
assuming that the disk-reflected photons are much less variable than the direct photons due to light bending, 
the spectral variation may be also explained \citep{fab03,min03}.

Here, If the spectral variation is normalized by the BH mass,
we should expect the same characteristic spectral variations in BHBs at timescales of sub-seconds.
To that end, both the energy-resolution of CCD detectors and a time-resolution of several ms are needed.
However, such an observation is difficult because read-out time of CCD detector is much longer (e.g.~2~s in Suzaku XIS Normal mode with 1/4 window option) in the standard imaging mode.

To achieve both high energy-resolution and high time-resolution in the present paper,
we use Suzaku XIS Parallel-sum clocking (P-sum) mode \citep{mitsuda07}.
In P-sum mode, events are summed along the Y-direction in the CCD
and the sum is treated as a single row in the subsequent process.
Thus, a higher timing resolution is achieved (8~s~/1024~$\simeq$~7.8~ms) than Normal mode
at a sacrifice of the imaging information along the Y-axis. 
P-sum mode had not been fully calibrated, but recently the Suzaku XIS team improved the calibration significantly\footnote{See \url{http://www.astro.isas.ac.jp/suzaku/analysis/xis/psum_recipe/Psum-recipe-20150326.pdf} for details.}.
Thus, we can analyze the spectra obtained with P-sum mode and study variations of the X-ray ``broad iron spectral features'' of BHBs at timescales of sub-seconds in details.

To investigate BHB's short time spectral variation,
we analyzed the archival P-sum mode data of GRS 1915$+$105.
GRS 1915$+$105 is a low-mass BHB at a distance of $8.6_{-1.6}^{+2.0}$~kpc, and the BH mass is estimated to be $12.4_{-1.8}^{+2.0}$~M$_\odot$ \citep{rei14}.
The companion star is estimated to be a K--M III star of 1.0--1.5~M$_\odot$ with an orbital period of 33.5 days \citep{gre01a}. 
It is the first Galactic superluminal source with radio relativistic jets \citep{mir94} and called as a ``microquasar'', which has properties similar to the quasars.
GRS 1915$+$105 shows unique and dramatic temporal/spectral variations that seem to be quite different from canonical BHBs (\citealt{fen04} for a review).
When the Suzaku observation was performed, it was in ``State C'', corresponding to the ``low-hard'' state of canonical BHBs \citep{bel00}.
In ``State C'', its X-ray spectrum is known to have the ``broad iron spectral feature'' around $\sim7$~keV \citep{mar02,nei09,blu09}.

In this paper, we have analyzed the Suzaku archival data of GRS 1915$+$105 
and investigated variation of the broad iron spectral feature 
for the first time with the CCD resolution down to timescales of milliseconds.
In Section 2, we describe the data and the data reduction.
Next, in Section 3, we show results of temporal/spectral analysis, and report that the variation amplitude of GRS 1915$+$105 has no feature at around the iron K-energy band in contrast to AGNs.
In Section 4, we show that the ``partial covering'' model can explain the observation more naturally and comprehensively than the relativistic disk reflection model.
Finally, we show our conclusion in Section 5.

\section{Observation and data reduction}
We used Suzaku data obtained on May 7--9, 2007 (ID$=$402071010).
The exposure and duration are 65.7 and 124.1~ks, respectively.
XIS1 was operated in Normal mode (1/4~window $+$ 1~s burst) and XIS0 and XIS3 were in P-sum mode.
We used the HEADAS 6.16 software package and the CALDB ver.~20141001.
The XIS1 data were screened with {\tt XSELECT} using the standard criteria \citep{koyama07}.
For the data reduction of XIS0 and XIS3 (P-sum mode), 
in cooperation with the XIS team, 
we reconstructed the calibration database and the analysis recipe 
using all the calibration datasets taken with P-sum mode.
The result was released by the XIS team$^1$.

\subsection{XIS0, 3 data}
\begin{figure}[htbp]
  \begin{center}
    \includegraphics[width=120mm]{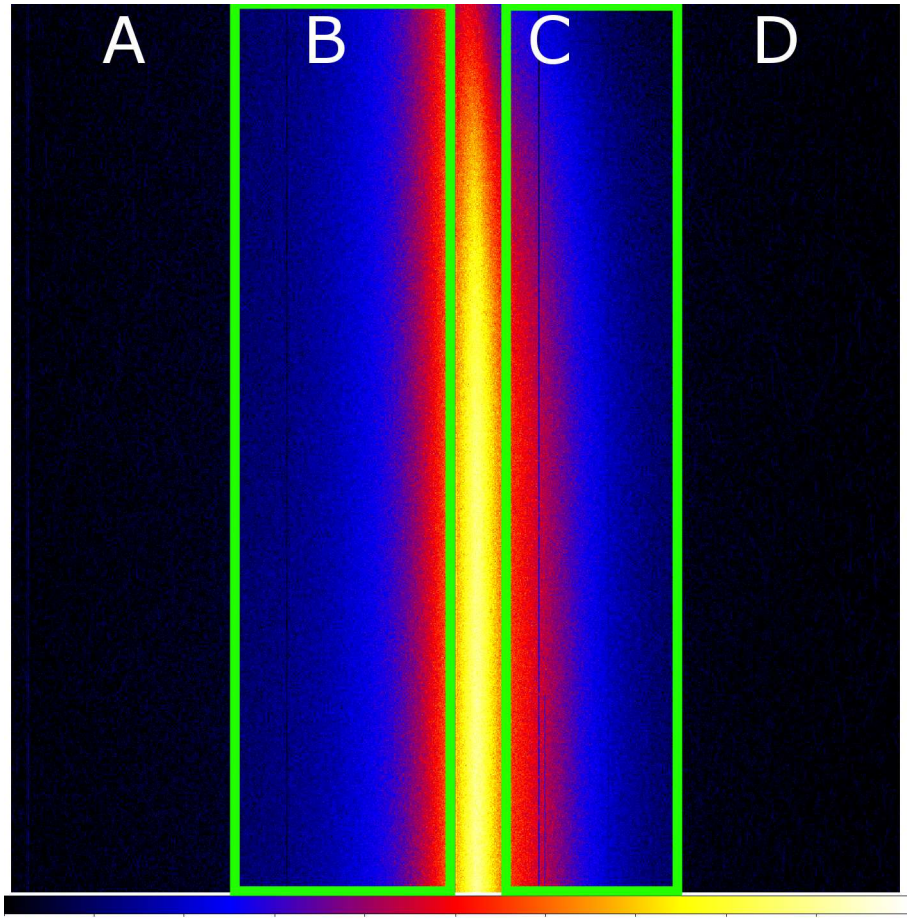}
    \end{center}
  \caption{
   P-sum mode (XIS0) image whose horizontal and vertical axes correspond to the ACTX and RAWY coordinates, respectively. The green rectangles show the source regions avoiding the central piled-up columns (yellowish part).
 }\label{f4-xis0}
\end{figure}

\begin{figure}[htbp]
  \begin{center}
    \includegraphics[width=120mm,angle=270]{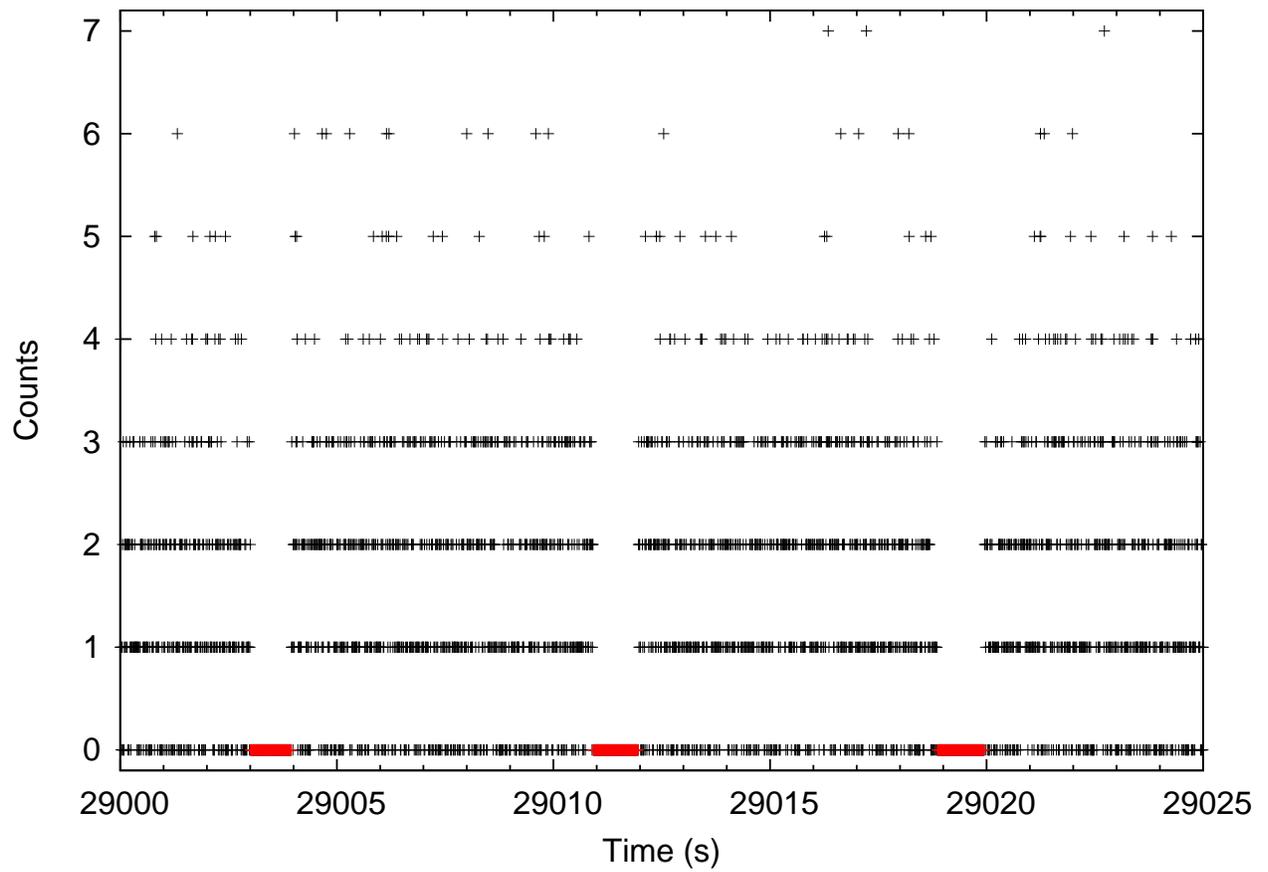}
    \end{center}
  \caption{Light curve of XIS0 Segment C with the shortest bin size of 7.8~ms. The red bins indicate the periods when telemetry saturation takes place.
 }\label{f4-ltcrv}
\end{figure}

Each XIS sensor has four segments (A, B, C, and D).
In P-sum mode, we only used data in segment B and C
because segment A and D are poorly calibrated.
We further removed the data suffered from the telemetry saturation and the pile-up,
as they are non-negligible for such a bright source.
Figure \ref{f4-xis0} shows the XIS0 image whose horizontal and vertical axes are ACTX and RAWY, respectively.
Events are read out in the order of segment B and A, or segment C and D,
thus segments A and D have much fewer counts than segments B and C due to telemetry saturation.
Besides, when telemetry saturation occurs within segments B or C,
the upper region of CCD has fewer counts than the lower region.
Figure \ref{f4-ltcrv} shows a part of the light curve of XIS0 segment C.
Count rates during telemetry saturation are found to be 0.
In order to remove the influence of telemetry saturation,
we removed the time-bins in which the counts are continuously 0 for a certain period of time.
For example, XIS0 segment C has an average count per bin ($=7.8$~ms) of 0.98, and the number of the time-bin is 8,406,107.
Thus, assuming that the distribution of counts follows the Poisson distribution,
the probability that the counting rate within a single time-bin is 0 is calculated to be $P_{\lambda=0.98}(0)=0.375$.
We calculated the minimum $n$ to fulfill $P_{\lambda=0.98}(0)^n\times8406107\leq0.01$, to find that $n=21$.
Namely, the chance probability for 21 continuous null bins is less than 1\%, when telemetry saturation does not take place.
Thus, we removed the period when the counts are continuously 0 over 21 time-bins.
We repeated this procedure in each segment.

In order to remove the pile-up regions, we created spectra by masking the central columns along the signal peak.
As a result, we found that the pile-up effect can be ignored 
when 80 columns are ignored in segment C.
Segment B was not effected by pile-up.
Figure \ref{f4-xis0} shows the source region with the pile-up mask.

\subsection{XIS1 data}
As for XIS1 (Normal mode with 1/4~window option), the central region is heavily piled-up.
Thus, we masked the central region around the signal peak.
We set the source region as a rectangle of 350 pix $\times$ 250 pix minus a circle of 80 pix radius centered on the signal peak, 
so that the spectral slope does not change above 7~keV when the radius is increased further \citep{yam12}.
We set the background region as a rectangle of 900 pix $\times$ 250 pix minus a rectangle of 600 pix $\times$ 250 pix.

\subsection{HXD data}
As for the background of HXD, 
the non X-ray background (NXB) model is provided by the HXD instrument team in the form of simulated event files \citep{fuk09}.
In HXD/PIN, we calculated the cosmic X-ray background (CXB) component using the PIN response for a flat emission distribution\footnote{\url{http://heasarc.gsfc.nasa.gov/docs/suzaku/analysis/pin_cxb.html}},
assuming the CXB spectrum measured by HEAO-1 \citep{bol87}.
The NXB model of HXD/GSO has some systematic uncertainty\footnote{\url{http://www.astro.isas.ac.jp/suzaku/doc/suzakumemo/suzakumemo-2008-01.pdf}}, so
we added 1\% systematic error to the GSO NXB model.
The CXB component of GSO was ignored because it is much weaker than the NXB.
We used response files provided by the HXD instrumental team.
Besides, in HXD/GSO, we also used an additional ancillary response file (ARF) for reducing the uncertainty of the linearity around Gd-K edge\footnote{\url{http://heasarc.gsfc.nasa.gov/docs/suzaku/analysis/gso_newarf.html}}.

\section{Data analysis and results}

\subsection{Time-averaged spectrum}\label{secave}

We first show the time-averaged spectra of XIS and HXD.
We did not use the XIS data below 2~keV because there are few photon counts due to the heavy interstellar absorption on the Galactic plane, 
and used up to 12~keV\footnote{See \url{http://www.astro.isas.jaxa.jp/suzaku/doc/suzakumemo/suzaku_memo_2015-03.pdf} for the XIS calibration above 10~keV.}.
We used the PIN data from 16~keV to 70~keV and the GSO data from 70~keV to 100~keV.
Hereafter, errors are quoted as the statistical 90\% level confidence range unless otherwise noted.

\subsubsection{Phenomenological model}
\begin{figure}[htbp]
  \begin{center}
    \includegraphics[width=120mm,angle=270]{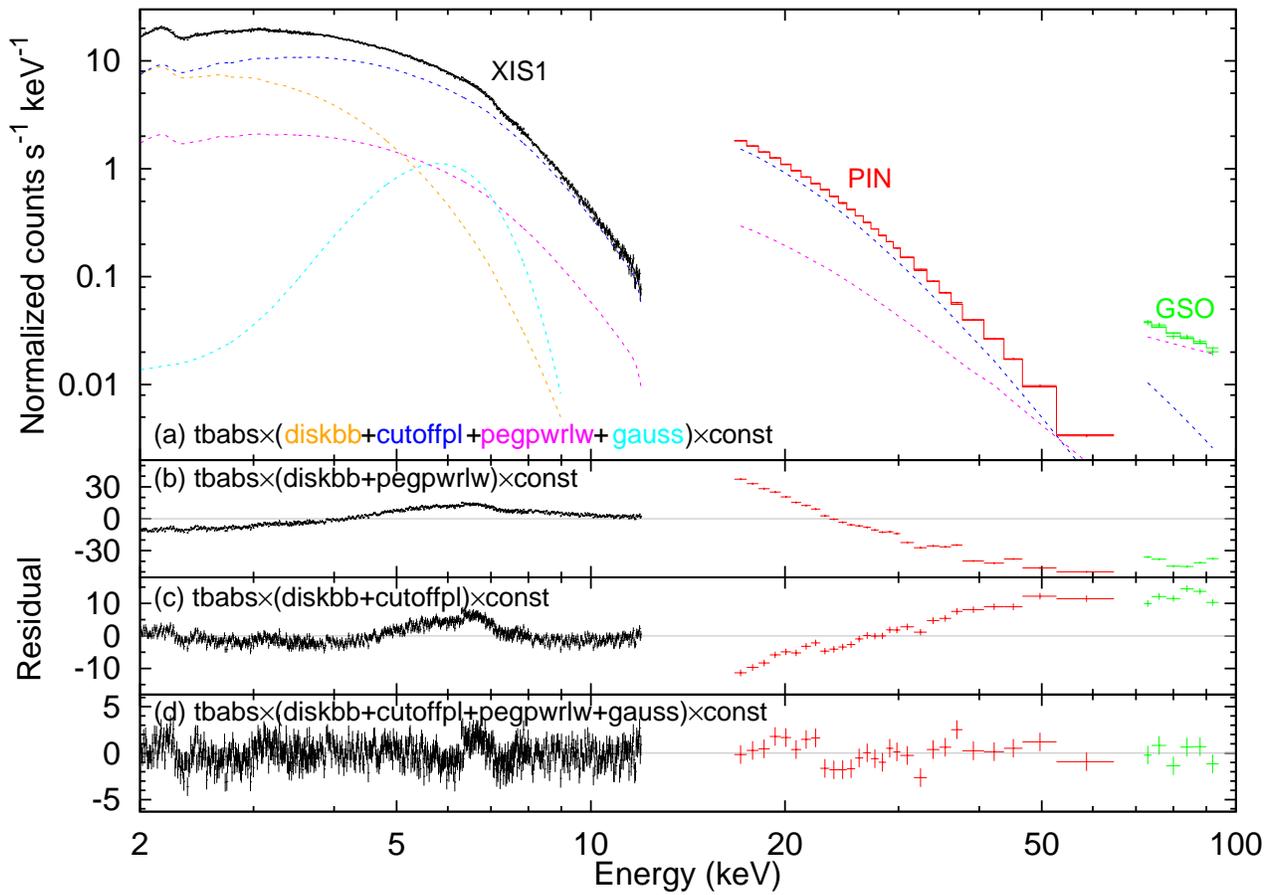}
  \end{center}
  \caption{Phenomenological spectral fitting. (a) shows the spectrum and the model, and (b)--(d) show the residuals when fitted by different models.
  }\label{f5-suzaku_cutoffpl}
\end{figure}

We fitted the spectrum with phenomenological models first.
We used XSPEC ver.\,12.8.1 for spectral fitting.
At first, we used the simplest model {\tt tbabs}$\times$({\tt diskbb}+{\tt pegpwrlw})$\times${\tt const}, 
where {\tt tbabs} is the X-ray absorption by the interstellar medium \citep{wil00}, 
{\tt diskbb} is a multi-color disk (MCD) component \citep{mit84,mak86}, 
{\tt pegpwrlw} is a powerlaw model, and
{\tt const} is the normalization factors\footnote{\url{http://www.astro.isas.ac.jp/suzaku/doc/suzakumemo/suzakumemo-2007-11.pdf}}.
The photoionized cross-section of {\tt tbabs} was calculated with \citet{bcmc} and \citet{yan98}.
The MCD component was added, which is not statistically necessary to fit the time-averaged spectrum,
but was required to explain the time variation (see details in Section \ref{sec:MCD}).
Thus, we added the MCD component with fixed parameters derived from the spectral variation analysis (Table \ref{t5-invariant}).
The resulting $N_H$ value of {\tt tbabs} was larger than the Galactic absorption estimated by the Leiden-Argentine-Bonn 21~cm survey, $1.39\times10^{22}\,\mathrm{cm}^{-2}$ \citep{kal05}, which is considered to be due to an additional absorption associated with GRS 1915$+$105.
The continuum in the HXD band was not fitted well in Figure \ref{f5-suzaku_cutoffpl}(b), thus
we used {\tt cutoffpl} instead of {\tt pegpwrlw}, where {\tt cutoffpl} shows the powerlaw model with high-energy exponential rolloff.
The reduced $\chi^2$ improved, but residuals still remained in the HXD band (Figure \ref{f5-suzaku_cutoffpl}c).
Subsequently, we added {\tt pegpwrlw} to explain the HXD component and added {\tt gauss} to explain the iron K-feature phenomenologically, where {\tt gauss} shows a gaussian function.
We found that the model {\tt tbabs}$\times$({\tt diskbb}$+${\tt cutoffpl}$+${\tt pegpwrlw}$+${\tt gauss})$\times${\tt const} can explain the whole spectra satisfactorily (Figure \ref{f5-suzaku_cutoffpl}d).
Table \ref{t5-cutoffpl} shows the fitting parameters.
The {\tt pegpwrlw} component was seen in some previous papers (e.g.~\citealt{gro98}), and it is argued as a non-thermal inverse Compton scattering component \citep{zdz01,tit09,ued10}.

\subsubsection{Relativistic disk reflection model}
\begin{figure}[htbp]
  \begin{center}
    \includegraphics[width=100mm,angle=270]{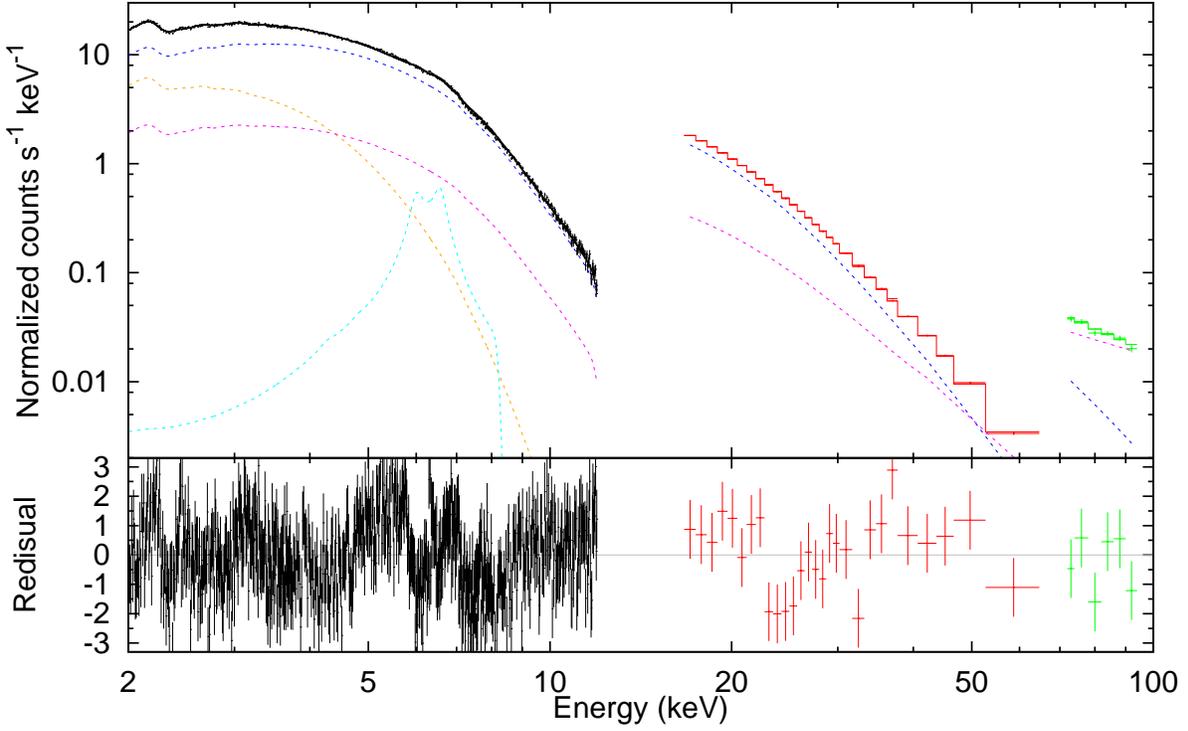}
  \end{center}
  \caption{
Spectral fitting with a relativistic disk reflection model. The orange, blue, magenta, and cyan lines show {\tt diskbb}, {\tt kerrconv}$\times${\tt pexriv}$_\mathrm{th}$, {\tt kerrconv}$\times${\tt pexriv}$_\mathrm{nth}$, and {\tt kerrdisk}, respectively.
  }\label{f6-disklinespect}
\end{figure}

Next, we fitted the spectrum with a relativistic disk reflection model.
In this picture, some photons of the MCD component are directly seen, and others are scattered in a thermal/non-thermal hybrid corona, some of which are reflected at around the inner radius of the rotating accretion disk.
\citet{blu09} fitted the spectrum made from the same Suzaku dataset used in this paper with the relativistic reflection model,
although they did not use GSO data which we took into account.
We used a model {\tt tbabs}$\times$({\tt diskbb}+{\tt kerrdisk}+{\tt kerrconv}$\times$({\tt pexriv}$_\mathrm{th}$ $+${\tt pexriv}$_\mathrm{nth}$))$\times${\tt const}, where
{\tt kerrdisk} is a fluorescent iron line from a relativistically rotating disk \citep{bre06},
{\tt kerrconv} is a convolution with the line shape from the {\tt kerrdisk} model \citep{bre06},
and {\tt pexriv} is a cut off powerlaw spectrum reflected from ionized materials \citep{mag95}.
The index ``th'' shows a ``thermal'' component, and ``nth'' shows a ``non-thermal'' component.
Figure~\ref{f6-disklinespect} and Table~\ref{t6-diskline} show the fitting results.
As a result, we can explain the time-averaged spectrum with the relativistic disk reflection model satisfactorily.

\subsubsection{Partial covering model}
Finally, we fitted the spectrum with a partial covering model.
In this picture, the MCD component and the thermal/non-thermal Compton component are partially covered by the intervening X-ray absorbers.
We used XSTAR Version 2.2.1bn21 \citep{kal04} to model the warm absorbers, 
assuming the solar abundance and the photon index of the ionizing spectrum to be 2.0. 
The temperature, pressure and density of the warm absorbers were assumed to be $10^5$ K, 0.03 dyn cm$^{-2}$, and $10^{12}$ cm$^{-3}$, respectively, following \citet{miy09}. 
We made a grid model by running XSTAR for different values of $\xi$ and $N_H$;
the $\log\xi$ values were from 0 to 5 (erg cm s$^{-1}$), and
the $N_H$ values were from $5\times10^{20}$ to $5\times10^{24}$ (cm$^{-2}$).
The number of steps for $\log \xi$ and  $N_H$ were 20 and 20, respectively, 
thus our grid model has $20\times20$ grid points.
We used a model {\tt tbabs}$\times$(({\tt partcov}$\times$``absorption table'')$\times$({\tt diskbb}+{\tt pexrav}$_\mathrm{th}$ $+${\tt pexrav}$_\mathrm{nth}$)$+${\tt gauss})$\times${\tt const},
where {\tt pexrav} shows an exponentially cut off power law spectrum reflected by neutral material \citep{mag95}.
Figure~\ref{f6-disklinespect} and Table~\ref{t6-diskline} show the fitting results.
 As a result, we can also explain the time-averaged spectrum with the partial covering model satisfactorily.

\subsection{Power spectrum}
Next, we investigated time variations of this object.
First, we conducted timing analysis using the generalized Lomb-Scargle method \citep{zec09}.
Figure \ref{f5-01} shows the power spectral density (PSD) using all events of XIS0 and 3.
We found a quasi-periodic oscillation (QPO) at $1.929\pm0.003$~Hz with a full width of a half maximum (FWHM) of $0.133\pm0.007$~Hz.
The Q value ($=$the central frequency$/$FWHM) is $14.5\pm0.7$, which belongs to type C QPO \citep{cas05}.
The presence of type C QPO, the fact that the light curve does not show characteristic variations as most classes of GRS 1915+105, and the ``broad iron spectral feature'' indicate that this object is in State C, or Class $\chi$ (\citealt{bel00} and references therein).

\subsection{Spectral variation (1) DVF method}\label{sec:DVFmethod}
In order to investigate spectral variation at various timescales,
we applied the ``Difference Variation Function'' method (DVF method) to the data.
DVF method is a spectro-temporal data-analysis method developed by \citet{ino11} for investigating X-ray spectral variations of MCG--6--30--15.
The principle of the method is as follows:
(1)~We determine the time-scale $\Delta T$, and created a light curve with a bin width of $\Delta T$. 
(2)~We define the ``bright spectrum'' and the ``faint spectrum'' from every two adjacent bins, so that the average count of the former is brighter than the latter (Figure \ref{f5-DVFltcrv}). 
(3) We average the bright and faint spectra from all of the samples of the bright and faint spectral pairs. 
For a given $\Delta T$, we have an average bright spectrum and an average faint spectrum. 
(4) We repeat the procedure for different intervals of $\Delta T$.
DVF method has a merit of holding the original photon statistics for spectral fitting even at a short $\Delta T$.

\subsubsection{QPO variations}\label{sec:QPO}
Figure \ref{f5-DVFspect:a} shows the bright spectra, the faint spectra, and the ratio of the spectra to the time-averaged spectra with DVF method at various timescales.
We used the light curve in the all XIS energy band.
In the bottom panels of Figure \ref{f5-DVFspect:a}, the XIS bright/faint spectra look harder/softer in the range of $0.1$~s~$\leq\Delta T\leq0.4$~s, respectively.
In the HXD band, the spectral variation seems flat both for the bright/faint spectra.
In order to evaluate these features quantitatively, we fitted the ratios of the spectra with linear functions.
Specifically, we fitted the lower panels of Figure \ref{f5-DVFspect:a} with $y=a\log_{10}(E\,\mathrm{(keV)})+b$ (XIS) or $y=c$ (HXD).
Figure \ref{f5-fitexample} shows an example of fitting, and 
Figure \ref{f5-DVFslope} shows the $\Delta T$ dependence of $a$ and $c$.
All the fitting results were acceptable (reduced $\chi^2<1.0$).
These figures show that the hardness in XIS and the counts in HXD vary in sync with the total X-ray flux with a peak of $\Delta T\simeq0.2$~s, which corresponds to the QPO frequency ($f=1/(2\Delta T)$).
Therefore, the peak at the frequency in Figure \ref{f5-DVFslope} is associated with the QPO.

In DVF method, variation amplitude ($F_\mathrm{var}$) is defined as
\begin{equation}
F_\mathrm{var}=\frac{a_B-a_F}{a_B+a_F},
\end{equation}
where $a_B$/$a_F$ are counts of the bright/faint phase, respectively \citep{ino11}.
Figure \ref{DVF_PS} shows the time-bin width dependence of $F_\mathrm{var}$.
We can see that $F_\mathrm{var}$ becomes larger with a shorter $\Delta T$, which is mostly due to the Poisson noise.
Assuming that the intrinsic X-ray luminosity is constant ($=c$),
the average count in a bin-width of $\Delta T$ is $c\Delta T$.
When the count follows the Poisson distribution,
the probability that the count is $k$ (non-negative integer) is $P_{c\Delta T}(k)$, where $P_\lambda(x)$ is the Poisson distribution with the average of $\lambda$.
Therefore, the expected variation amplitude is calculated to be
\begin{equation}
F_\mathrm{var,poisson}(\Delta T)=\sum_k\sum_{k^\prime}\frac{|k-k^\prime|}{k+k^\prime}P_{c\Delta T}(k)P_{c\Delta T}(k^\prime).
\end{equation}
The gray lines in Figure \ref{DVF_PS} show $F_\mathrm{var,poisson}(\Delta T)$.
Subtracting the Poisson noise, we see no excess at $\Delta T < 0.06$~s.
Namely, in this time domain, we did not detect any intrinsic variation.
We see clear excess at $0.06$~s$\leq\Delta T\leq0.4$~s,
which is the QPO variation discussed above.
At longer timescales, we recognize some excess, which shows intrinsic variation of this object.

\subsubsection{Variations of the continuum} \label{sec:MCD}
The continuum spectrum of GRS 1915$+$105 is mainly explained by two components; one is a MCD component below several keV, and the other is an inverse Compton component.
The latter is dominant in the HXD region, thus the flatness of the high/low spectral ratios in the HXD region for a wide range of different timescales (Figure \ref{f5-DVFspect:a})
is understood as variation of the inverse Compton component without changing its spectral shape.
If the continuum spectrum is comprised of a variable component and a non-variable component,
the variation should be small if the non-variable component is dominant, and the opposite if the variable component is dominant.
Thus, assuming that the variable component has no change in spectral shape, 
we can extract the non-variable component from the ratio of the bright/faint spectrum.
Figure \ref{f5-invariant} shows the non-variable component thus extracted.
This spectral component can be fitted with a {\tt diskbb} model convolved with a foreground absorption (Table \ref{t5-invariant}).
We thus found the buried non-variable MCD component with $T_\mathrm{in}=830$~eV.

\subsubsection{Variations of the iron K-feature}\label{sec:ironvariation}
Figure \ref{f6-DVF_bf1} shows energy dependences of the variation amplitudes calculated using the DVF method in the XIS energy band for each timescale.
In order to examine the possible existence of any spectral structure,
we fitted the variation amplitudes between 5~keV and 8~keV (the red areas in Figure \ref{f6-DVF_bf1}c--$\zeta$) with a constant model, 
and found that the fitting results were acceptable (reduced $\chi^2<1.0$) at all $\Delta T$.
This indicates that the variation amplitude has {\it no} structure around the iron K-feature at any investigated timescale.
This result is totally different from AGN spectral variations, 
which show a significant drop in their variation amplitude around the iron K-energy band using the similar method (e.g.\,\citealt{fab02a,mat03,miz15}).

What does it mean that the variation amplitude has no spectral structure? 
In the procedure of our calculation, we separated the entire X-ray spectrum into the bright/faint spectra, and studied the variation amplitude.
If the variation of the iron structure was a slight one, but not completely correlated with that of the continuum at a certain timescale,
we should see some residual structure in the variation amplitude.
For example, if the iron feature is less variable than the continuum, the bright spectrum has a weaker iron feature, and the faint one has a stronger one.
This is the reason that we commonly see the dip-like iron features in the variation amplitudes of AGNs (e.g.\,\citealt{fab02a,mat03,miz15}).
Suppose that the continuum and the iron structure vary completely in sync (i.e.\,no change in the spectral shape) in all the investigated time scales, then we would see no structure at the iron energy band in the variation amplitude.
On the contrary, if the iron structure and the continuum vary completely independently, no iron structure would be seen in the variation amplitude either,
bacause the iron features disappear as the bright/faint phase is defined solely by the continuum flux.
Therefore, in order to explain the ``flatness'' of the variation amplitude at the iron K-energy band, there can be two interpretations:
One is that the broad iron line profile varies exactly in sync with the continuum for all the examined timescales.
The other is that the broad iron line profile and the continuum vary completely independently.

\subsection{Spectral variation (2) time-sliced spectra}\label{sec:timeslice}

To examine whether the iron feature is variable or not,
we investigated time-sliced spectra.
First, we sliced the spectrum at the Suzaku orbital period ($\simeq5760$~s) and created 22 time-sliced spectra.
Second, we fitted the continuum at 2--5~keV and 8--10~keV with a power law, avoiding the iron line band, and calculated the normalized flux of the continuum.
Third, we calculated the ratio of the spectrum to the continuum and extracted the iron line profiles.
Figure \ref{f6-orbit1} shows the iron line profiles of the time-sliced spectra thus calculated.
We can see that the iron line profiles {\em do} vary.
Figure \ref{f6-EW} shows variation of the spectral parameters to characterize the iron line profiles (either equivalent width (EW) or partial covering fraction) 
and the continuum flux.
We found that both the iron spectral feature and the continuum flux are variable at timescales of hours regardless of the spectral models.
We calculated correlation coefficients between the continuum flux and the EW/partial covering fraction, and found that they are $-0.16$/0.14, respectively, which quantitatively shows that there are no correlation.
We had proposed two interpretations in the end of \S\ref{sec:ironvariation}, and we concluded that the broad iron line profile and the continuum vary completely independently.

\section{Discussion}

\subsection{Origin of the spectral variation}\label{sec4-1}
We have analyzed the Suzaku P-sum mode data of GRS 1915$+$105, and
 found two major results;
One is that spectral variations of GRS 1915$+$105 around the iron K-energy band are featureless at all the timescales of $0.06-63000$~s with DVF method (Result I),
and the other is that the iron line flux and the continuum flux are independently variable at timescales of hours (Result II).

In the ``relativistic disk reflection'' model, most variation of the iron feature is caused by the change of the source height, or variation of the distance between the compact ``lamp-post'' corona and the accretion disk \citep{min04}.
The geometrical configuration is the same for AGNs and BHBs, and its physical size is scaled with the BH mass.
Hence, in BHBs, similar variability to AGNs is expected in a few milliseconds \citep{fab03,min03}.
Result I shows that, if the iron-line emitting region is compact, the upper limit of the size is $c\Delta T\simeq0.06c\simeq400\,r_s$, where $c$ is the light velocity.
This constraint is not so strong because the size of the iron-line emitting region is expected to be an order of Schwarzschild radius \citep{mil07}.
On the other hand, because relativistic disk lines are expected to arise close to the BH itself,
variations of the iron feature are expected to follow variations in the hard X-ray continuum \citep{mil07}.
However, this expectation contradicts Results II.
In order to explain the variability of the iron line profile in this model,
physical parameters of the accretion disk, presumably a source height or emissivity, need to vary at timescales of hours.
We have currently no established physical explanation for this variability.

In the ``partial covering'' model,
the outflow absorbers can vary independently from the continuum flux.
Assuming that the length of the absorbers, $\Delta r$, is similar to the distance from the source \citep{ued09},
the distance from the source, $r$, is calculated to be 
\begin{eqnarray}
r&=&\frac{L}{\xi(nr)}\nonumber\\
&\sim& \frac{L}{\xi(n\Delta r)}\nonumber\\ 
&=&\frac{L}{\xi N_H} \nonumber\\
&\sim&10^{12}\,\mathrm{cm}\,\,\,\sim10^{5-6}\,r_s, \label{eq:VPC}
\end{eqnarray}
where the intrinsic luminosity $L=1.9\times10^{38}$~erg s$^{-1}$, the ionization degree of the absorbers $\xi=10^{1.9}$, and the column density of the absorbers $N_H=7.2\times10^{23}$~cm$^{-2}$ (Table \ref{t6-VPC}).
This $r$ value is consistent with the previous results (e.g.~\citealt{ued09}),
and shows that the absorbers exist at around the outer radius of the disk, $\sim10^5\,r_s$ \citep{rem06}.
The variation timescale of the absorbers is calculated to be about $\sim10$~ks assuming a Kepler motion.
This timescale can naturally explain Result II.
As shown in \S\ref{sec:timeslice}, the continuum flux and the partial covering fraction vary independently, which does not contradict Result I (see \S\ref{sec:ironvariation}).
In this manner, the partial covering model can naturally explain the observed spectral variation.

\subsection{Difference of outflows in AGN and BHB}
Next, we need to understand the difference of the spectral variations of AGN and BHB in the framework of the partial covering model.
In AGN, assuming that the continuum flux can be regarded as constant, the spectral variation can be explained (e.g.\,\citealt{miy12}).
This assumption implies that the variation timescale of partial absorbers are much shorter than that of intrinsic luminosity.
In fact, \citet{miz14,miz15} proposed that the variation timescale of partial absorbers in 1H0707--495 is hours, while that of intrinsic luminosity is over days.
They also suggested that the partial absorbers are UV-line driven disk winds located at $\sim500\,r_s$.
On the other hand, in GRS 1915$+$105, the location of the partial absorbers is calculated to be about $\sim10^{5-6}\,r_s$ (see \S\ref{sec4-1}).
Namely, the variation timescale of the partial absorbers normalized by the BH mass in GRS 1915$+$105 is much longer than that of 1H0707--495.

In AGN, UV photons from an accretion disk are considered to drive a ``UV-line driven'' outflow (e.g.\,\citealt{pro00}).
The outflow is considered to be launched where the UV photons are the strongest, thus the outflow absorbers are considered to be at the middle of the disk \citep{nom13}.
In BHB, in contrast, the accretion disk is so hot that absorbers are fully-ionized at the middle of the disk, thus the ``UV-line driven'' outflow does not take place.
Instead, a ``thermally-driven'' outflow where the thermal energy is larger than the binding energy is considered to be dominant \citep{pro02}.
In the ``thermally-driven'' outflow, absorbers are at the outer skirt of the disk, where the binding energy is relatively weak \citep{beg83}.
Therefore, we argue that the difference of locations of the partial absorbers between AGN and BHB is a natural consequence of the difference of the outflow driving mechanisms (Figure \ref{f:outflow}).

In Figure~\ref{f6-EW}, we can see that variation of the covering fraction is $\sim10$~\%, thus
the number of the absorbers in the line-of-sight is estimated to be about $(10\%)^{-2}=100$,
assuming that the variation is due to fluctuation of the partial absorbers.
The total covering fraction is about 0.4, so a cross-section of the single absorber is considered to be about $1/400$ of the X-ray source.
If the size of the X-ray source is $\simeq20\,r_s$, the size of the absorber is $\simeq1\,r_s$.

\section{Conclusion}
We have analyzed the Suzaku data of GRS 1915$+$105 taken with XIS P-sum mode,
where the CCD energy resolution and time resolution of 7.8~ms are both available.
Our main conclusions are as follows:
\begin{enumerate}

\item The time-averaged ``broad iron spectral feature'' of GRS 1915+105 can be fitted by both the relativistic disk reflection model and the partial covering model.
\item The spectral variation of GRS 1915$+$105 at iron K-energy band is found to be featureless at any timescales from 0.06~s to 63000~s.
This is totally different from spectral variations of AGN, where variation amplitude significantly drops at the iron K-energy band at a timescale of $\sim10^4$~s.
\item The iron line feature and the intrinsic luminosity are found to be independently variable at timescales of hours in GRS 1915$+$105.
\item The partial covering model can explain the difference of the spectral variations between AGN and BHB naturally and comprehensively, where locations of the partial absorbers normalized by the Schwarzschild radius are different between AGN and BHB.
\item The differences between AGN and BHB is considered to be due to difference of the outflow types; the UV-line driven wind (AGN) or the thermally-driven wind (BHB).
\end{enumerate}

\bigskip
\bigskip
The authors would like to thank S.~Nakahira for his fruitful comment, H.~Takahashi for his advice about the HXD instrument, and the Suzaku XIS team for their calibration of P-sum mode.
This work has made use of public Suzaku data obtained through the Data ARchives and Transmission System (DARTS), 
provided by the Institute of Space and Astronautical Science (ISAS) at Japan Aerospace Exploration Agency (JAXA).
For data reduction, we used software provided by HEASARC at NASA/Goddard Space Flight Center (GSFC).
M.\,M. is financially supported by the Japan Society for the Promotion of Science (JSPS) KAKENHI Grant Number 15J07567.
M.\,T. is financially supported by MEXT/JSPS KAKENHI Grant Numbers 24105007 and 15H03642.

\begin{table}[!ht]
 \begin{center}
  \caption{Parameters of phenomenological spectral fitting.
  }\label{t5-cutoffpl}
    \begin{threeparttable}
  \begin{tabular}{lll}
   \hline
\hline
{\tt tbabs}& $N_H$ ($\times10^{22}$~cm$^{-2}$)& $6.47_{-0.07}^{+0.11}$\\
\hline
{\tt diskbb}\tnote{$\ast$1}& $T_\mathrm{in}$~(eV) & 825 (fix) \\
& Norm.\tnote{$\ast$2} & 620 (fix)\tnote{$\ast$3}\\
\hline
{\tt cutoffpl} & $\Gamma$ & $1.28_{-0.13}^{+0.11}$ \\
& $E_c$~(keV) & $15.6_{-1.4}^{+2.1}$ \\
& Norm.\tnote{$\ast$4} & $1.2_{-0.4}^{+0.5}$\\
\hline
{\tt pegpwrlw} & $\Gamma$ &$1.9_{-0.6}^{+0.2}$ \\
& Norm.\tnote{$\ast$5} & $4.5_{-2.7}^{+1.7} \,\times10^3$\\
\hline
 {\tt gauss} & $E_{l}$~(keV) & $6.00\pm0.05$ \\
& $\sigma$~(keV) & $1.13\pm0.06$\\
& Norm.\tnote{$\ast$6}& $5.3_{-0.3}^{+0.5}\,\times10^{-2}$ \\
\hline
\multicolumn{2}{l}{Reduced $\chi^2$}& 1.42 (d.o.f.$=691$) \\
\hline
  \end{tabular}
\begin{tablenotes}\footnotesize
    \item[$\ast$1] The parameters of {\tt diskbb} are fixed at the values of Figure \ref{t5-invariant}. See \S\ref{sec:MCD} for details.
    \item[$\ast$2] $\left(\frac{r_\mathrm{in}/\mathrm{km}}{D/10\mathrm{kpc}}\right)^2\cos i$, where $r_\mathrm{in}$ is the inner radius, $D$ is the distance, and $i$ is the inclination angle.
    \item[$\ast$3] This value is 1/1.026 times of that in Table \ref{t5-invariant} because the normalization factor of XIS1 is 1.026 against XIS0.
    \item[$\ast$4] Photons~keV~$^{-1}$~cm$^{-2}$~s$^{-1}$ at 1~keV
    \item[$\ast$5] Flux $\times10^{-12}$~erg~cm$^{-2}$~s$^{-1}$ between 2--100~keV
    \item[$\ast$6] Photons~cm$^{-2}$~s$^{-1}$ in the line.
    \end{tablenotes}
  \end{threeparttable}
  \end{center}
\end{table}
\begin{table}[!ht]
 \begin{center}
  \caption{Parameters of relativistic disk reflection spectral fitting.
  }\label{t6-diskline}
    \begin{threeparttable}
  \begin{tabular}{lll}
   \hline
\hline
{\tt tbabs}& $N_H$ ($\times10^{22}$~cm$^{-2}$)& $6.36\pm0.01 $\\
\hline
{\tt diskbb}& $T_\mathrm{in}$~(eV) & 825 (fix) \\
& Norm.\tnote{$\ast$1} & 434 (fix)\\
\hline
{\tt pexriv}$_\mathrm{th}$ & $\Gamma$ & $1.59_{-0.03}^{+0.05} $ \\
&  $E_c$~(keV) & $17.7_{-0.4}^{+0.8} $ \\
& $rel_\mathrm{refl}$ & $0.57_{-0.01}^{+0.02}$ \\
& $\cos i$ & 0.342 (fix)\\
& Norm.\tnote{$\ast$2} & $1.74_{-0.01}^{+0.12} $\\
\hline
{\tt pexriv}$_\mathrm{nth}$\tnote{$\ast$3} &  $\Gamma$ &$1.91_{-0.15}^{+0.01} $ \\
&  $E_c$~(keV) & 0 (fix)\tnote{$\ast$4} \\
& Norm.\tnote{$\ast$3} & $0.43\pm0.01$\\
\hline
{\tt kerrdisk} & $E_l$~(keV) & $6.30^{+0.01}$\\
& emissivity index & $1.58_{-0.03}^{+0.05}$\\
& $a$ & $0.9955_{-0.0001}^{+0.0010}$\\
& Norm.\tnote{$\ast$5} & $1.30_{-0.08}^{+0.05}\,\times10^{-2}$ \\
\hline
\multicolumn{2}{l}{Reduced $\chi^2$}& 1.65 (d.o.f.$=689$) \\
\hline
  \end{tabular}
\begin{tablenotes}\footnotesize
    \item[$\ast$1] $\left(\frac{r_\mathrm{in}/\mathrm{km}}{D/10\mathrm{kpc}}\right)^2\cos i$, where $r_\mathrm{in}$ is the inner radius, $D$ is the distance, and $i$ the an inclination angle.
    \item[$\ast$2] Photons~keV~$^{-1}$~cm$^{-2}$~s$^{-1}$ at 1~keV without reflection
    \item[$\ast$3] Parameters are not tabulated if they are the same with those of {\tt pexriv}$_\mathrm{th}$
    \item[$\ast$4] This means no cutoff.
    \item[$\ast$5] Photons~cm$^{-2}$~s$^{-1}$ in a line.
    \item[$\ast$6] Parameters of {\tt kerrconv} are the same as those of {\tt kerrdisk}.
    \end{tablenotes}
  \end{threeparttable}
  \end{center}
\end{table}

\begin{table}[!ht]
 \begin{center}
  \caption{Parameters of partial covering spectral fitting.
  }\label{t6-VPC}
    \begin{threeparttable}
  \begin{tabular}{lll}
   \hline
\hline
{\tt tbabs}& $N_H$ ($\times10^{22}$~cm$^{-2}$)& $6.90\pm0.04$\\
\hline
{\tt diskbb}& $T_\mathrm{in}$~(eV) & 825 (fix) \\
& Norm.\tnote{$\ast$1} & 620 (fix)\\
\hline
{\tt pexrav$_\mathrm{th}$} & $\Gamma$ & $2.00_{-0.04}^{+0.03}$ \\
&  $E_c$~(keV) & $28.8\pm1.2$ \\
& $rel_\mathrm{refl}$ & $<0.072$ \\
& $\cos i$ & 0.342 (fix)\\
& Norm.\tnote{$\ast$2} & $6.5_{-1.3}^{+0.7}$\\
\hline
{\tt pexrav$_\mathrm{nth}$}\tnote{$\ast$3}  & $\Gamma$ &$1.88\pm0.02$ \\
&  $E_c$~(keV) & 0 (fix)\tnote{$\ast$4} \\
& Norm.\tnote{$\ast$3} & $0.34\pm0.1$\\
\hline
Absorbers & $N_H$~($\times10^{22}$~cm$^{-2}$) & $7.2_{-0.5}^{+0.2}\,\times10^{1}$ \\
&$\log \xi$& $1.90_{-0.28}^{+0.04}$\\
&Covering factor& $0.42_{-0.02}^{+0.01}$\\
&Redshift& 0 (fix)\\
\hline
{\tt gauss} & $E_{l}$~(keV) & $6.51\pm0.03$ \\
& $\sigma$~(keV) & $0.2$ (fix)\\
& Norm.\tnote{$\ast$5}& $3.2\pm0.4\,\times10^{-3}$ \\
\hline
\multicolumn{2}{l}{Reduced $\chi^2$}& 1.44 (d.o.f.$=688$) \\
\hline
  \end{tabular}
\begin{tablenotes}\footnotesize
    \item[$\ast$1] $\left(\frac{r_\mathrm{in}/\mathrm{km}}{D/10\mathrm{kpc}}\right)^2\cos i$ where $r_\mathrm{in}$ is an inner radius, $D$ is a distance, and $i$ is an inclination angle.
    \item[$\ast$2] Photons~keV~$^{-1}$~cm$^{-2}$~s$^{-1}$ at 1~keV without reflection
    \item[$\ast$3] Unshown parameters are the same as those of {\tt pexriv}$_\mathrm{th}$
    \item[$\ast$4] This means no cutoff.
    \item[$\ast$5] Photons~cm$^{-2}$~s$^{-1}$ in a line.
    \end{tablenotes}
  \end{threeparttable}
  \end{center}
\end{table}
\begin{table}[!ht]
 \begin{center}
  \caption{Fitting parameter of the non-variable component
  }\label{t5-invariant}
    \begin{threeparttable}
  \begin{tabular}{lll}
   \hline
\hline
{\tt TBabs}& $N_H$ ($\times10^{22}$~cm$^{-2}$)& 6.52 (fix)\tnote{$\ast$2}\\
\hline
{\tt diskbb} & $T_\mathrm{in}$ & $8.3_{-0.9}^{+1.1} \times 10^2$~eV\\
& Normalization\tnote{$\ast$3} & $636_{-6}^{+7}$ \\
\hline
  \end{tabular}
\begin{tablenotes}\footnotesize
    \item[$\ast$1] Errors are quoted at the statistical 68\% level.
    \item[$\ast$2] $N_H$ is fixed at the value of Figure \ref{t5-cutoffpl}. 
    \item[$\ast$3] $\left(\frac{r_\mathrm{in}/\mathrm{km}}{D/10\mathrm{kpc}}\right)^2\cos i$ where $r_\mathrm{in}$ is an inner radius, $D$ is a distance, and $i$ is an inclination.
    \end{tablenotes}
  \end{threeparttable}
  \end{center}
\end{table}


\begin{figure}[htbp]
  \begin{center}
    \includegraphics[width=100mm,angle=270]{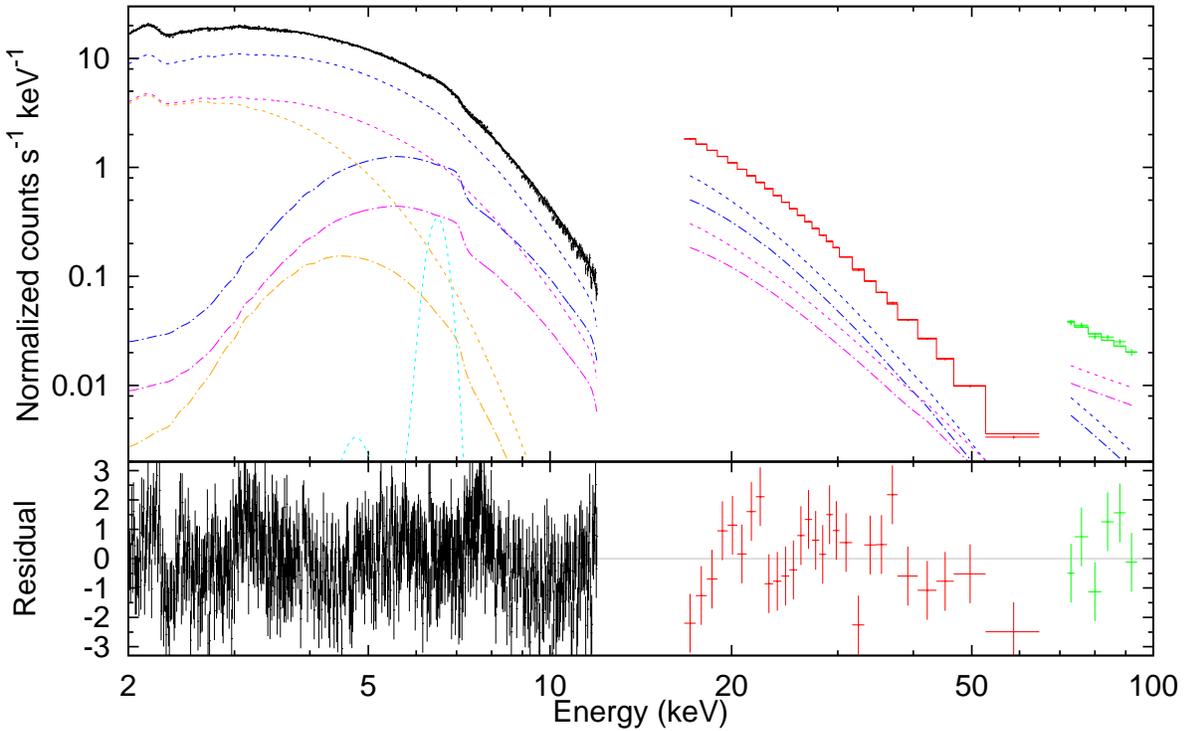}
  \end{center}
  \caption{Spectral fitting with a partial covering model. The orange, blue, magenta, and cyan lines show {\tt diskbb}, {\tt cutoffpl}, {\tt pegpwrlw}, and {\tt gauss}, respectively.
  }\label{f6-VPCspect}
\end{figure}

\begin{figure}[htbp]
  \begin{center}
    \includegraphics[width=100mm,angle=270]{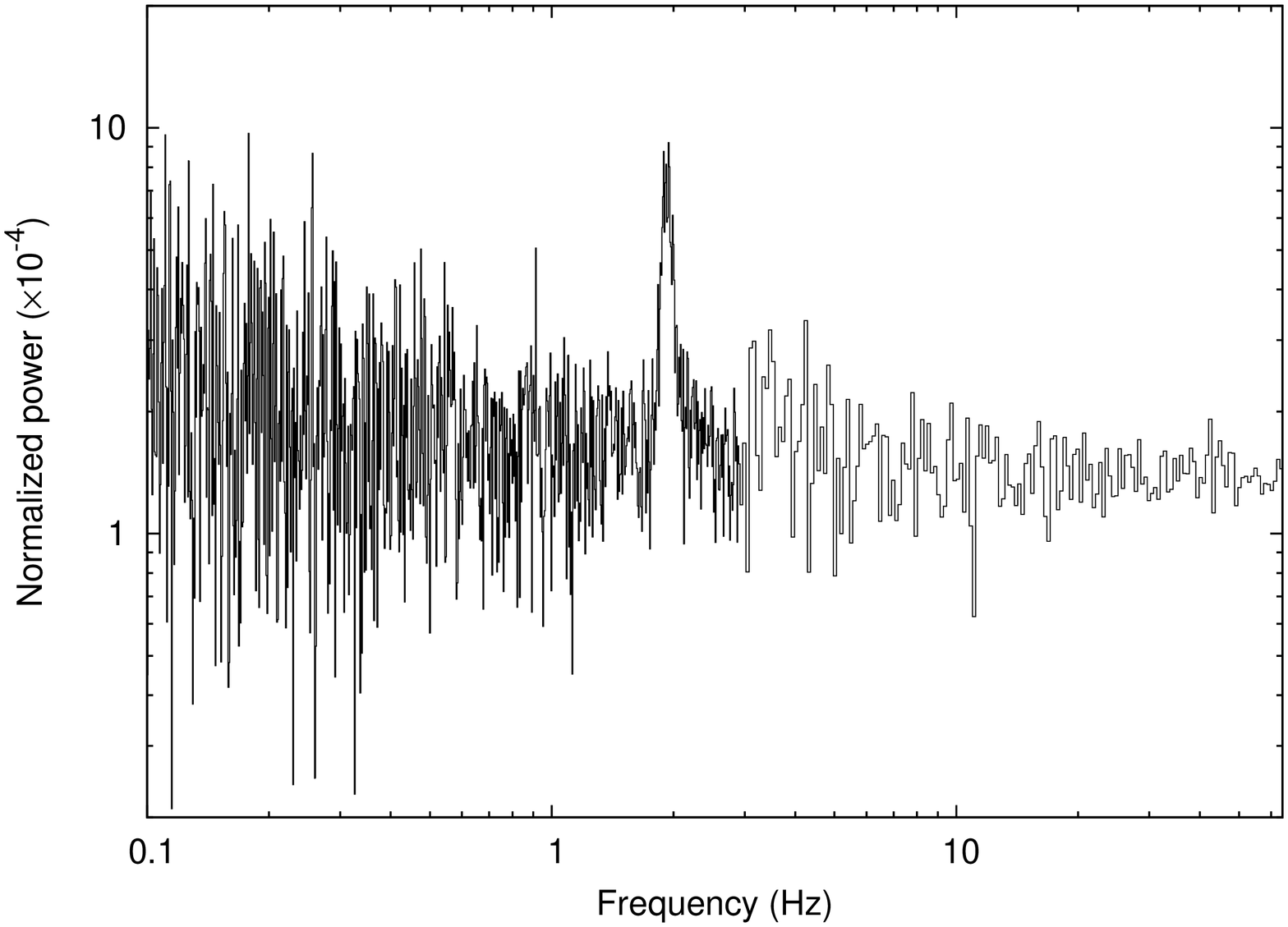}\\
    \includegraphics[width=100mm,angle=270]{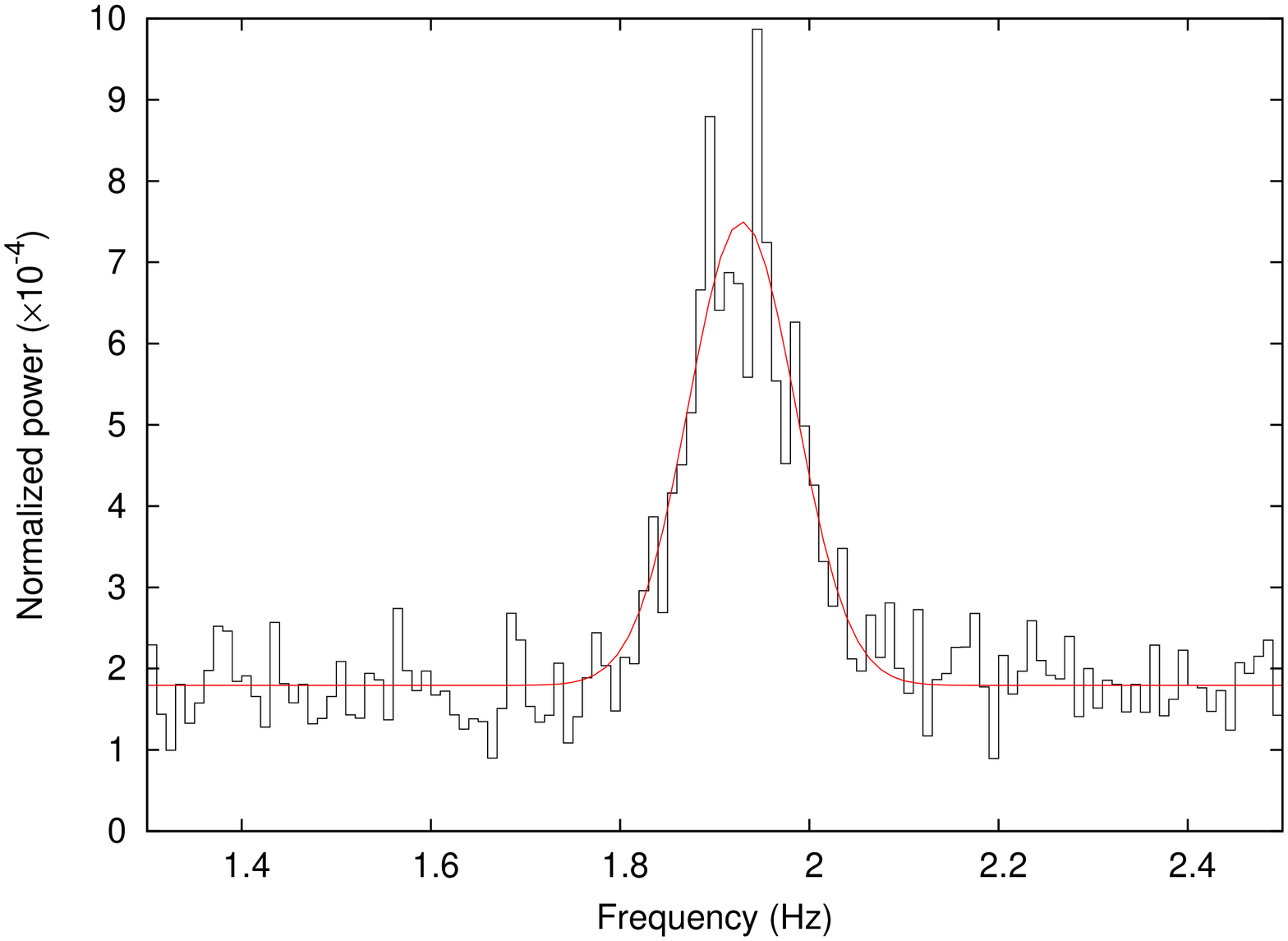}
    \end{center}
  \caption{
   (Upper panel) PSD of GRS 1915$+$105. (Lower panel) Expansion of a part of the PSD, indicating a QPO at 1.93~Hz. The red line shows the gaussian fitting.
 }\label{f5-01}
\end{figure}

\begin{figure}[htbp]
  \begin{center}
    \includegraphics[width=120mm,angle=270]{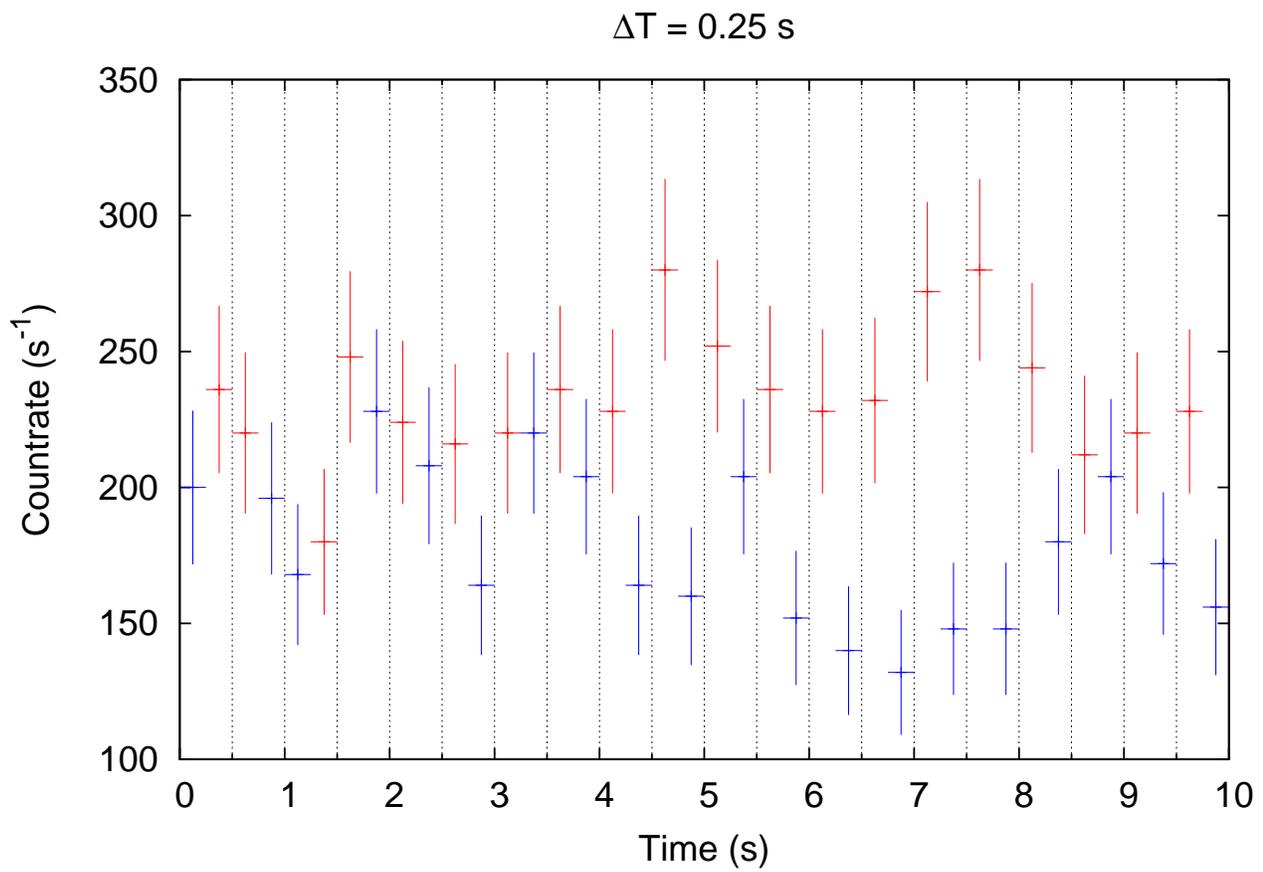}
    \end{center}
  \caption{The P-sum light curve with DVF method ($\Delta T=$0.25~s). The red bins show the bright phase, and the blue bins show the faint phase.
 }\label{f5-DVFltcrv}
\end{figure}

\begin{figure*}
\centering
\subfigure{
        \resizebox{5.5cm}{!}{\includegraphics[angle=270]{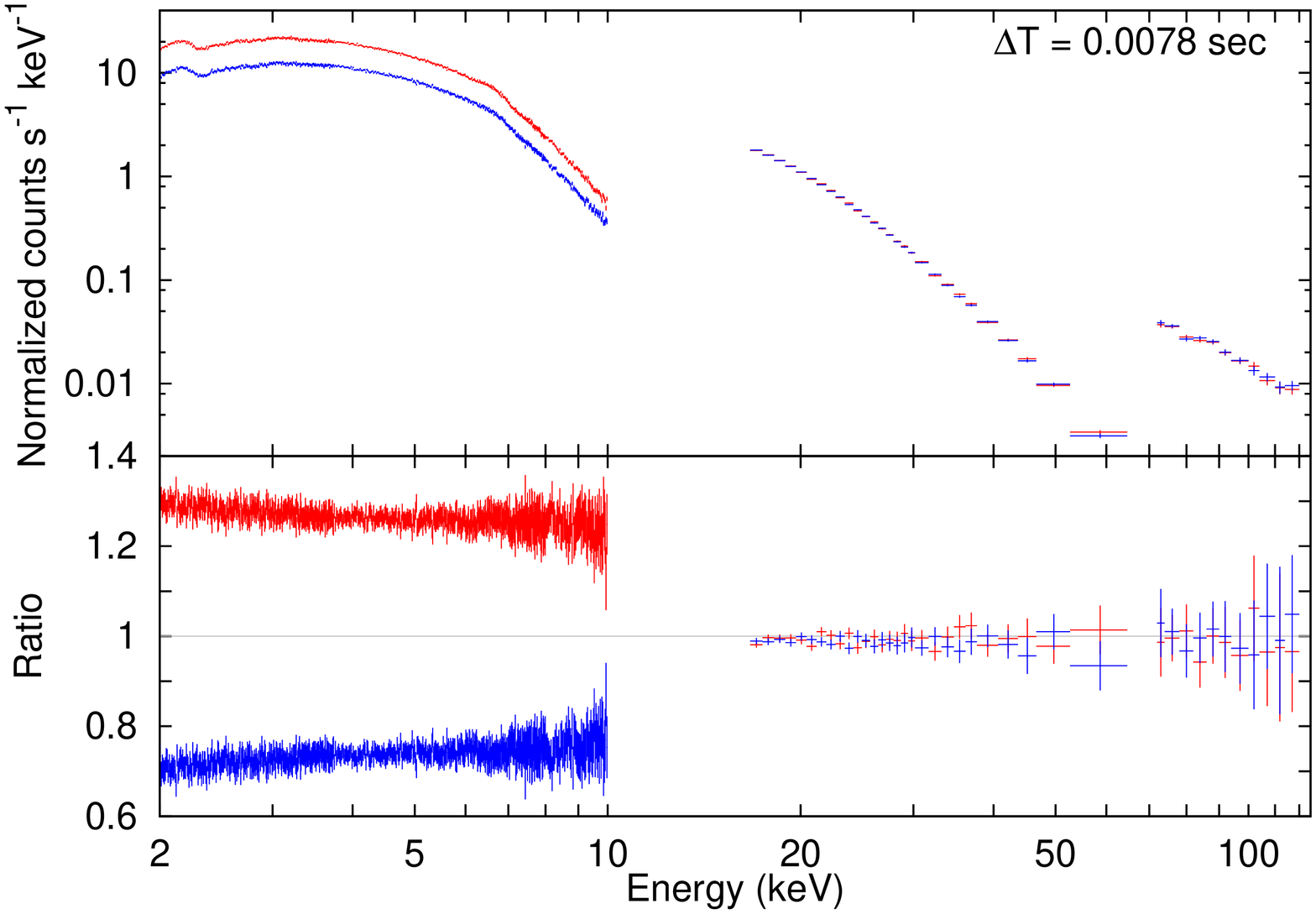}}
        \resizebox{5.5cm}{!}{\includegraphics[angle=270]{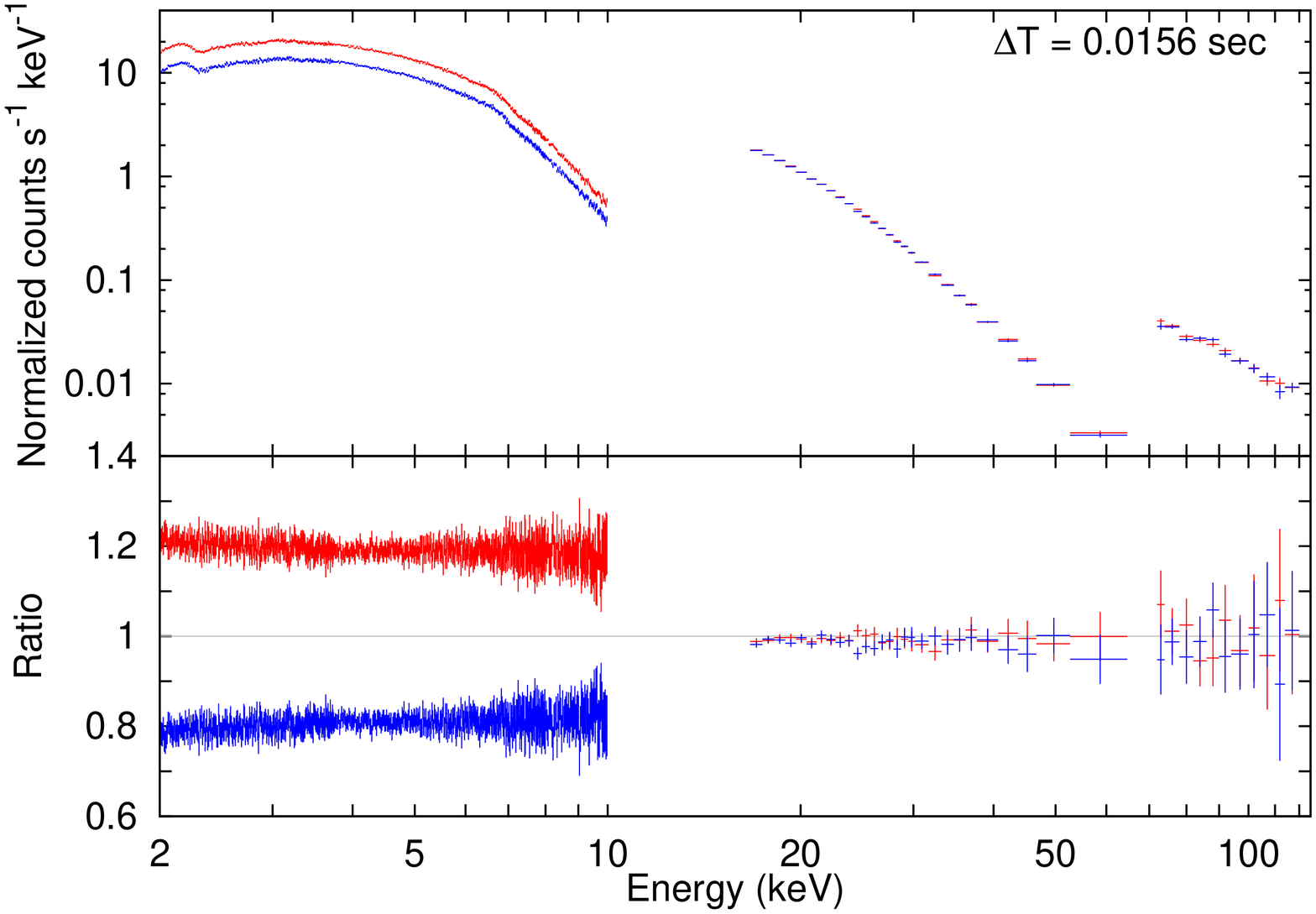}}
        \resizebox{5.5cm}{!}{\includegraphics[angle=270]{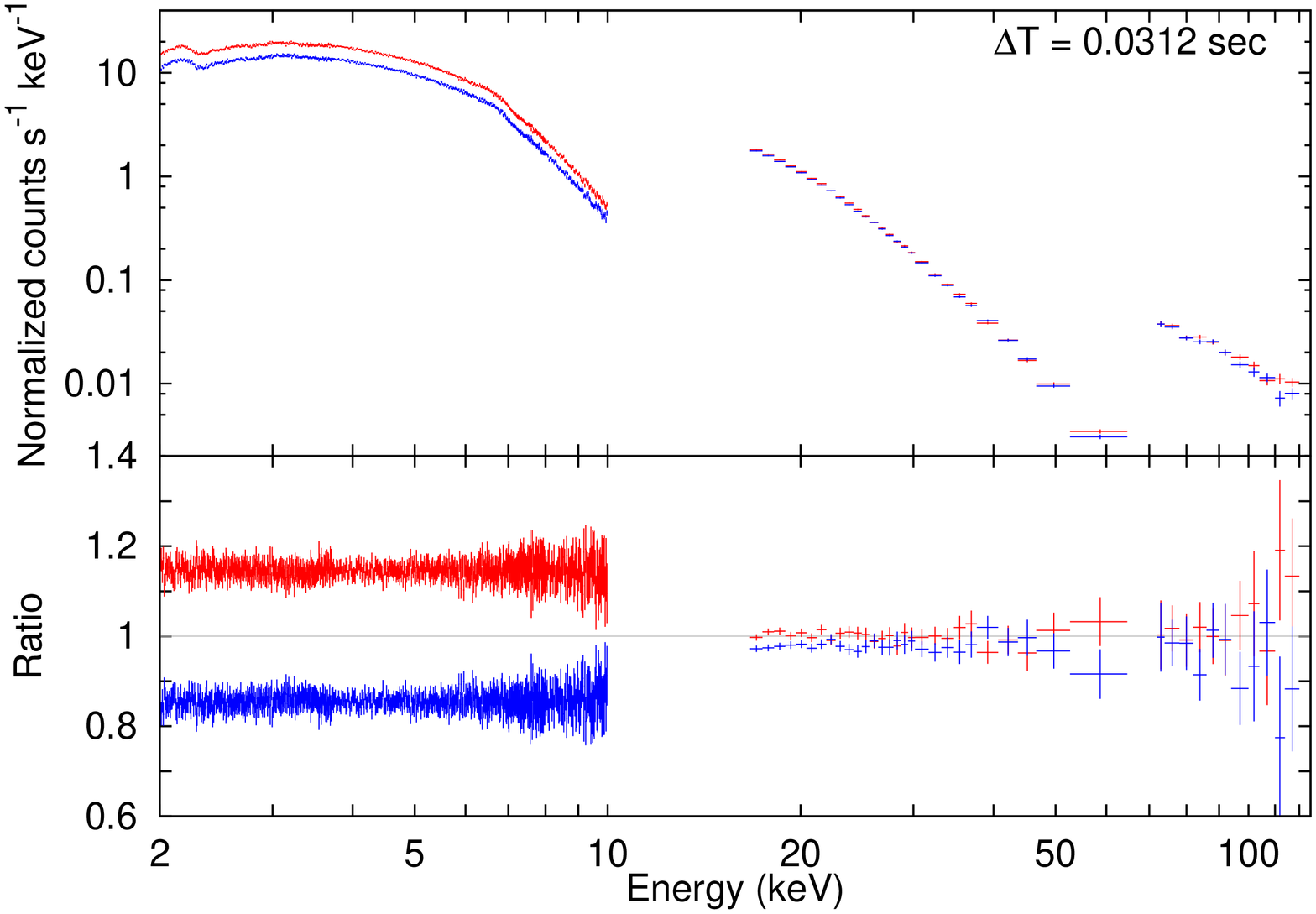}}
}
\subfigure{
        \resizebox{5.5cm}{!}{\includegraphics[angle=270]{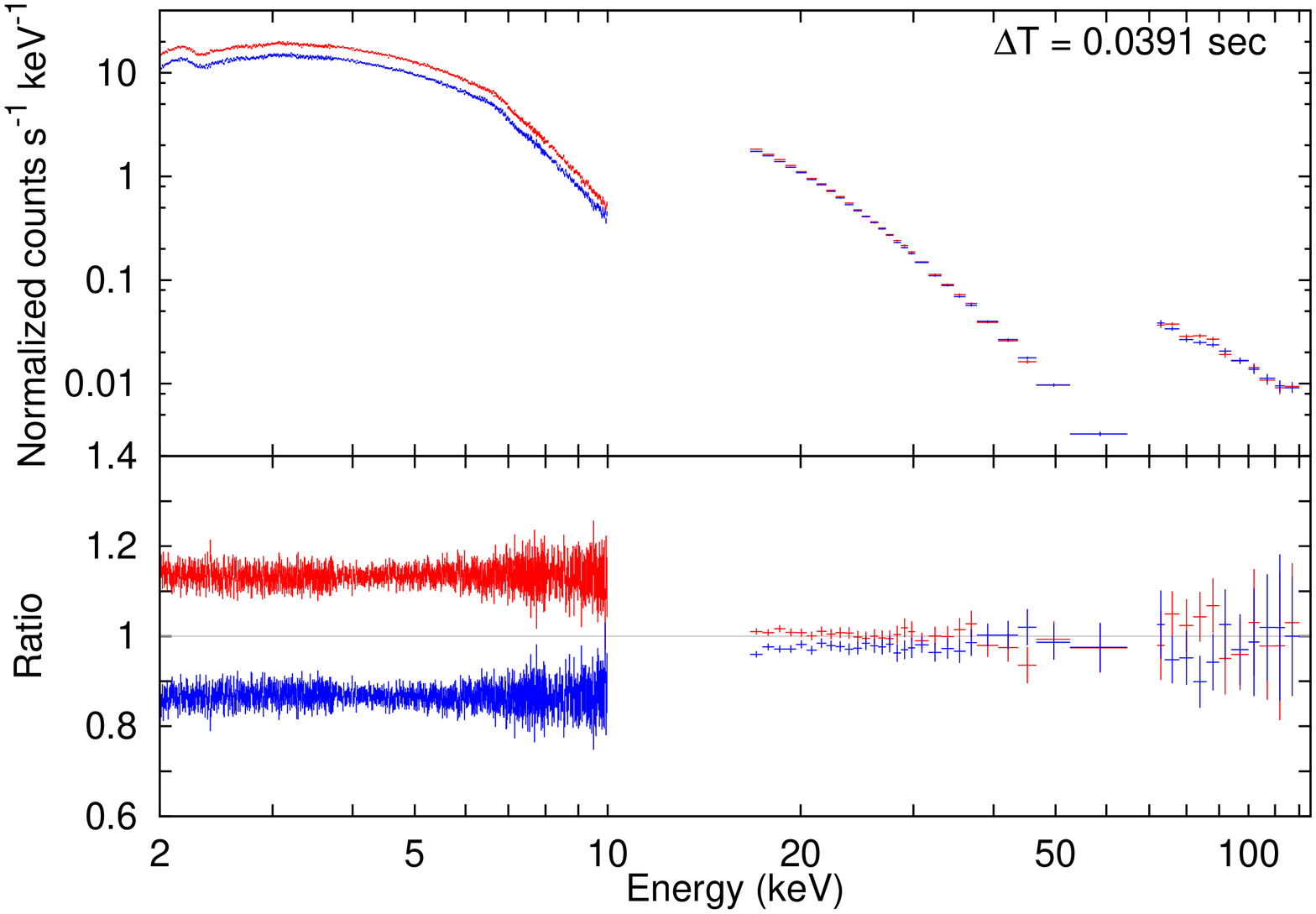}}
        \resizebox{5.5cm}{!}{\includegraphics[angle=270]{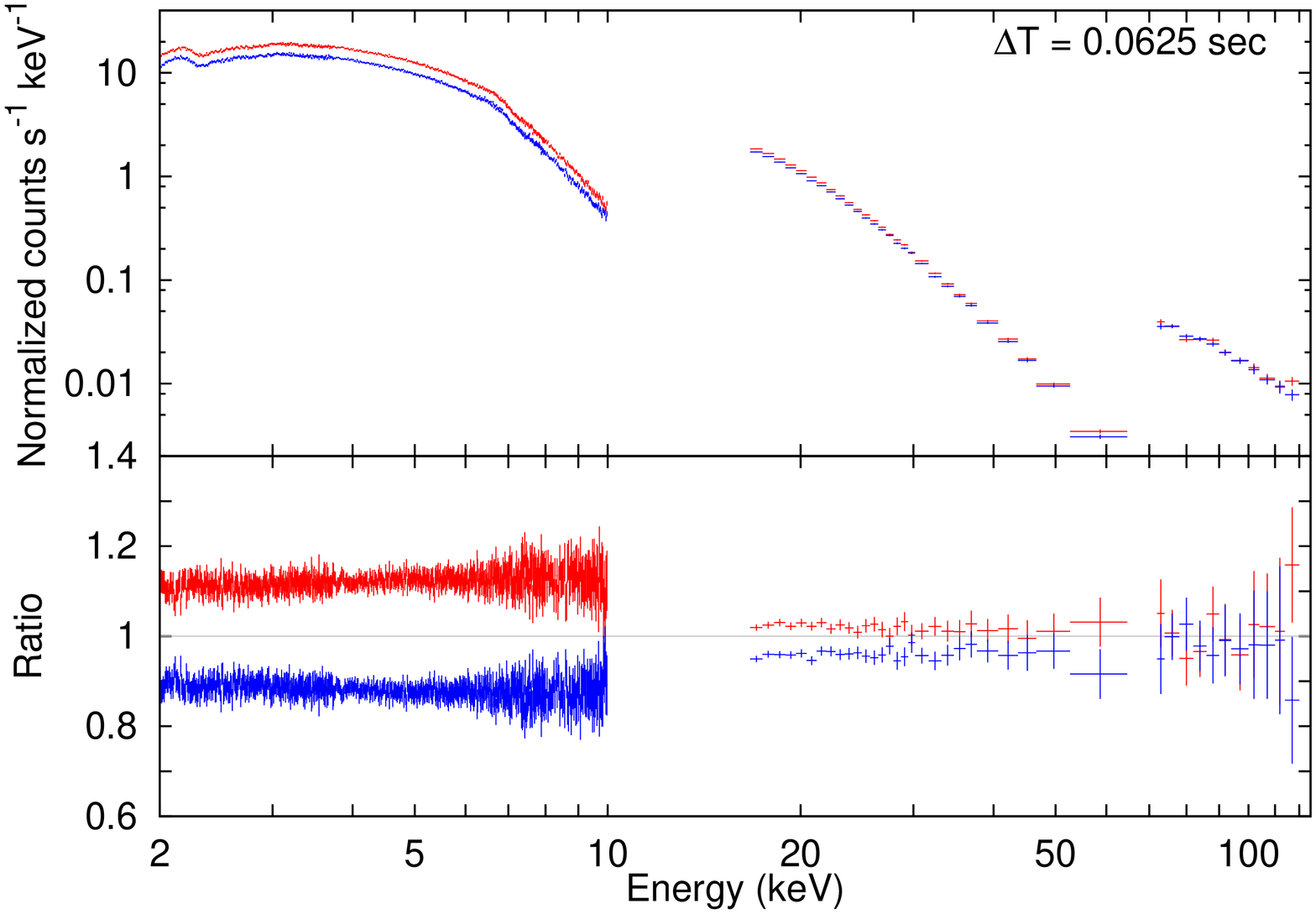}}
        \resizebox{5.5cm}{!}{\includegraphics[angle=270]{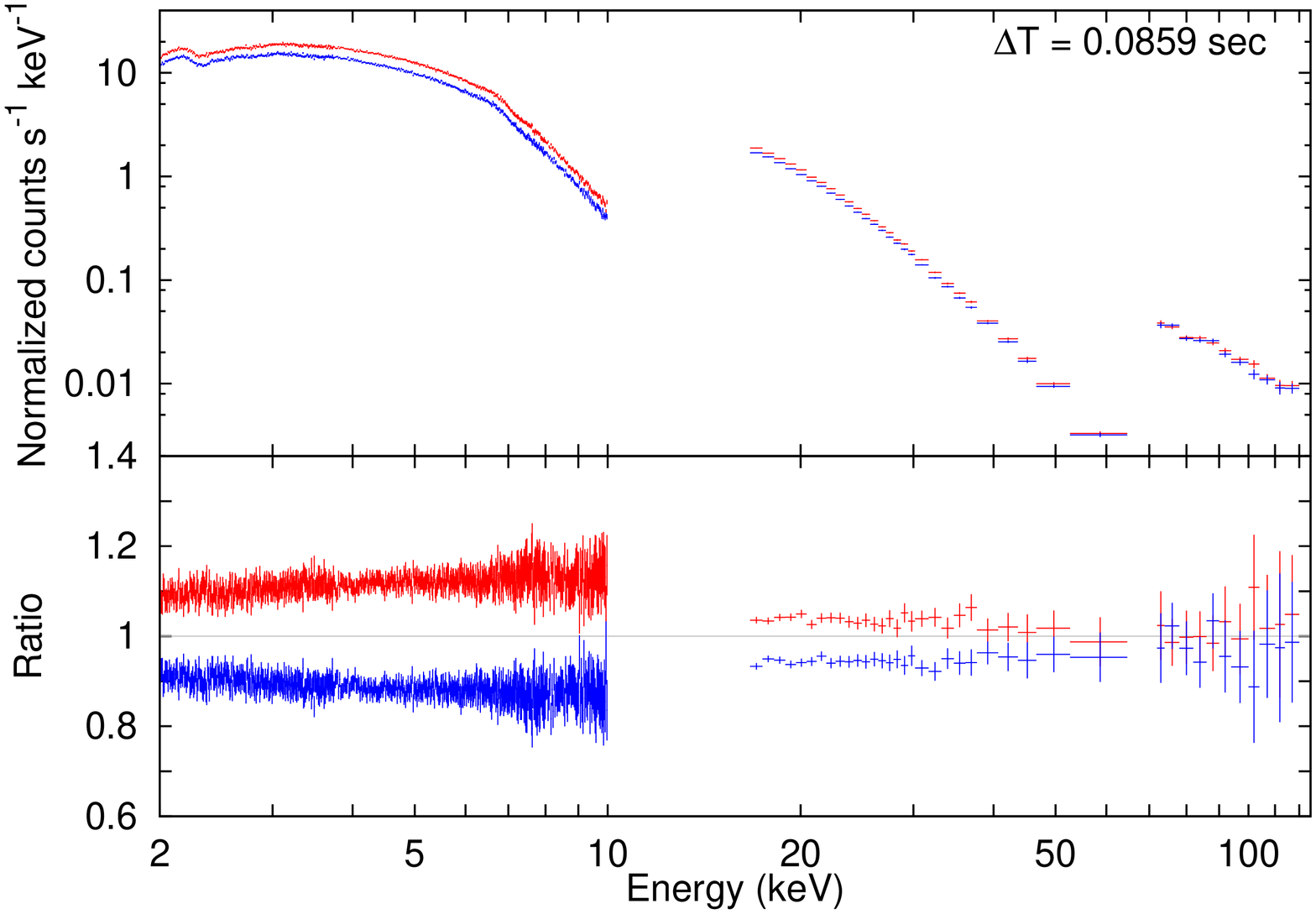}}
}
\subfigure{
        \resizebox{5.5cm}{!}{\includegraphics[angle=270]{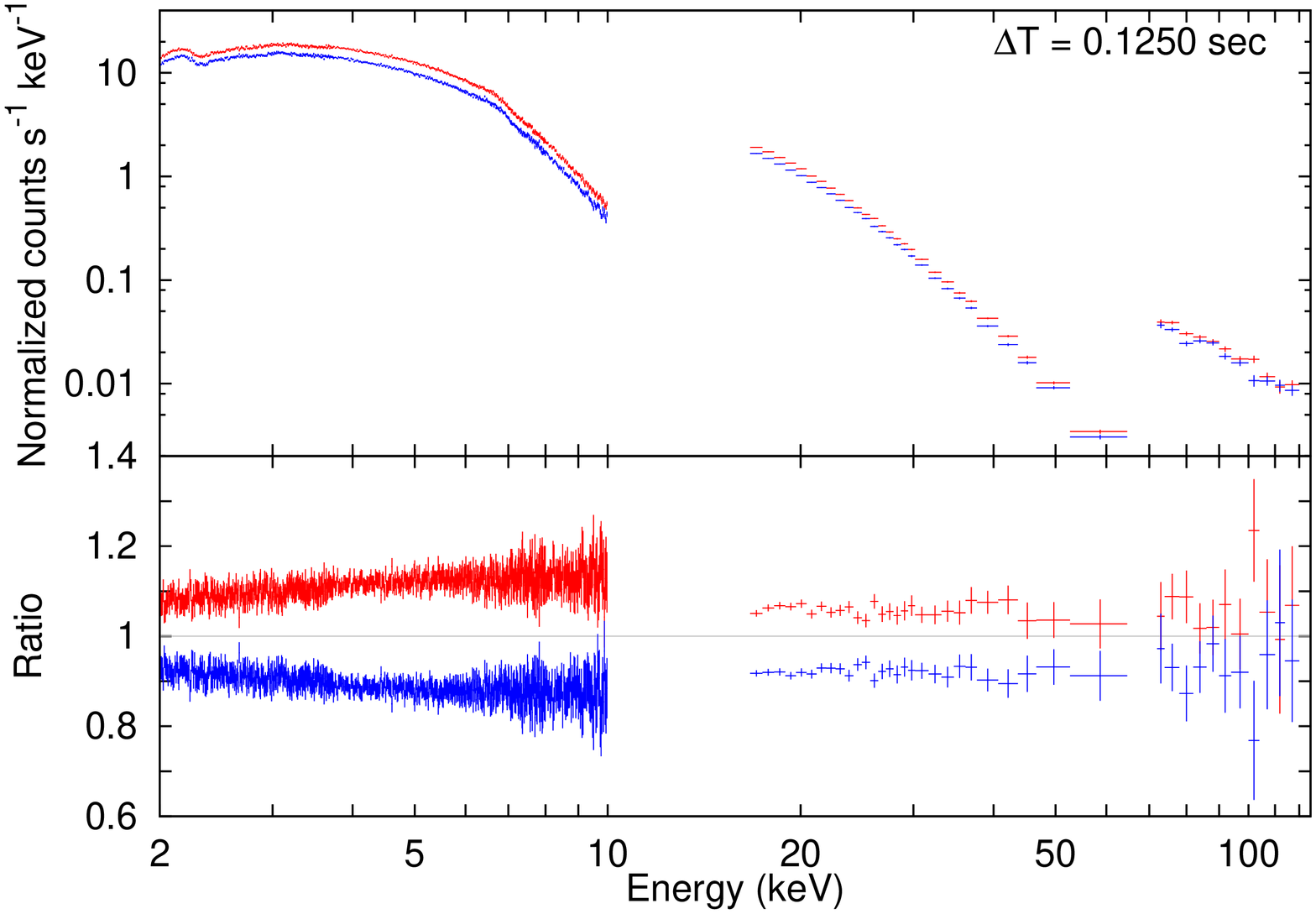}}
        \resizebox{5.5cm}{!}{\includegraphics[angle=270]{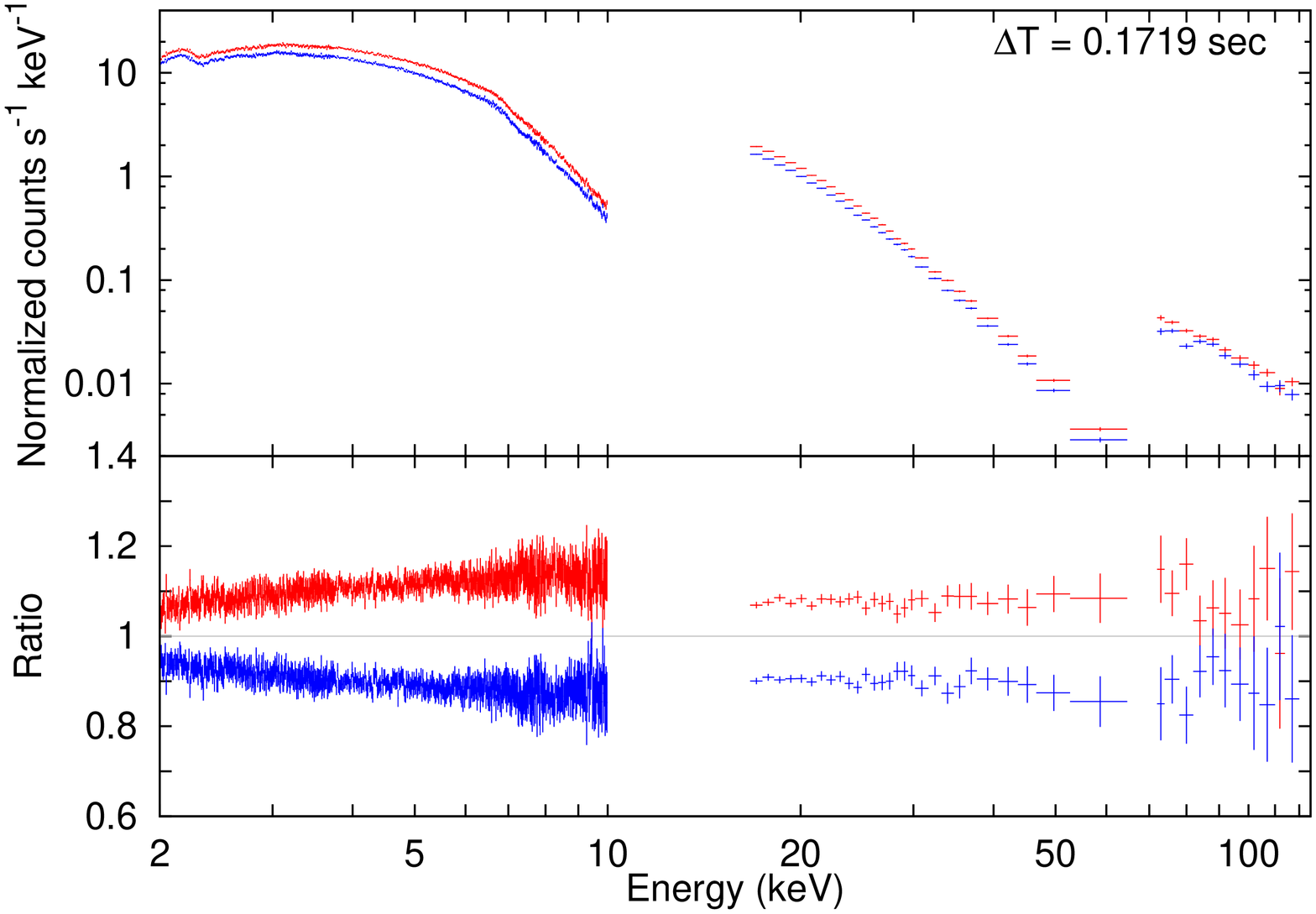}}
        \resizebox{5.5cm}{!}{\includegraphics[angle=270]{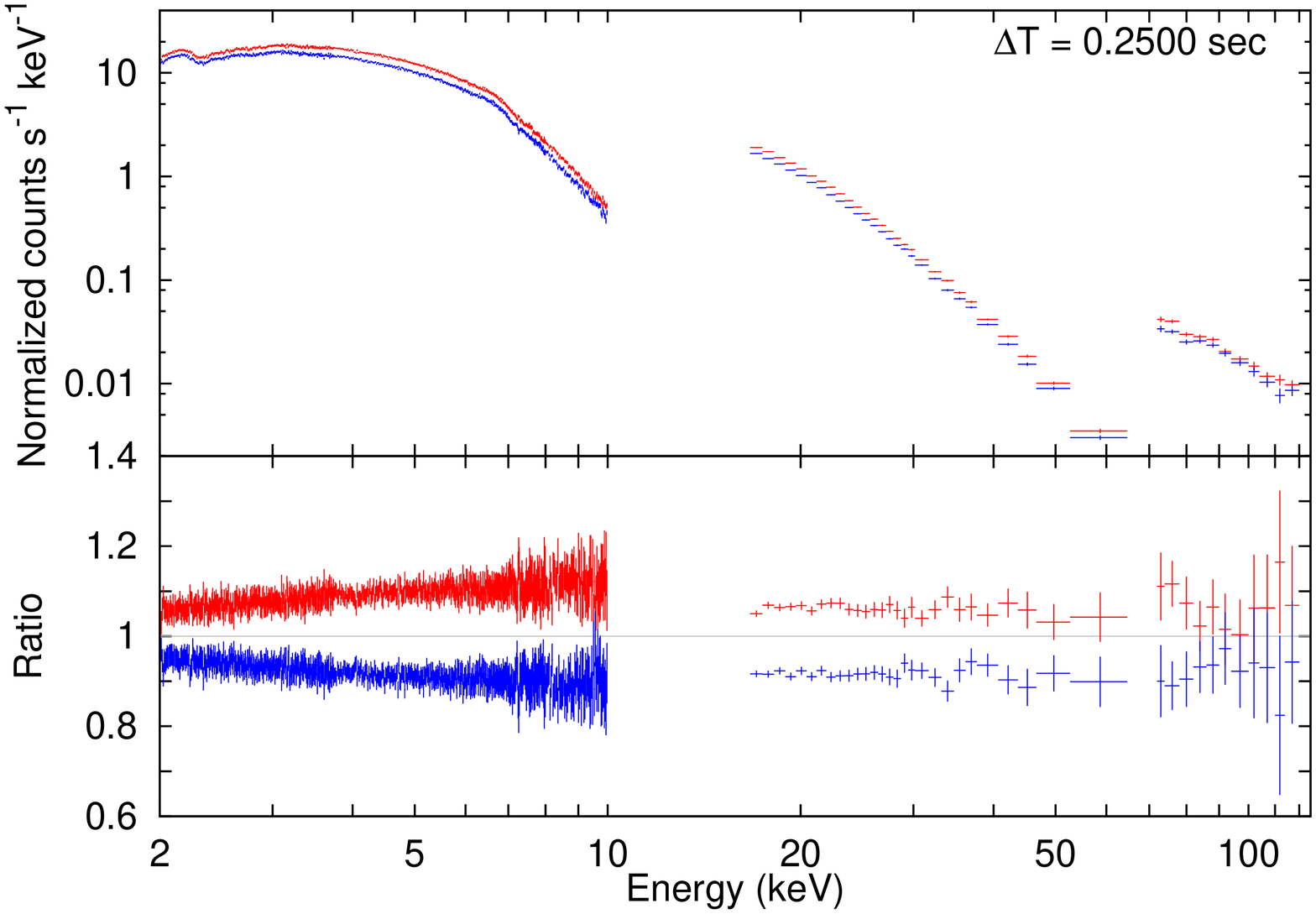}}
}
\subfigure{
        \resizebox{5.5cm}{!}{\includegraphics[angle=270]{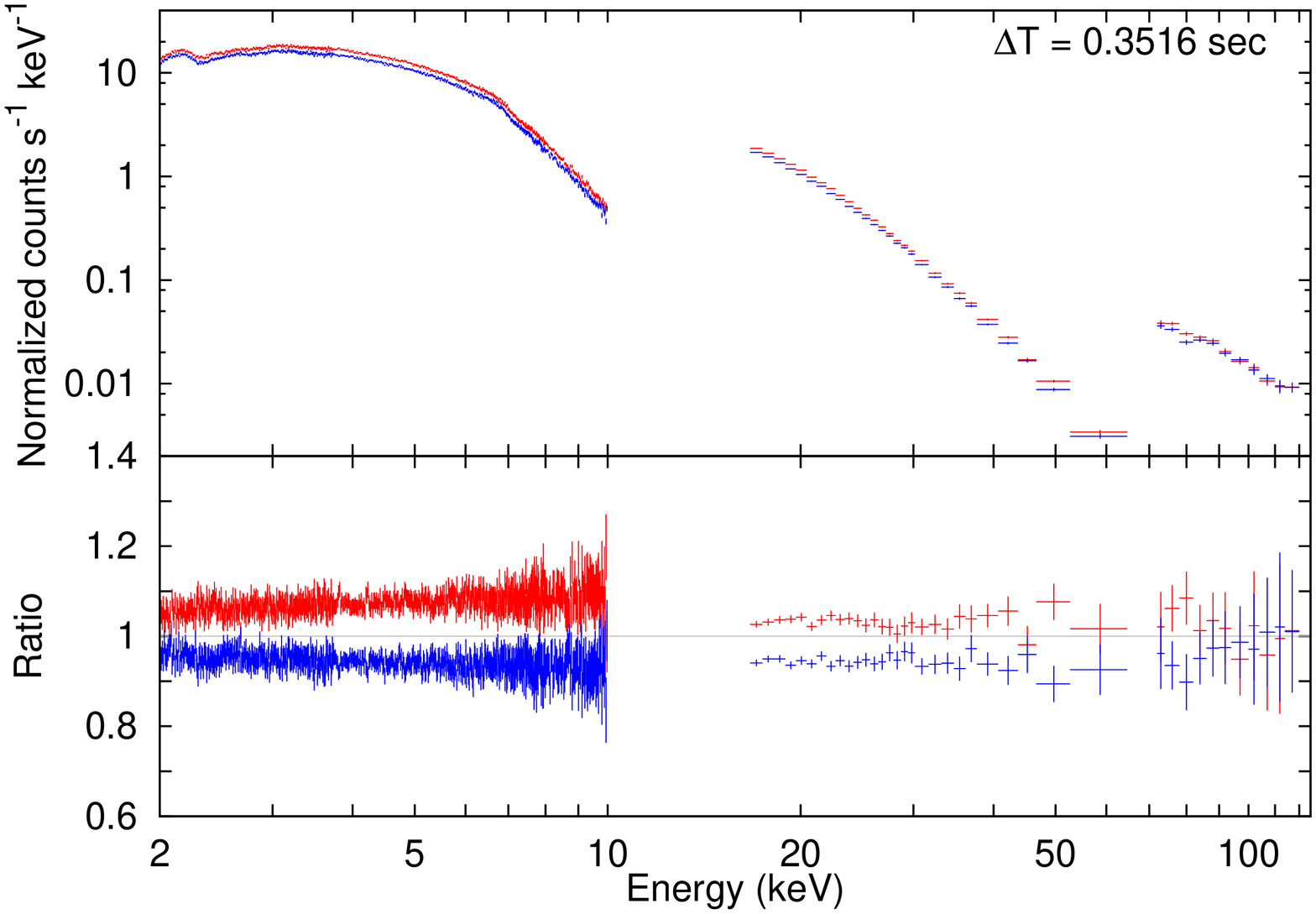}}
        \resizebox{5.5cm}{!}{\includegraphics[angle=270]{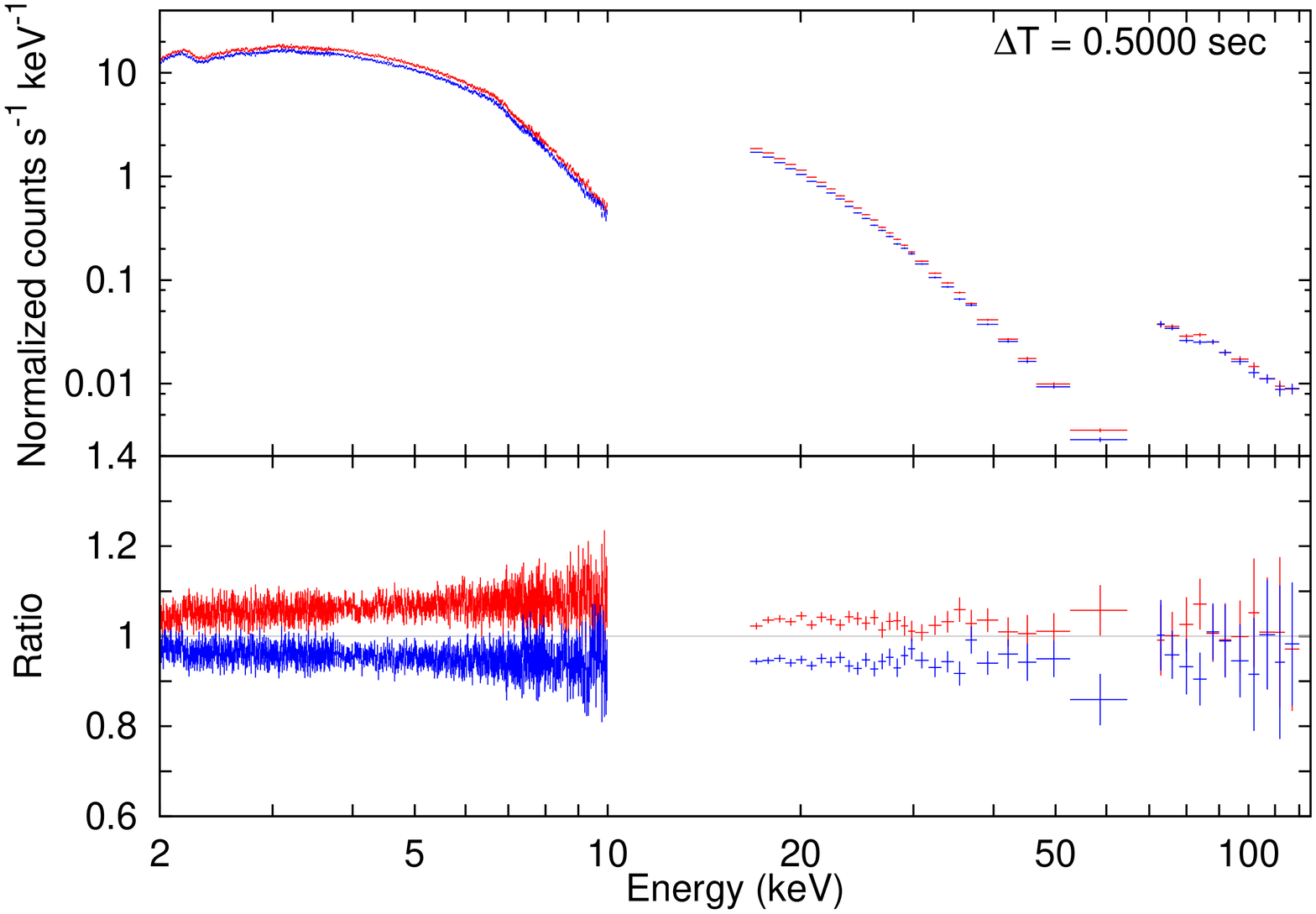}}
        \resizebox{5.5cm}{!}{\includegraphics[angle=270]{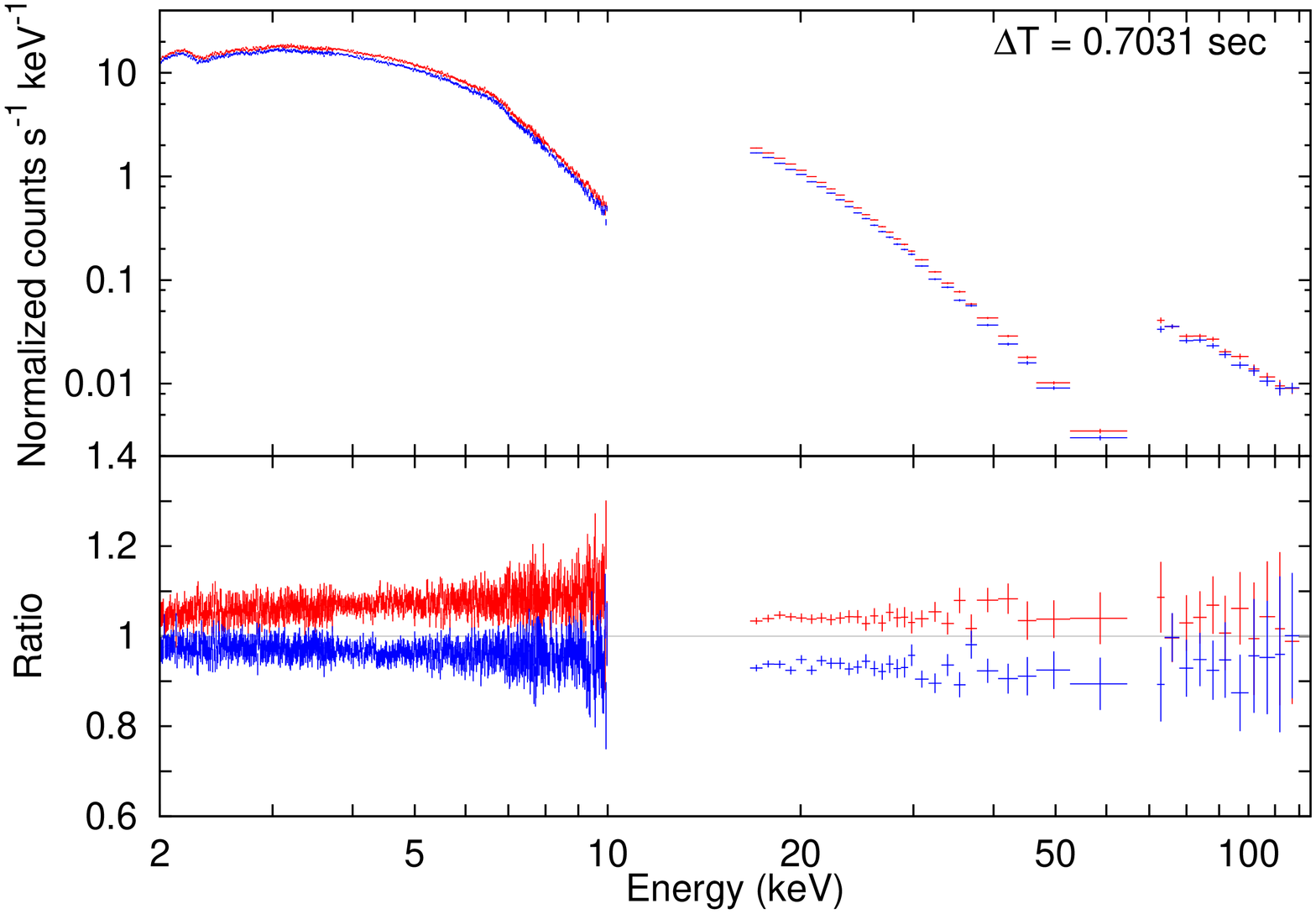}}
}
\subfigure{
        \resizebox{5.5cm}{!}{\includegraphics[angle=270]{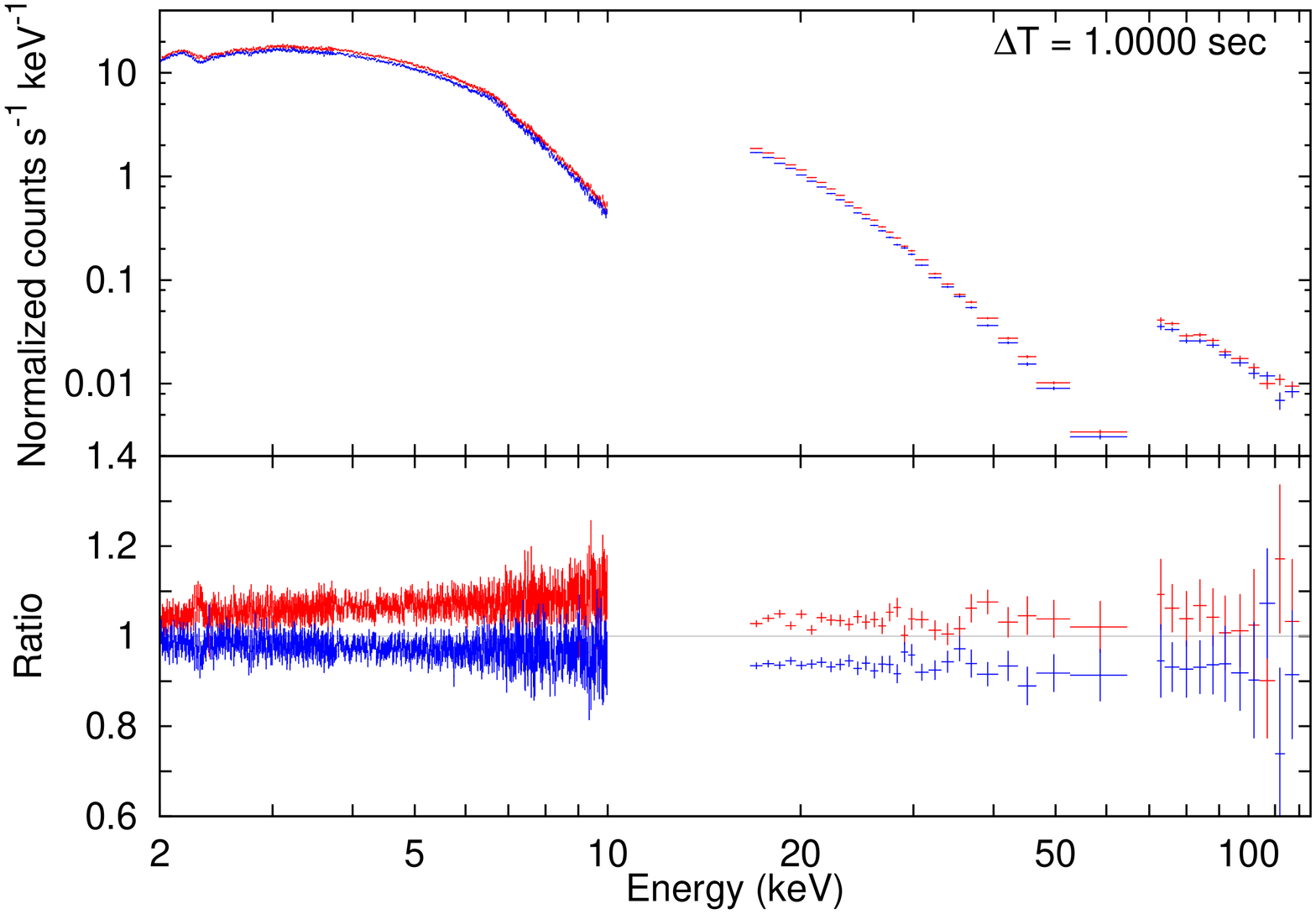}}
        \resizebox{5.5cm}{!}{\includegraphics[angle=270]{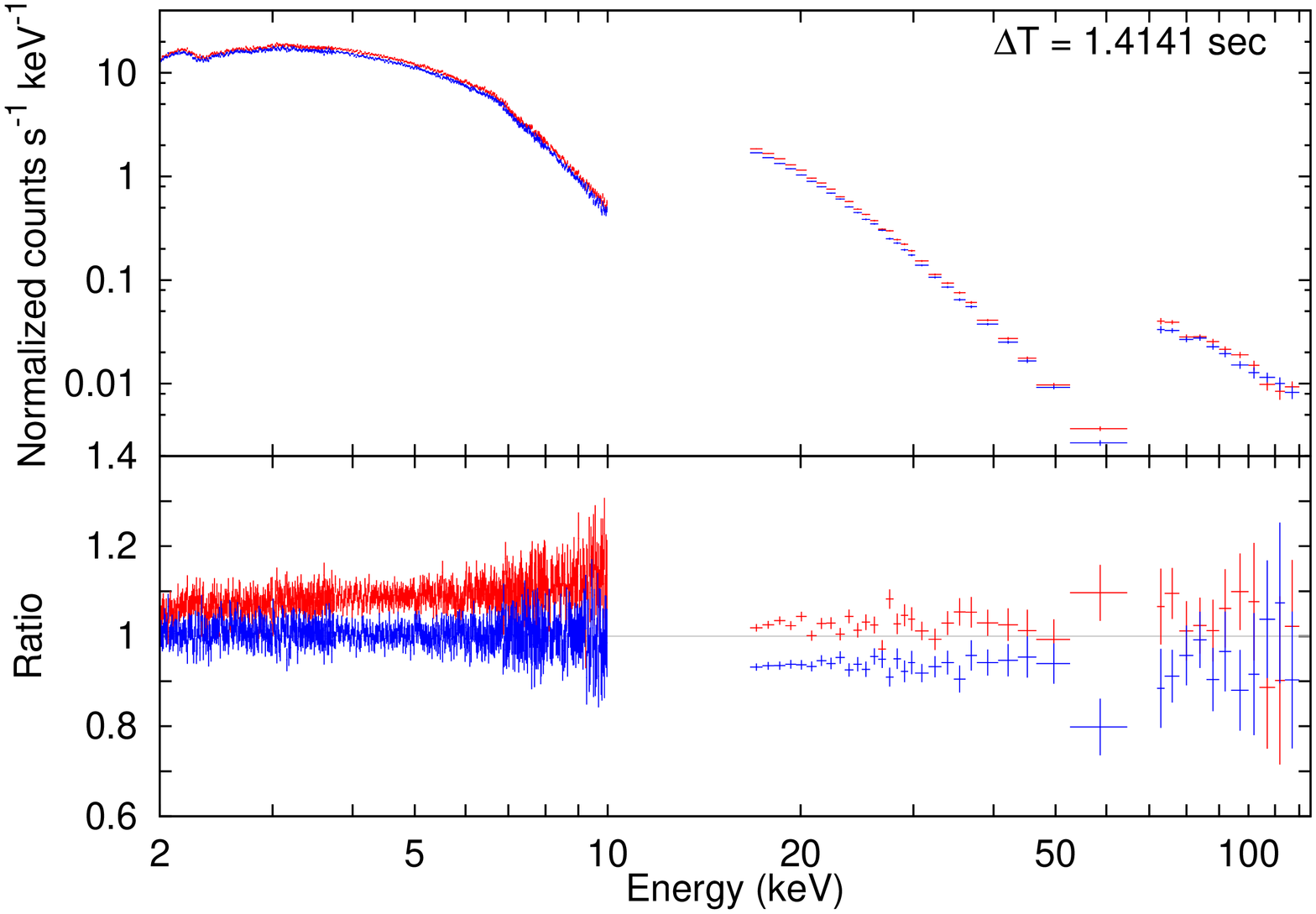}}
        \resizebox{5.5cm}{!}{\includegraphics[angle=270]{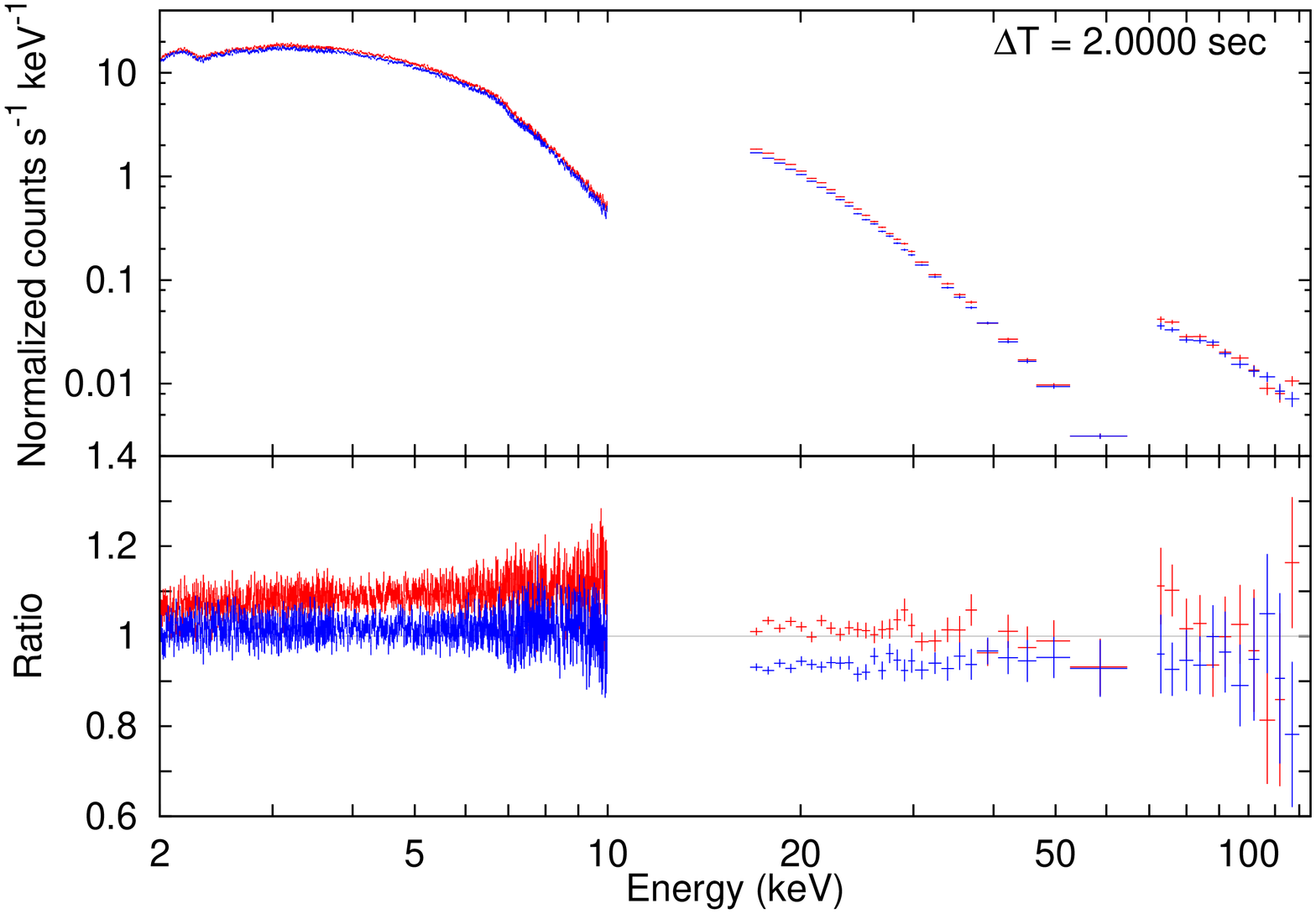}}
}
\caption{Results of spectro/temporal analysis with DVF method at timescales from 7.8~ms to 1024~s. 
For each timescale, the upper figures show the bright spectra (red) and the faint spectra (blue).
The lower figures show the ratio of the spectra to the time-averaged spectra.
P-sum mode data are used for $\Delta T \leq 2$~s, and
Normal mode data are used for $\Delta T \geq 4$~s.
As for P-sum mode data, only XIS0 Segment B data are shown for clarity.
  }
\label{f5-DVFspect:a}
\end{figure*}
\addtocounter{figure}{-1}
\begin{figure*}
\addtocounter{subfigure}{1}
\centering
\subfigure{
        \resizebox{5.5cm}{!}{\includegraphics[angle=270]{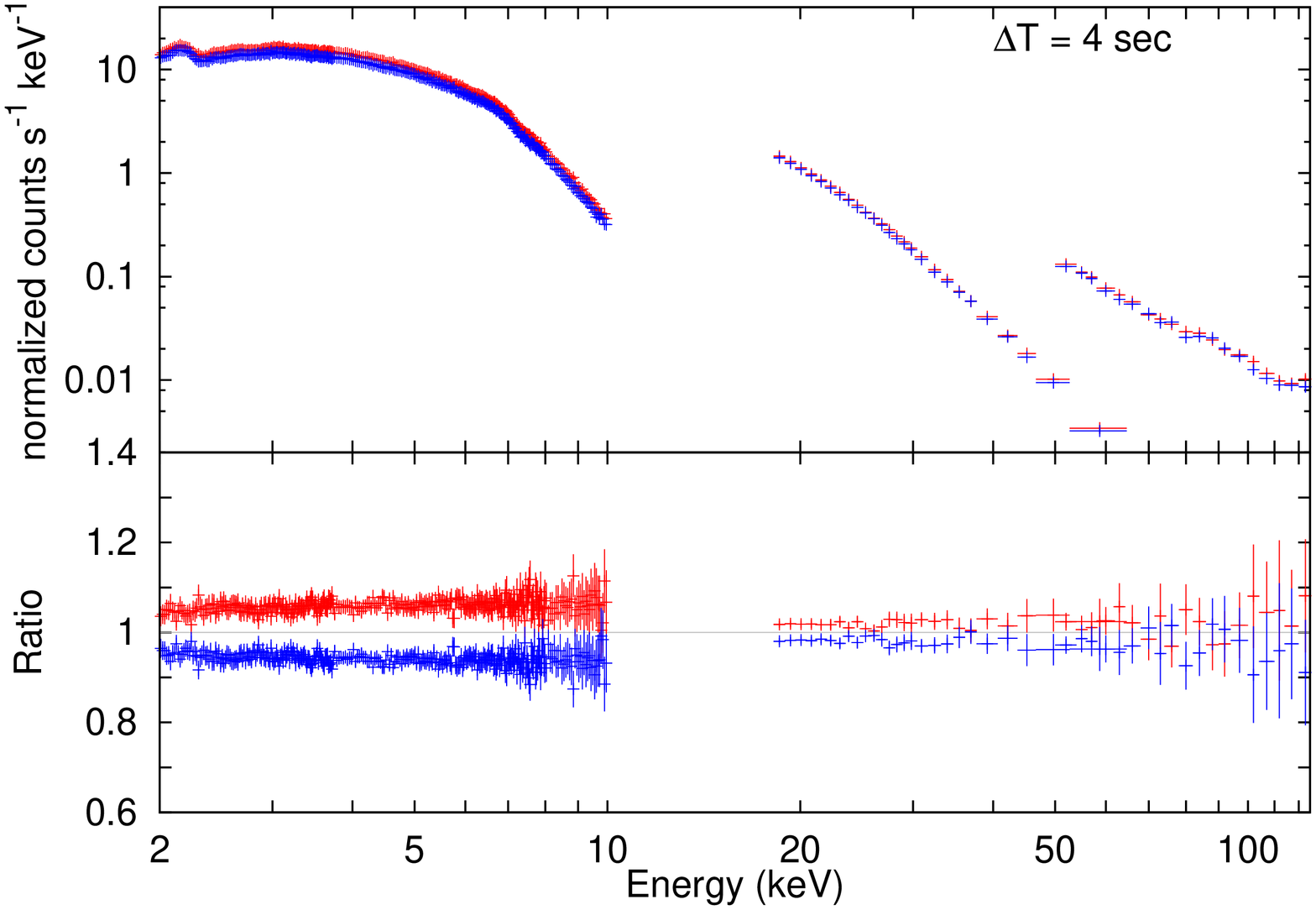}}
        \resizebox{5.5cm}{!}{\includegraphics[angle=270]{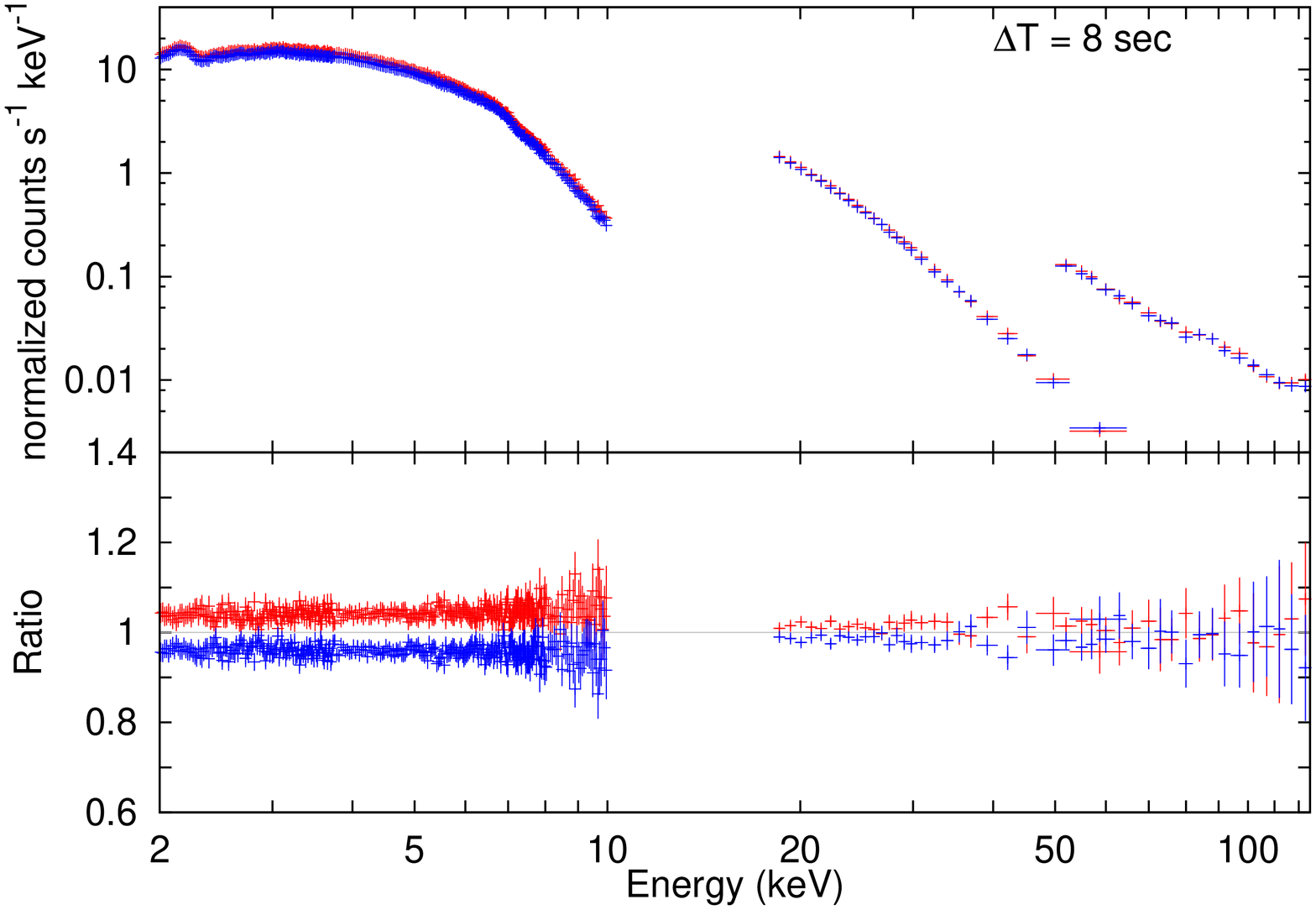}}
        \resizebox{5.5cm}{!}{\includegraphics[angle=270]{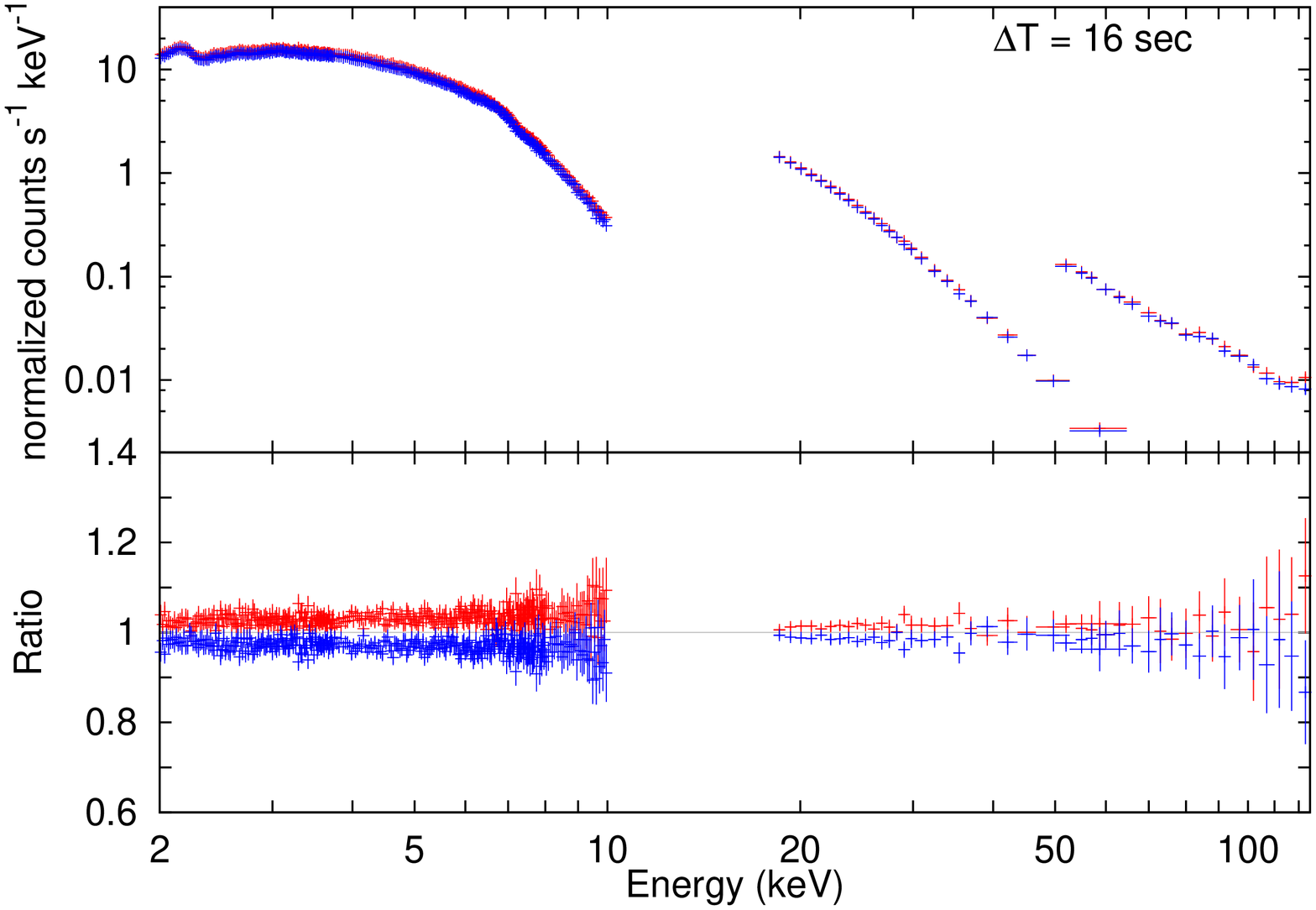}}
}
\subfigure{
        \resizebox{5.5cm}{!}{\includegraphics[angle=270]{./Figures/DVFspect_bin128.eps}}
        \resizebox{5.5cm}{!}{\includegraphics[angle=270]{./Figures/DVFspect_bin181.eps}}
        \resizebox{5.5cm}{!}{\includegraphics[angle=270]{./Figures/DVFspect_bin256.eps}}
}
\subfigure{
        \resizebox{5.5cm}{!}{\includegraphics[angle=270]{./Figures/DVFspect_4sec.eps}}
        \resizebox{5.5cm}{!}{\includegraphics[angle=270]{./Figures/DVFspect_8sec.eps}}
        \resizebox{5.5cm}{!}{\includegraphics[angle=270]{./Figures/DVFspect_16sec.eps}}
}
\subfigure{
        \resizebox{5.5cm}{!}{\includegraphics[angle=270]{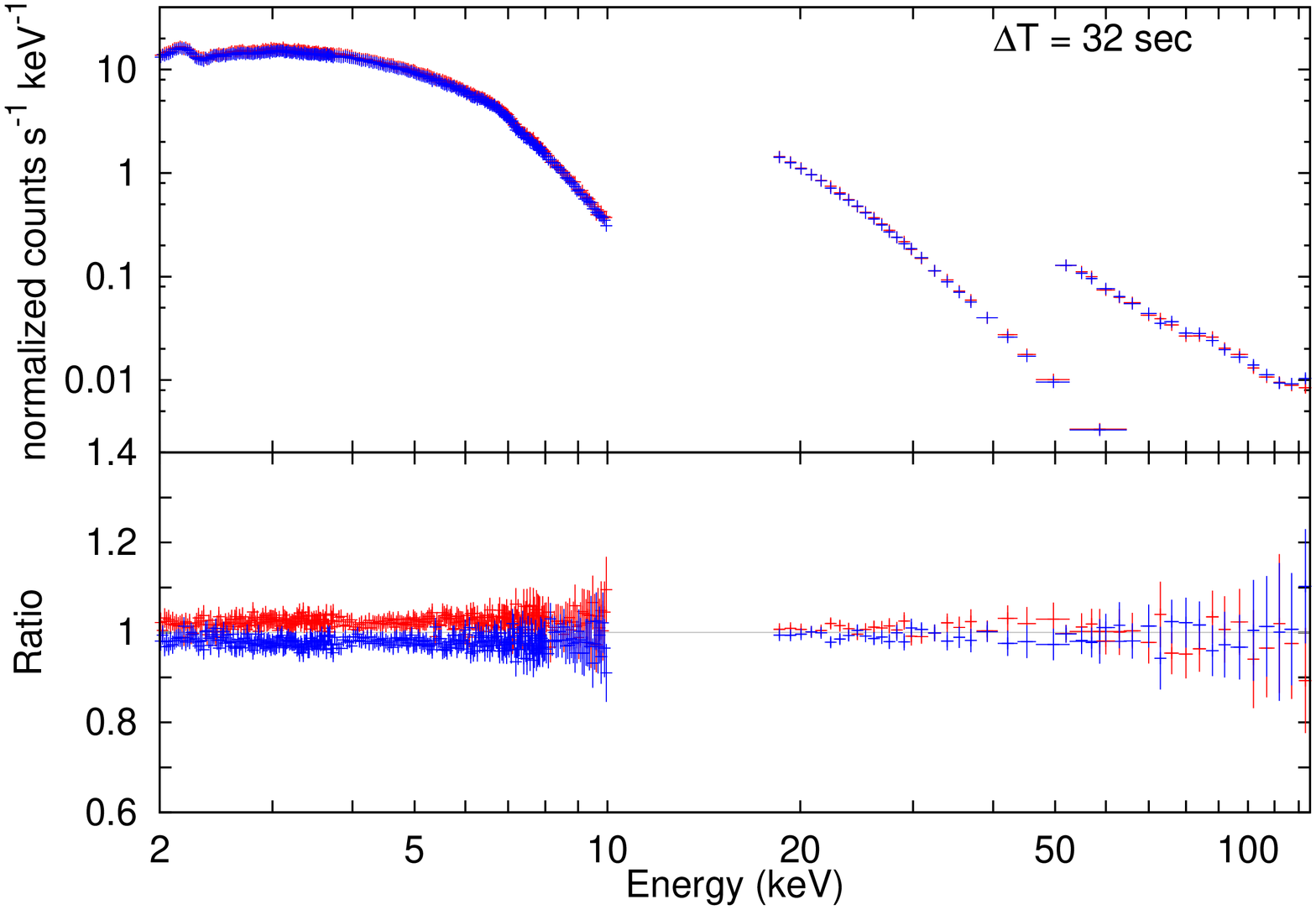}}
        \resizebox{5.5cm}{!}{\includegraphics[angle=270]{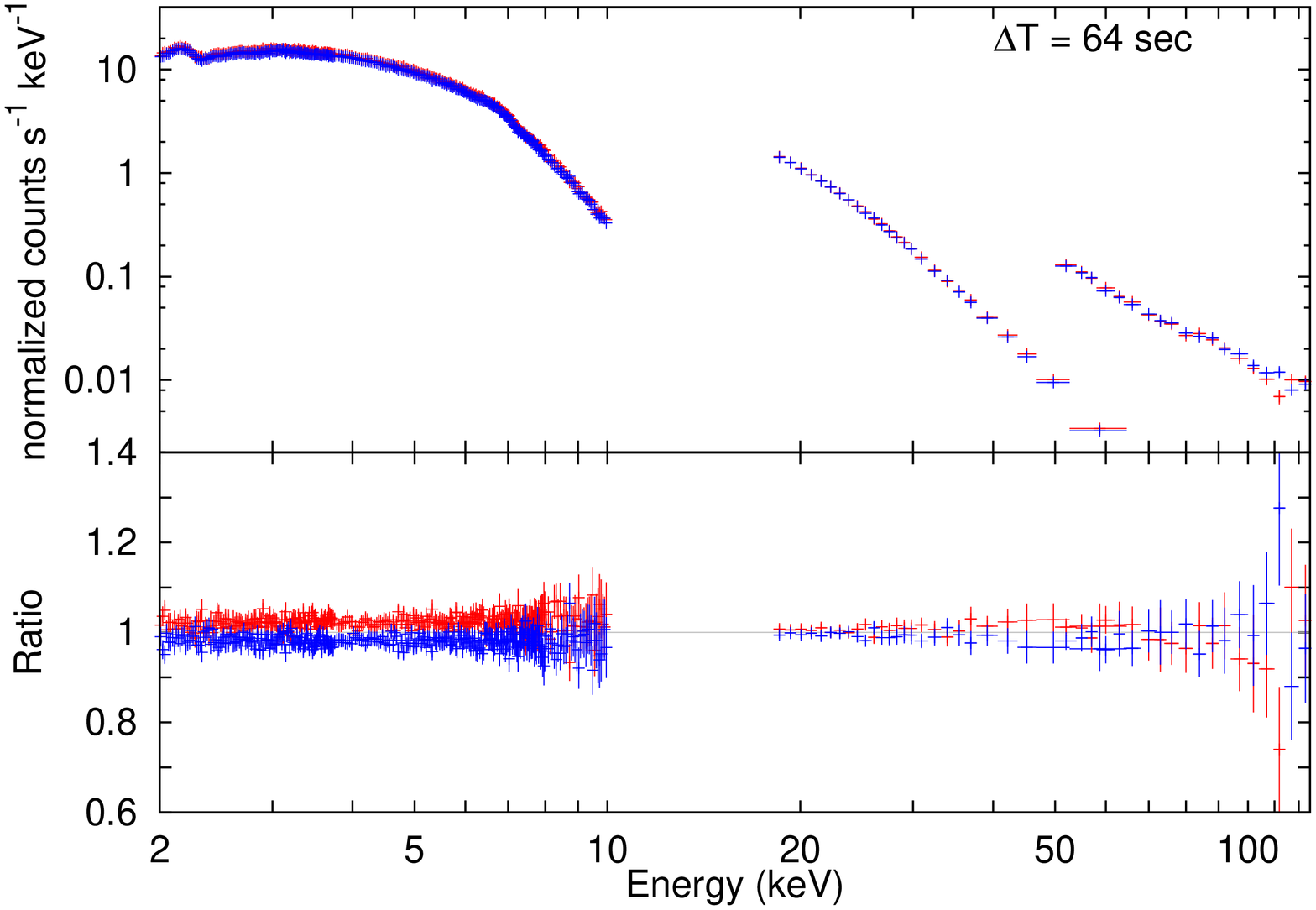}}
        \resizebox{5.5cm}{!}{\includegraphics[angle=270]{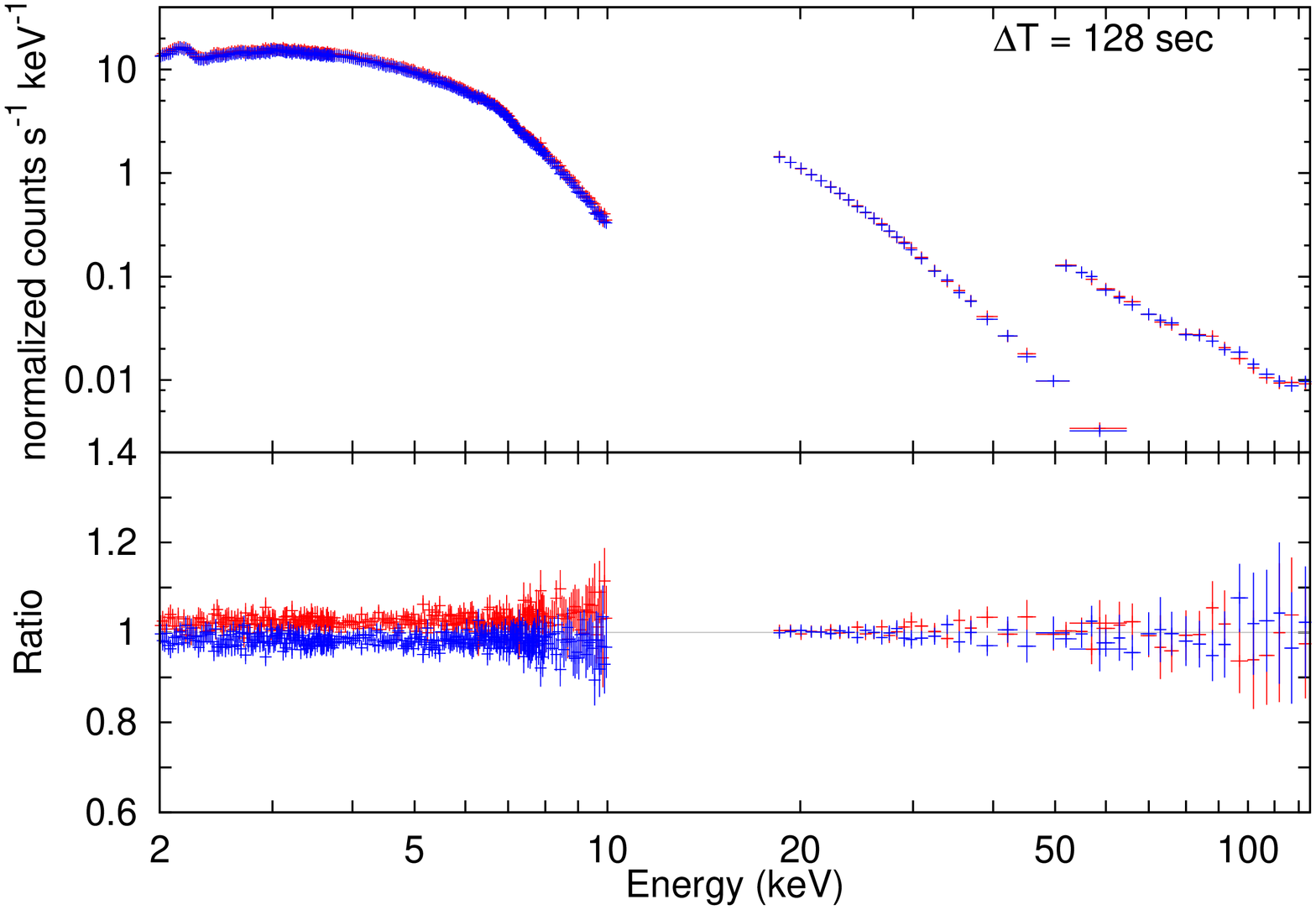}}
}
\subfigure{
         \resizebox{5.5cm}{!}{\includegraphics[angle=270]{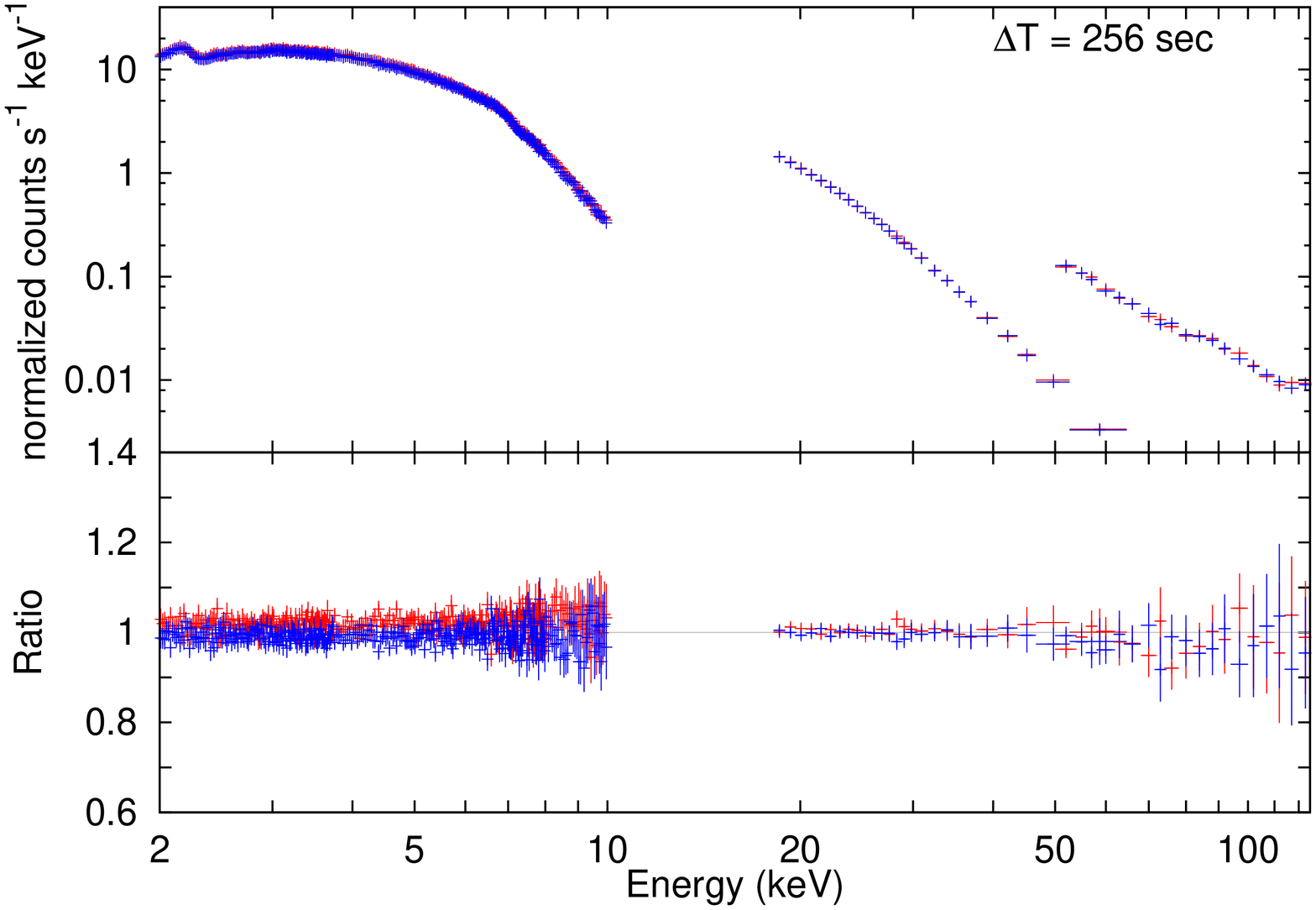}}
       \resizebox{5.5cm}{!}{\includegraphics[angle=270]{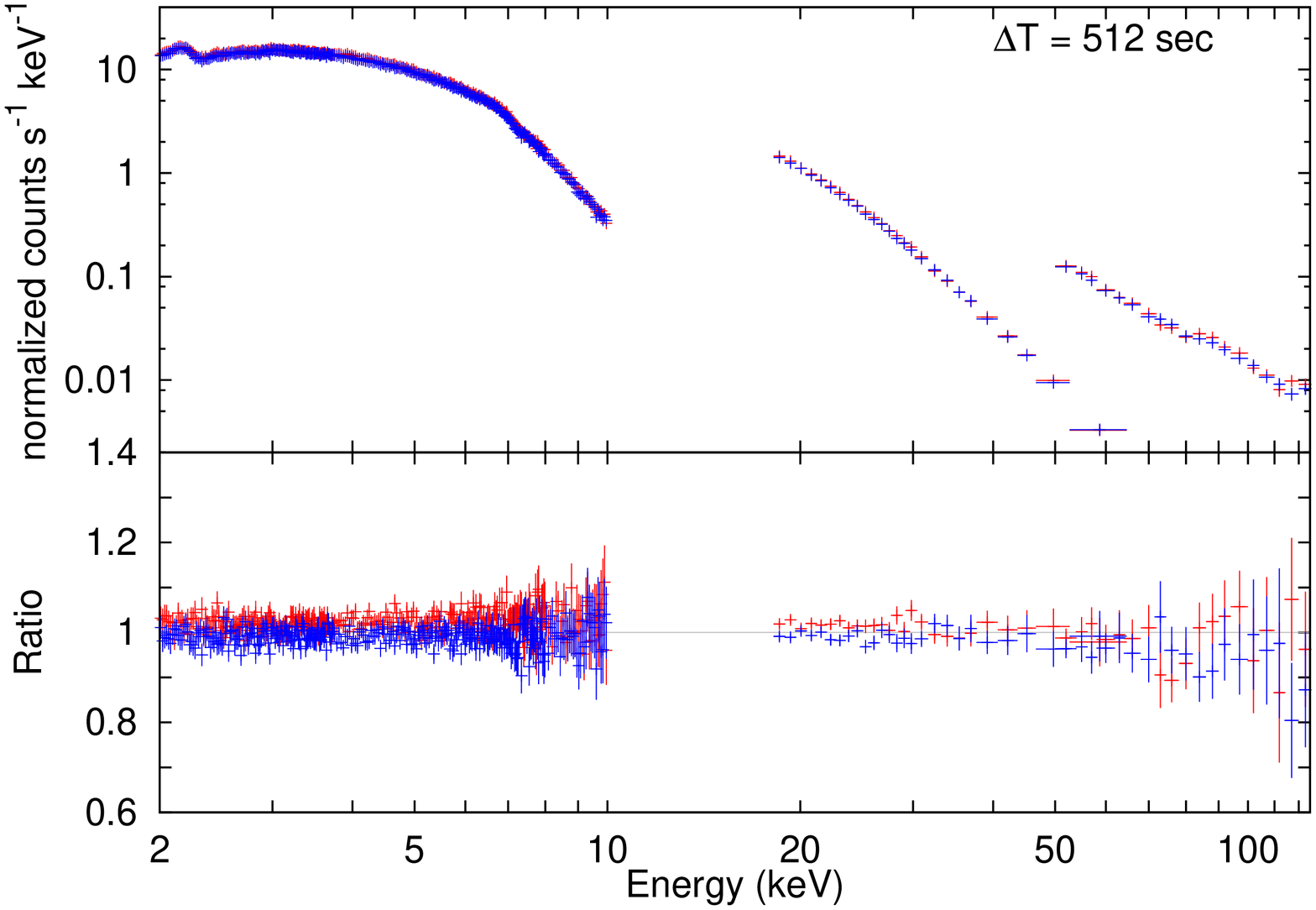}}
        \resizebox{5.5cm}{!}{\includegraphics[angle=270]{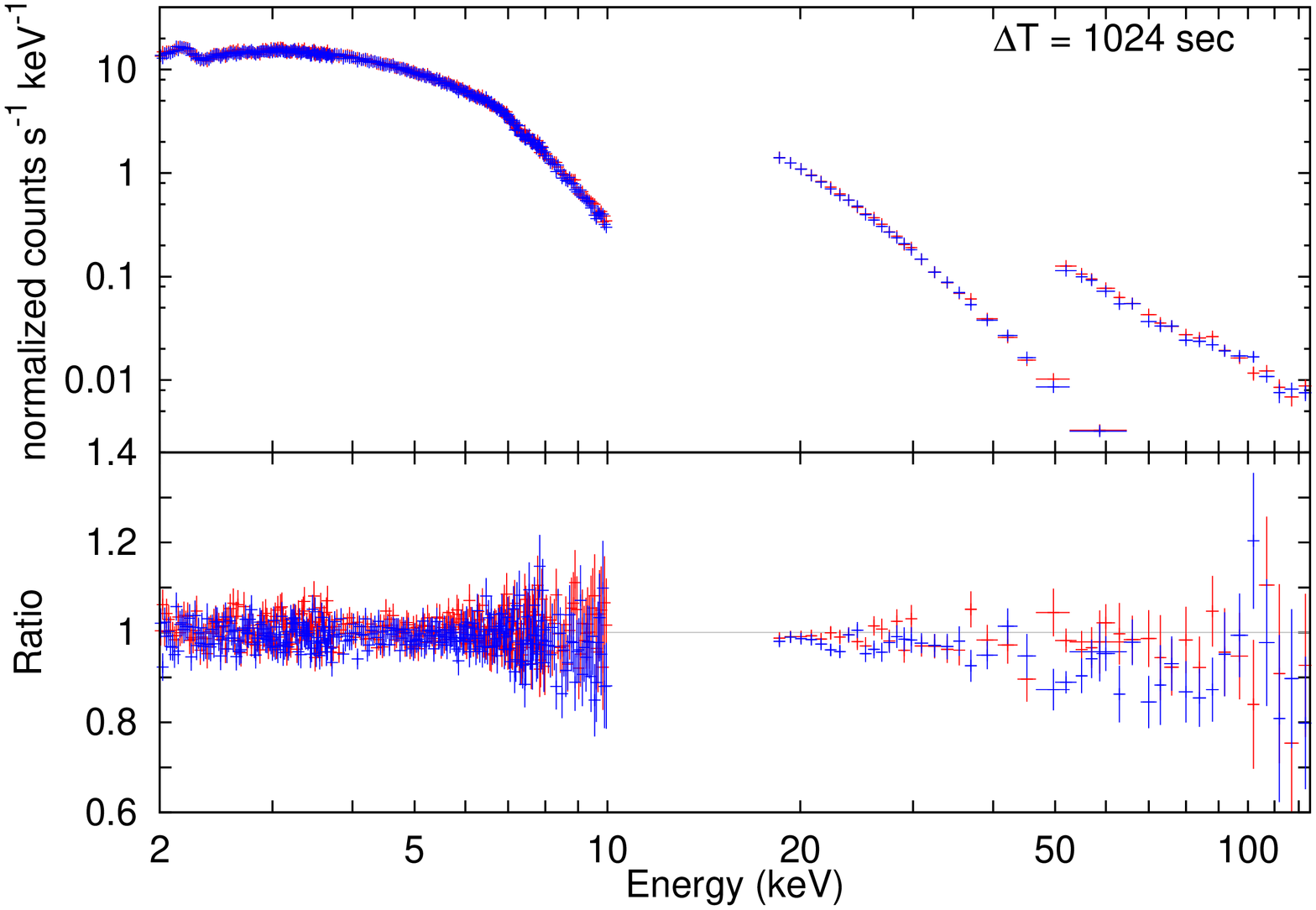}}
}
\caption{
        {\it Continued.}
}
\label{f5-DVFspect:b}
\end{figure*}
\begin{figure}[htbp]
  \begin{center}
    \includegraphics[width=100mm,angle=270]{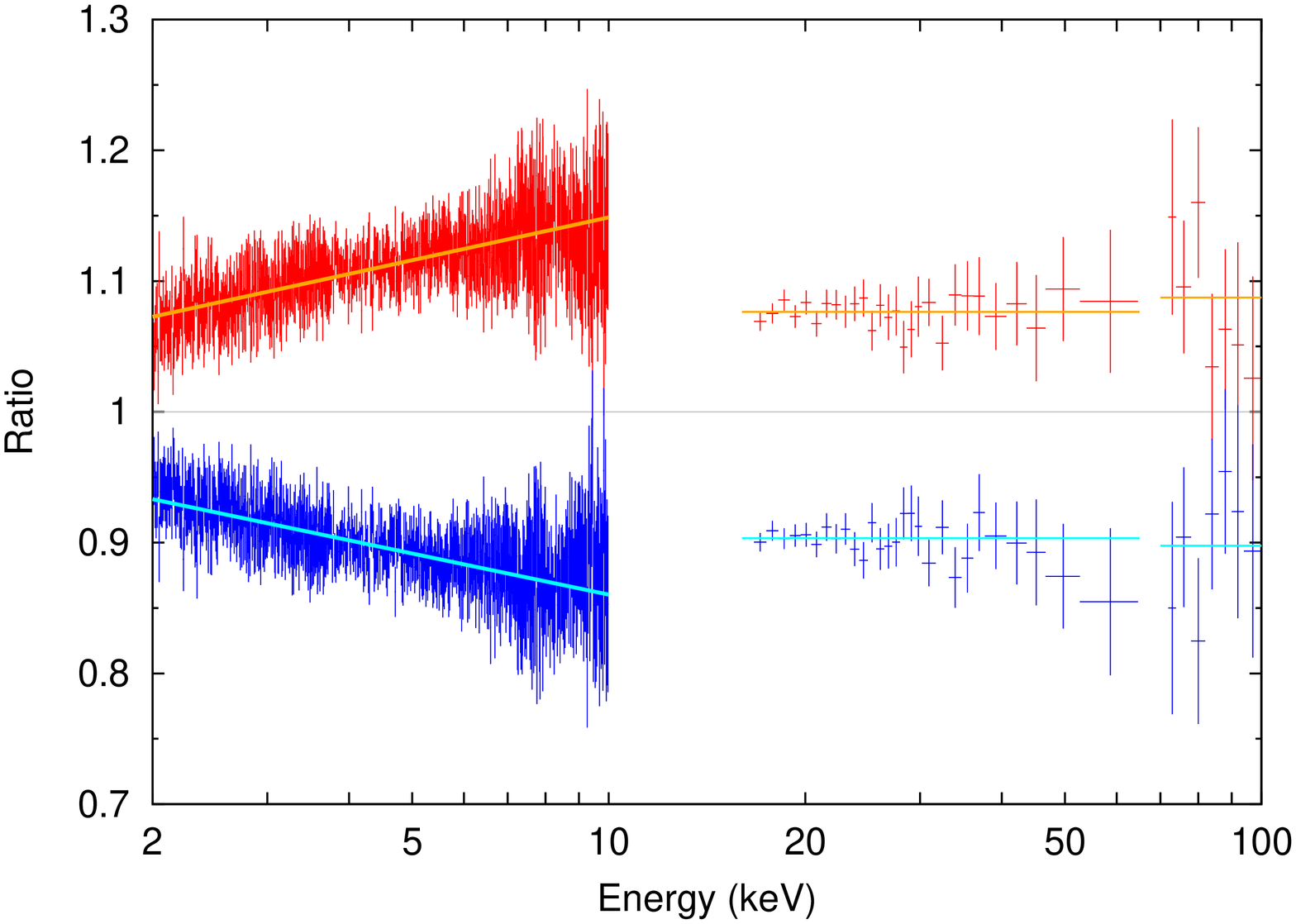}
  \end{center}
  \caption{
Fitting example of the spectral variation for $\Delta T=0.1719$~s. The red/blue bins show the ratio of the bright/faint spectra to the time-averaged spectrum, respectively (Same as the bottom panels of Figure \ref{f5-DVFspect:a}). The orange/cyan lines show the fitting linear functions, respectively. See the main text for the details of the fitting. 
  }\label{f5-fitexample}
\end{figure}
\begin{figure}[htbp]
  \begin{center}
    \includegraphics[width=100mm,angle=270]{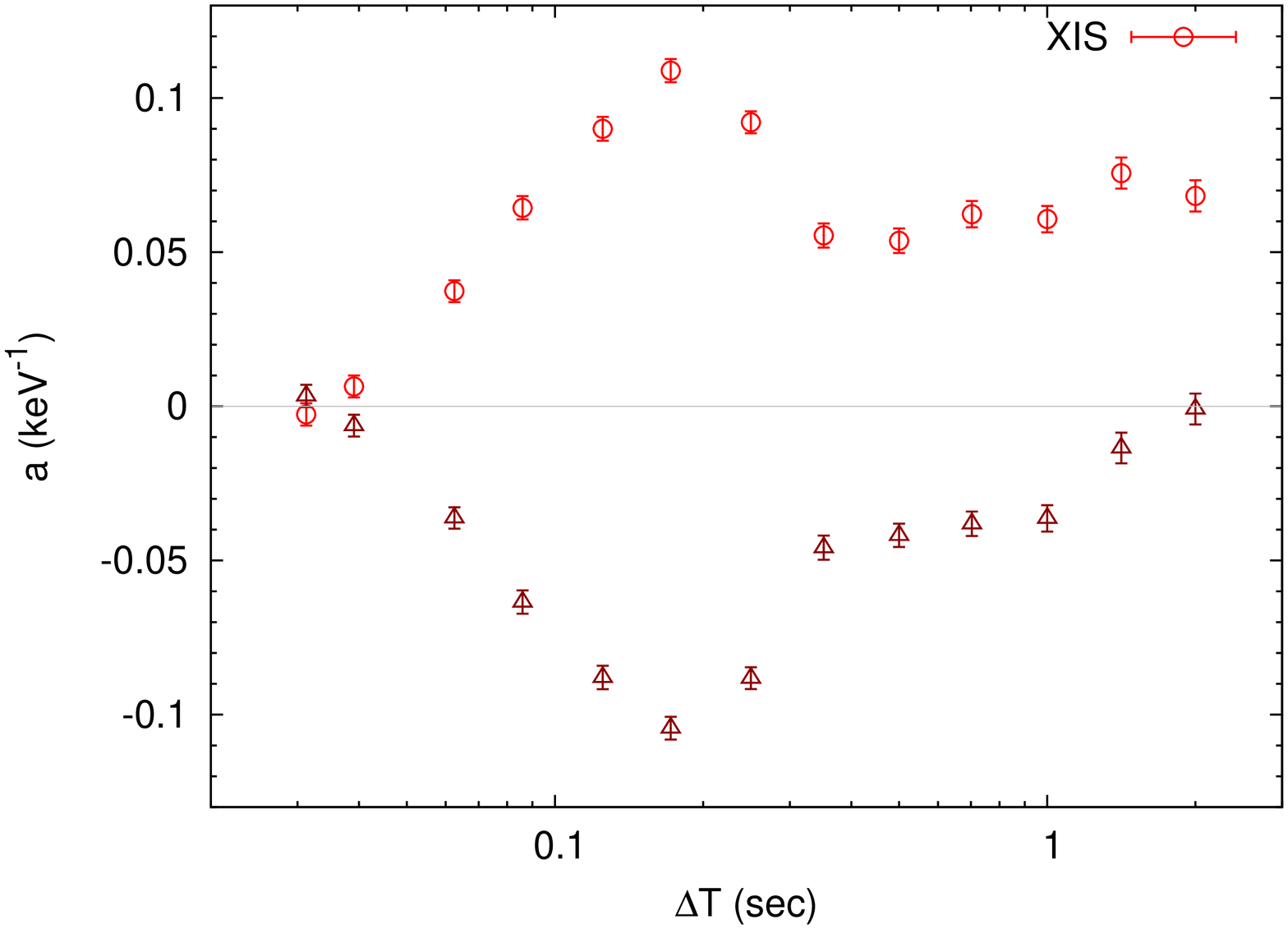}\\
    \includegraphics[width=100mm,angle=270]{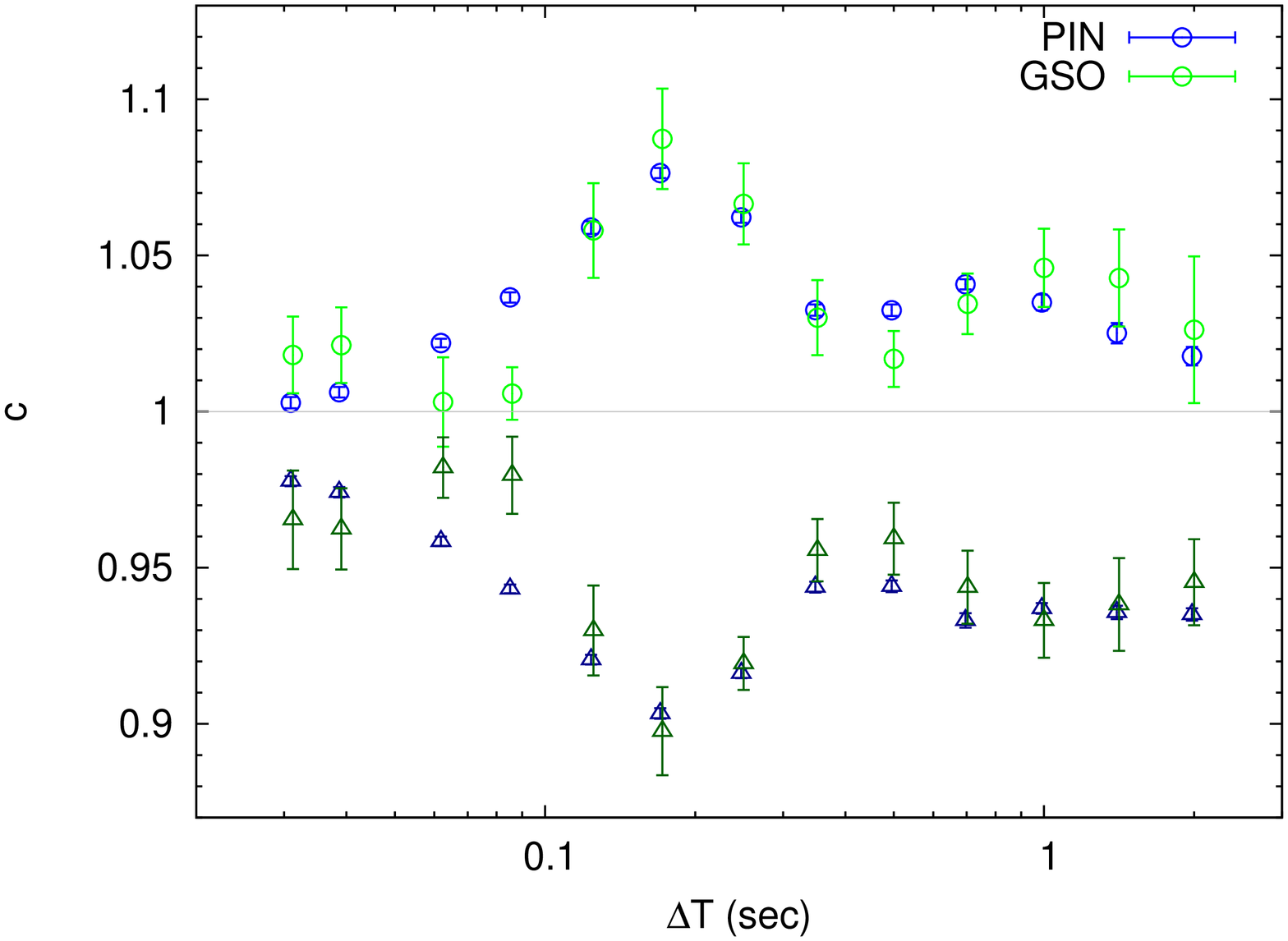}
  \end{center}
  \caption{
(Upper panel) XIS variation calculated with DVF method. The horizontal axis shows the timescale $\Delta T$, and the vertical axis shows $a$ in the main text (\S\ref{sec:QPO}). 
(Lower panel) HXD variation. The vertical axis shows $c$ in the main text. 
The red/blue/green bins with circular points show the bright phase of XIS/PIN/GSO, and those trianglar points show the faint phase of XIS/PIN/GSO, respectively.
  }\label{f5-DVFslope}
\end{figure}

\begin{figure}[htbp]
  \begin{center}
    \includegraphics[width=100mm,angle=270]{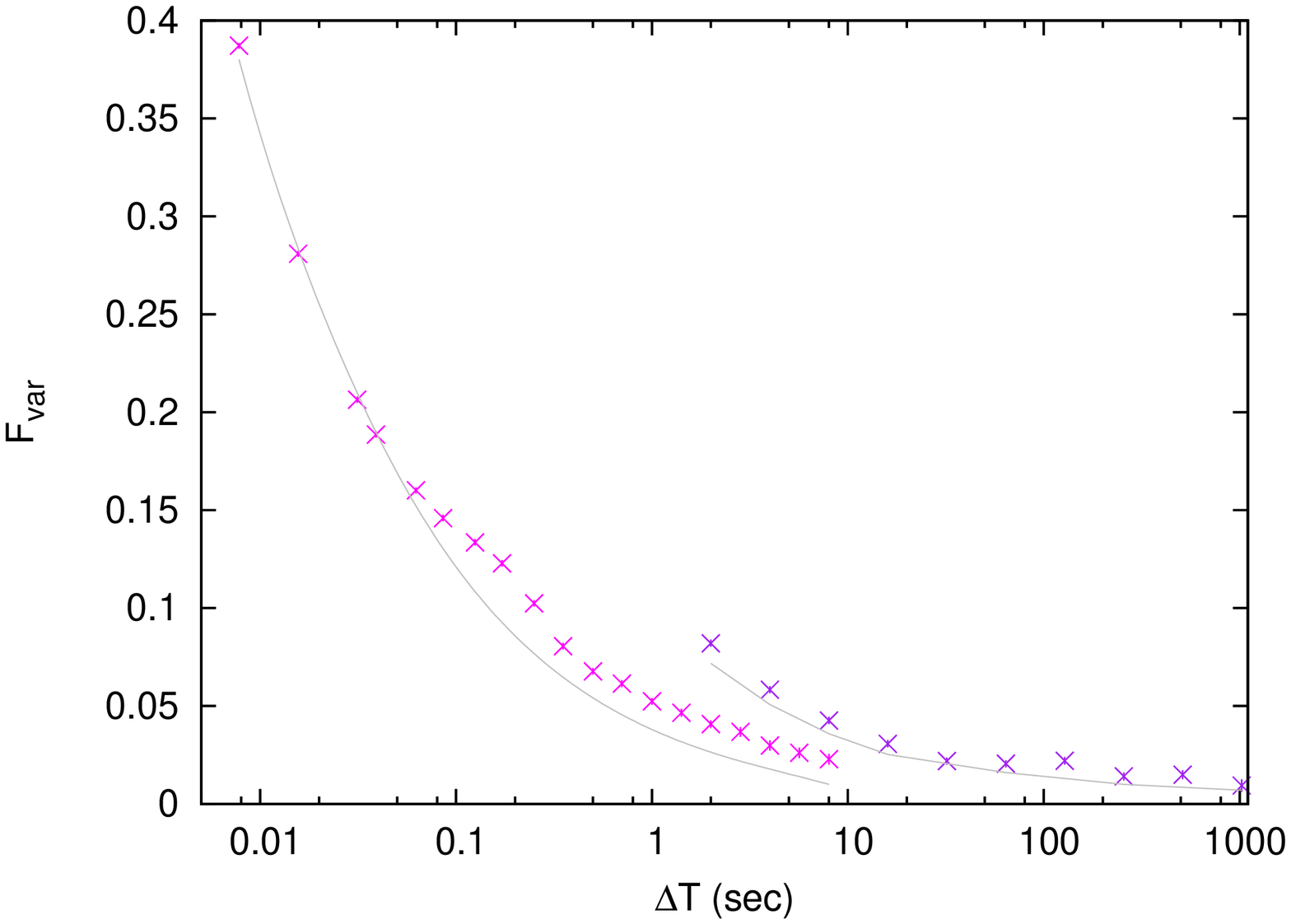}
  \end{center}
  \caption{Time-bin dependence of variation amplitude ($F_\mathrm{var}$). The magenta bins and the purple bins show $F_\mathrm{var}$ of XIS0 Segment B (P-sum mode) and XIS1 (Normal mode), respectively. Other data of P-sum mode are not shown for clarity. The gray lines show the effect of Poisson noise ($F_\mathrm{var,poisson}(\Delta T)$).
  }\label{DVF_PS}
\end{figure}

\begin{figure}[htbp]
    \begin{center}
      \includegraphics[width=100mm,angle=270]{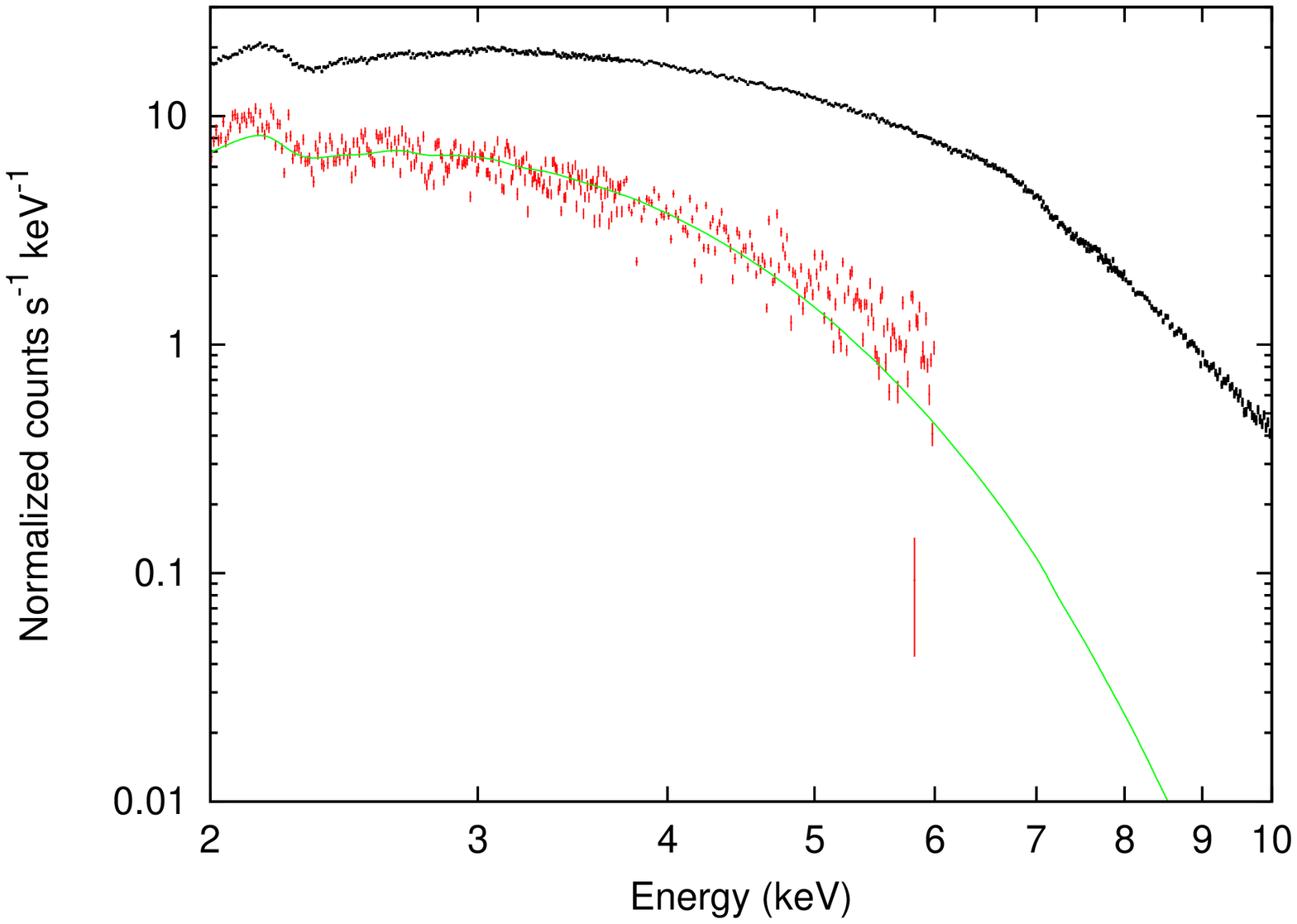}
    \end{center}
  \caption{
The non-variable component (red) within the XIS1 spectrum (black). The green line shows the MCD component with $T_\mathrm{in}=830$~eV, convolved with foreground absorption. The bins above 6~keV are not used for fitting because of few statistics.
    }\label{f5-invariant}
\end{figure}
\begin{landscape}
\begin{figure}[htbp]
  \begin{center}
\subfigure{
\resizebox{8.5cm}{!}{\includegraphics[angle=270]{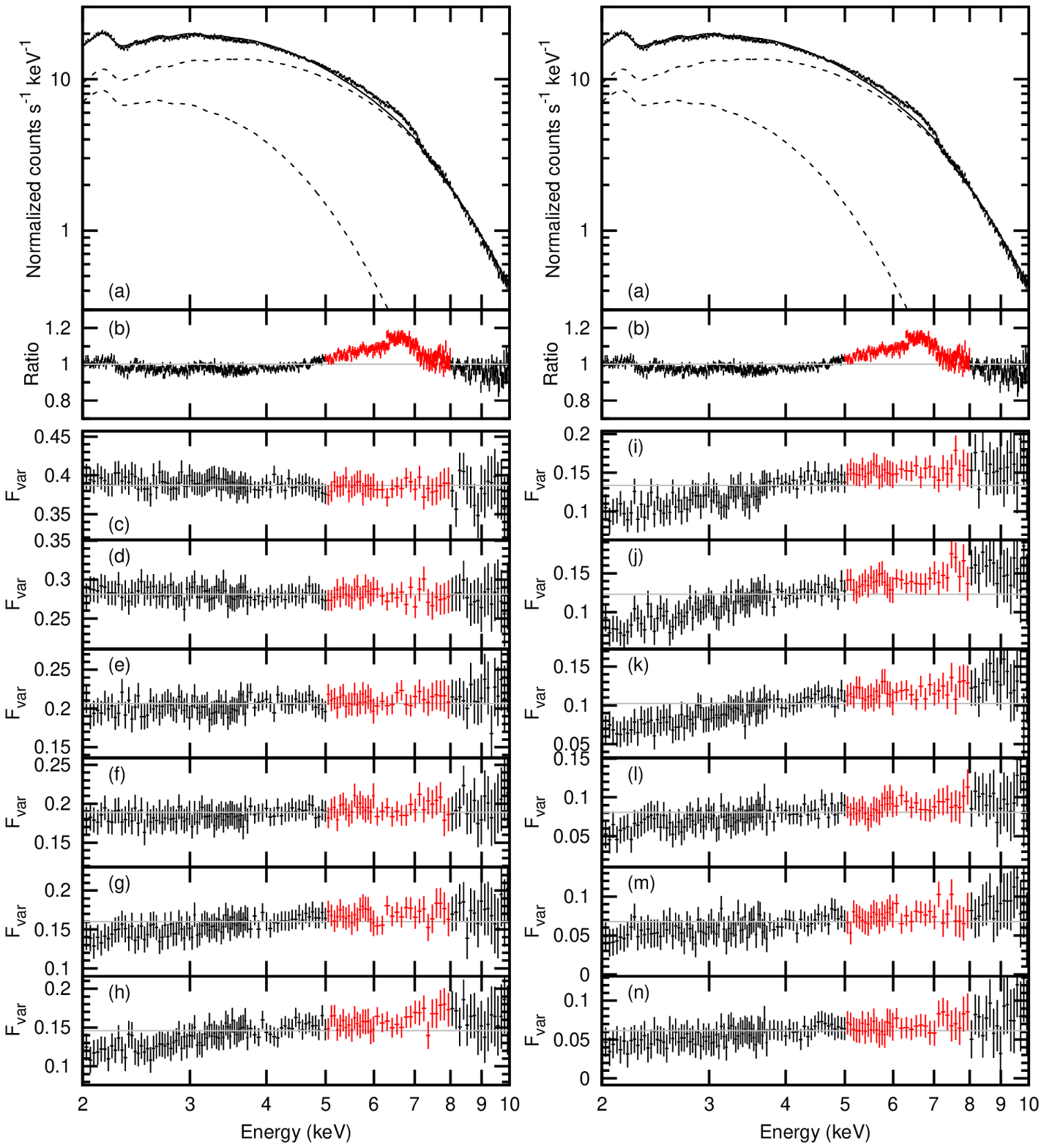}}
\resizebox{8.5cm}{!}{\includegraphics[angle=270]{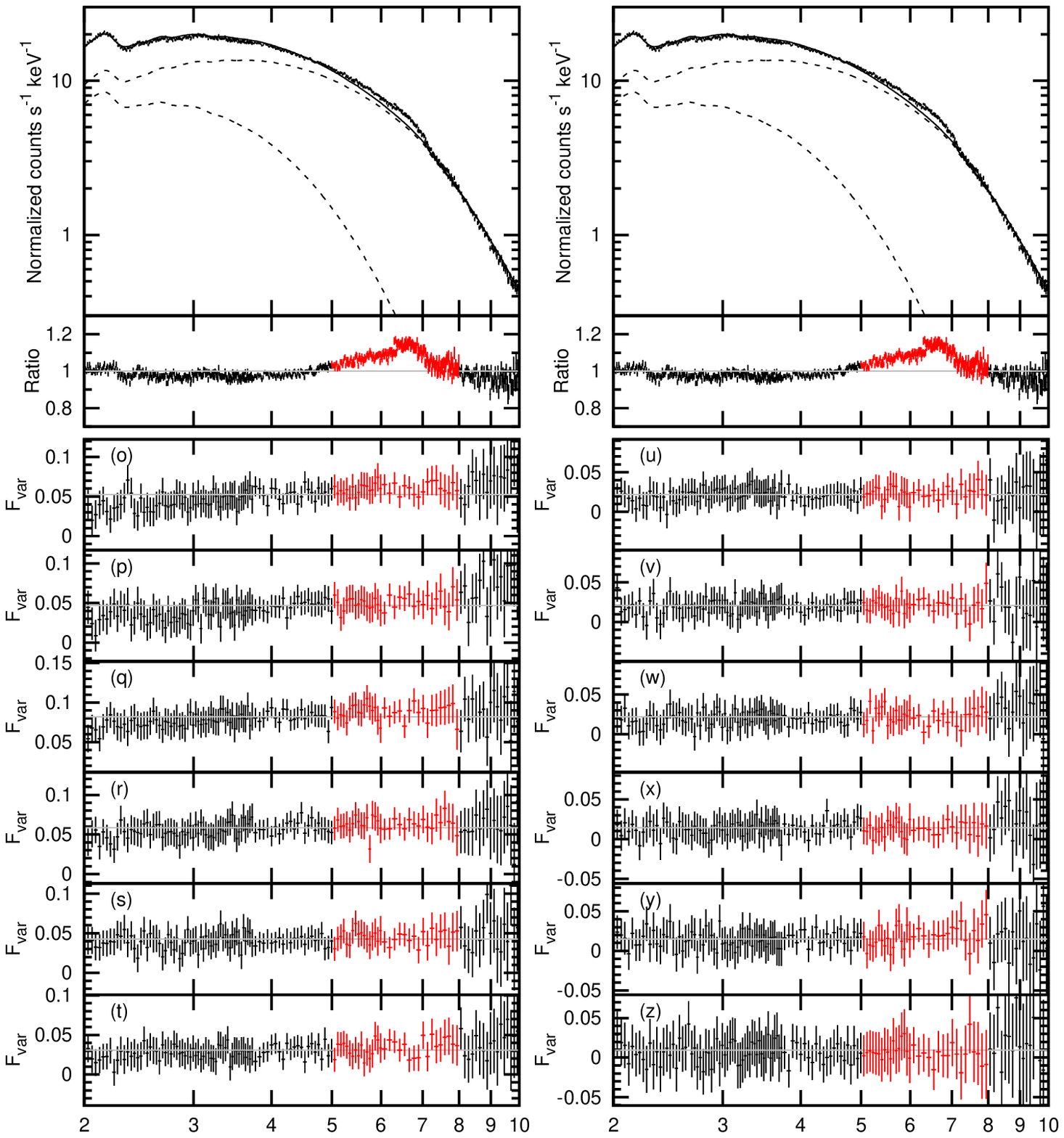}}
\resizebox{8.5cm}{!}{\includegraphics[angle=270]{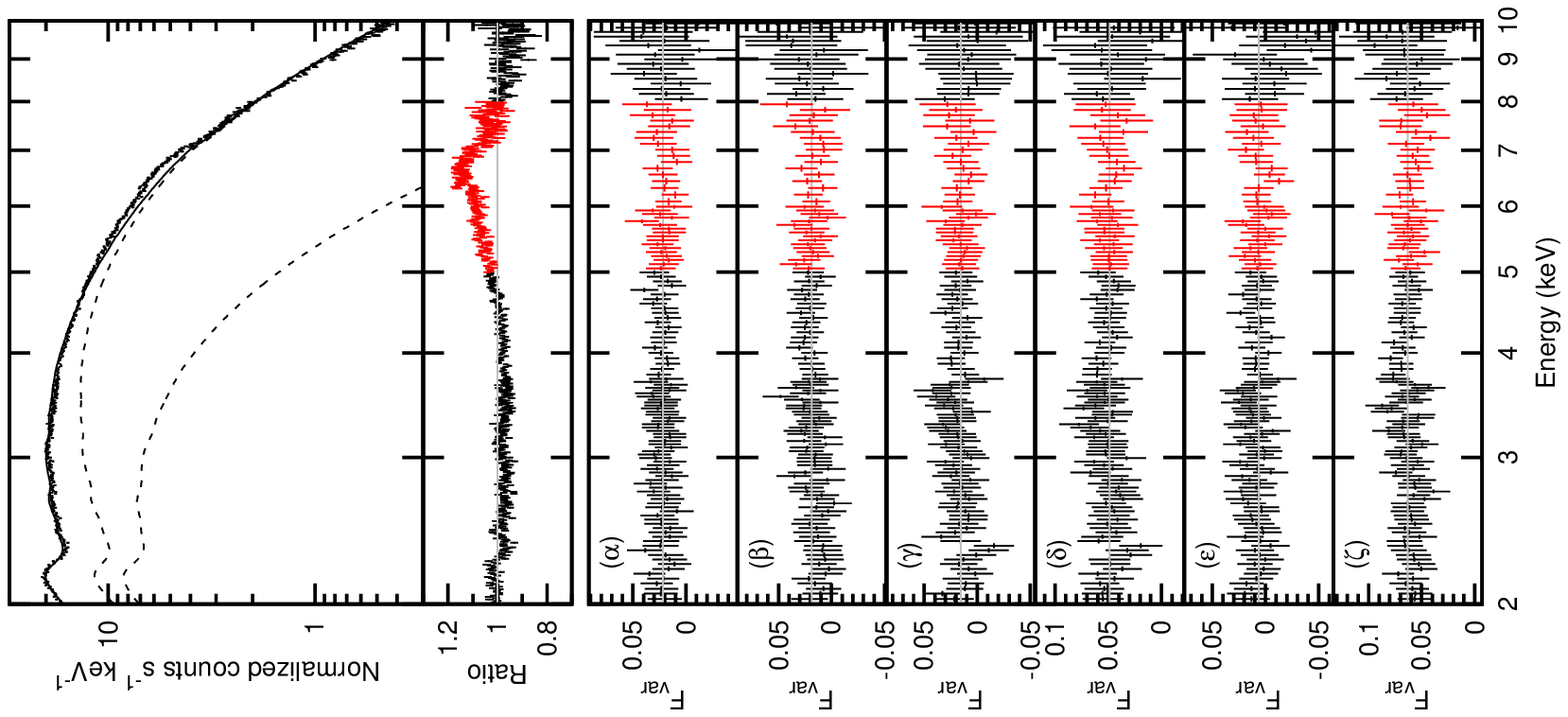}}
}
  \caption{(a)~The time-averaged spectrum obtained by XIS1 and the {\tt diskbb}$+${\tt cutoffpl} model shown in Figure~\ref{f5-suzaku_cutoffpl}~(c). (b)~Ratio of the spectrum to the model. (c)--($\zeta$)~Variation amplitude ($F\mathrm{var}$) calculated with DVF method. The red region indicates the iron K-spectral feature. The gray lines show averages of $F_\mathrm{var}$.
In $\Delta T < 2$~s, only the results of XIS0 Segment B are shown for the sake of visibility.
In $\Delta T \geq 2$~s, results of XIS1 are shown.
The diagrams (c)--($\zeta$) corresponds to the following:
$\Delta T=$0.016~s, 0.031~s, 0.039~s, 0.063~s, 0.086~s, 0.125~s, 0.172~s, 
0.25~s, 0.35~s, 0.50~s, 0.70~s, 1~s, 1.4~s, 2~s, 4~s, 8~s, 16~s, 32~s,
64~s, 128~s, 256~s, 512~s, 1024~s, 5760~s, 11520~s, 17280~s, 23040~s, 28800~s, and 63360~s.
The sensitivity is shown in Figure \ref{DVF_PS}.
  }\label{f6-DVF_bf1}
  \end{center}
\end{figure}
\end{landscape}

\begin{landscape}
\begin{figure}[htbp]
  \begin{center}
\subfigure{
\resizebox{10cm}{!}{\includegraphics[angle=270]{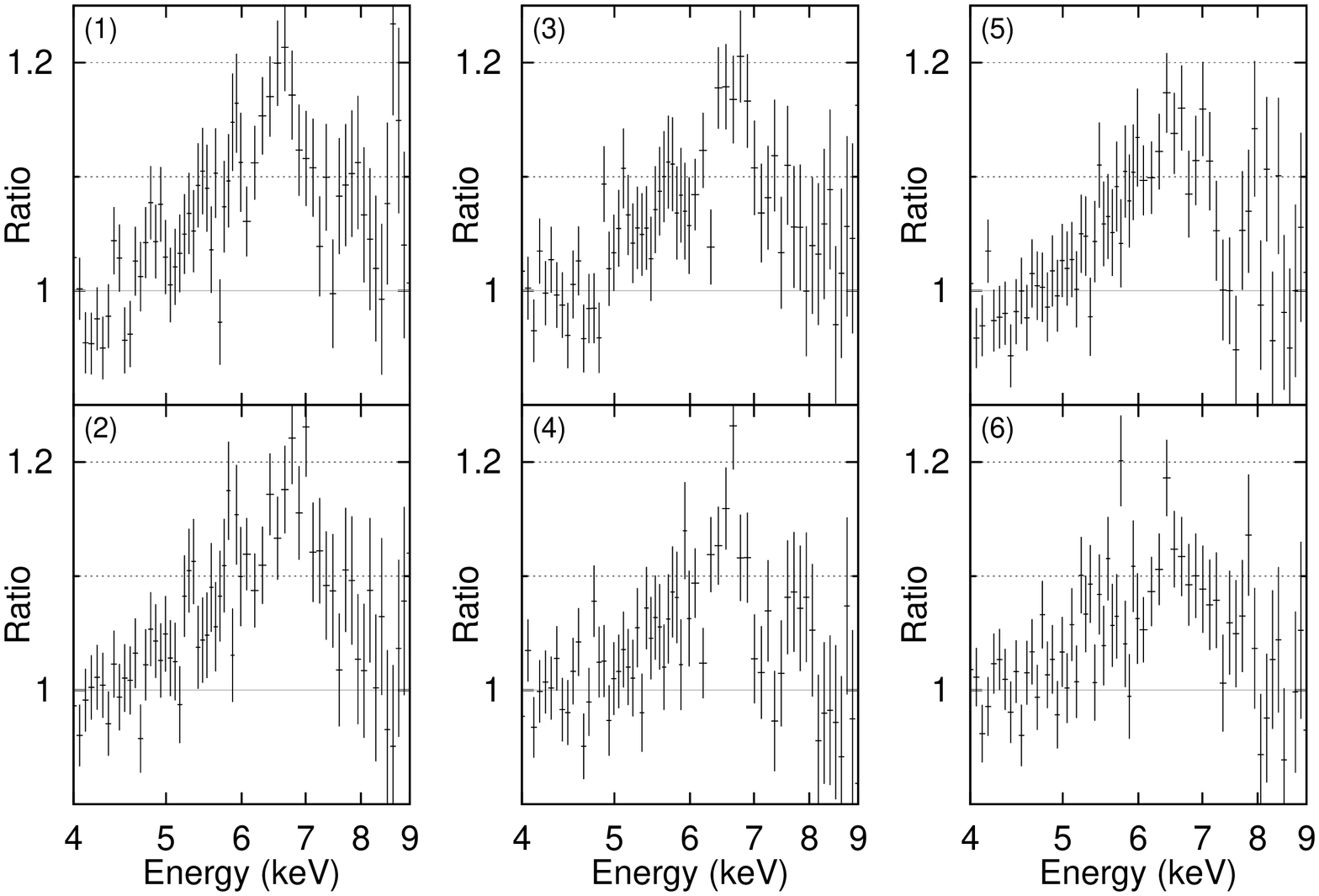}}
\resizebox{10cm}{!}{\includegraphics[angle=270]{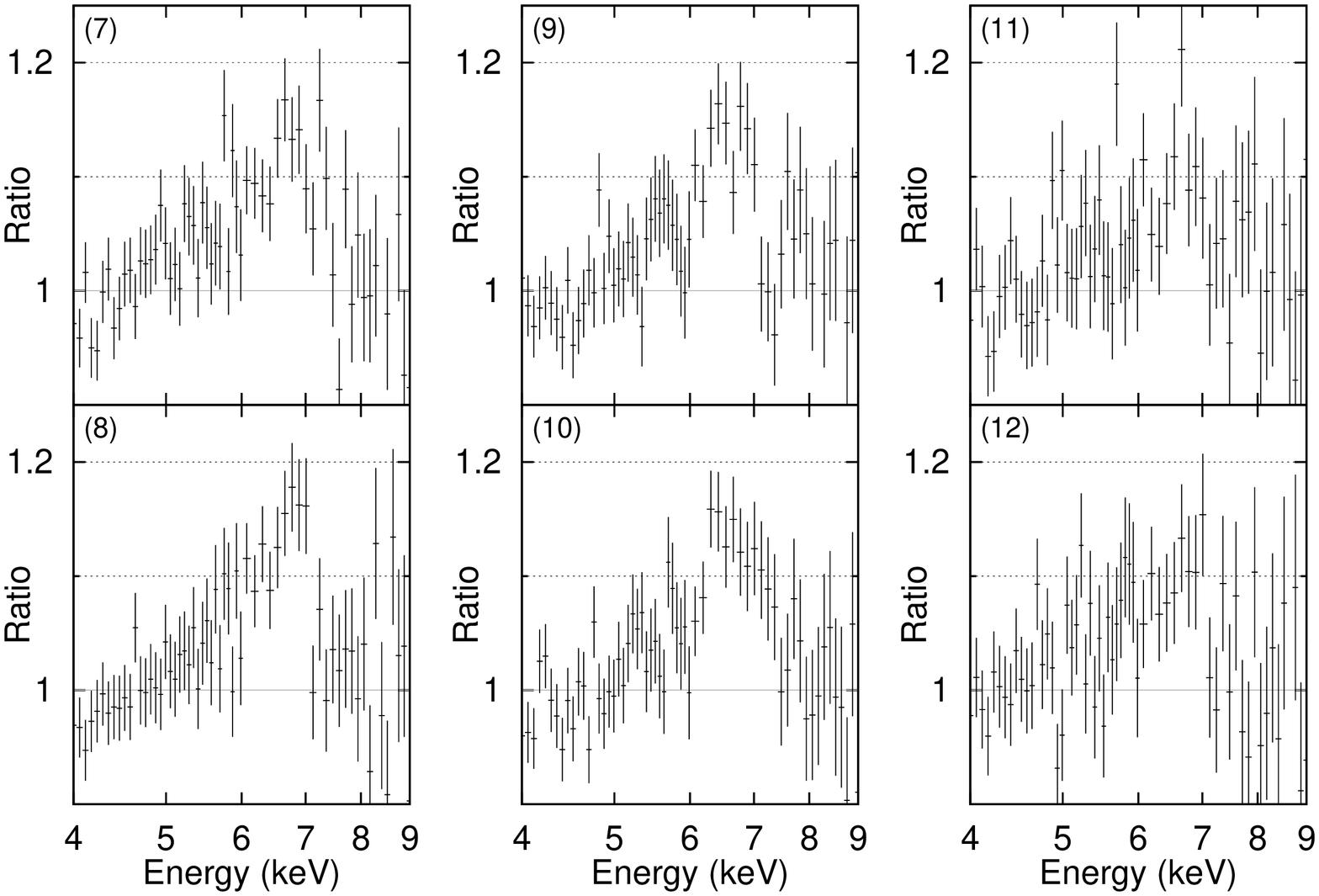}}
}
\subfigure{
\resizebox{10cm}{!}{\includegraphics[angle=270]{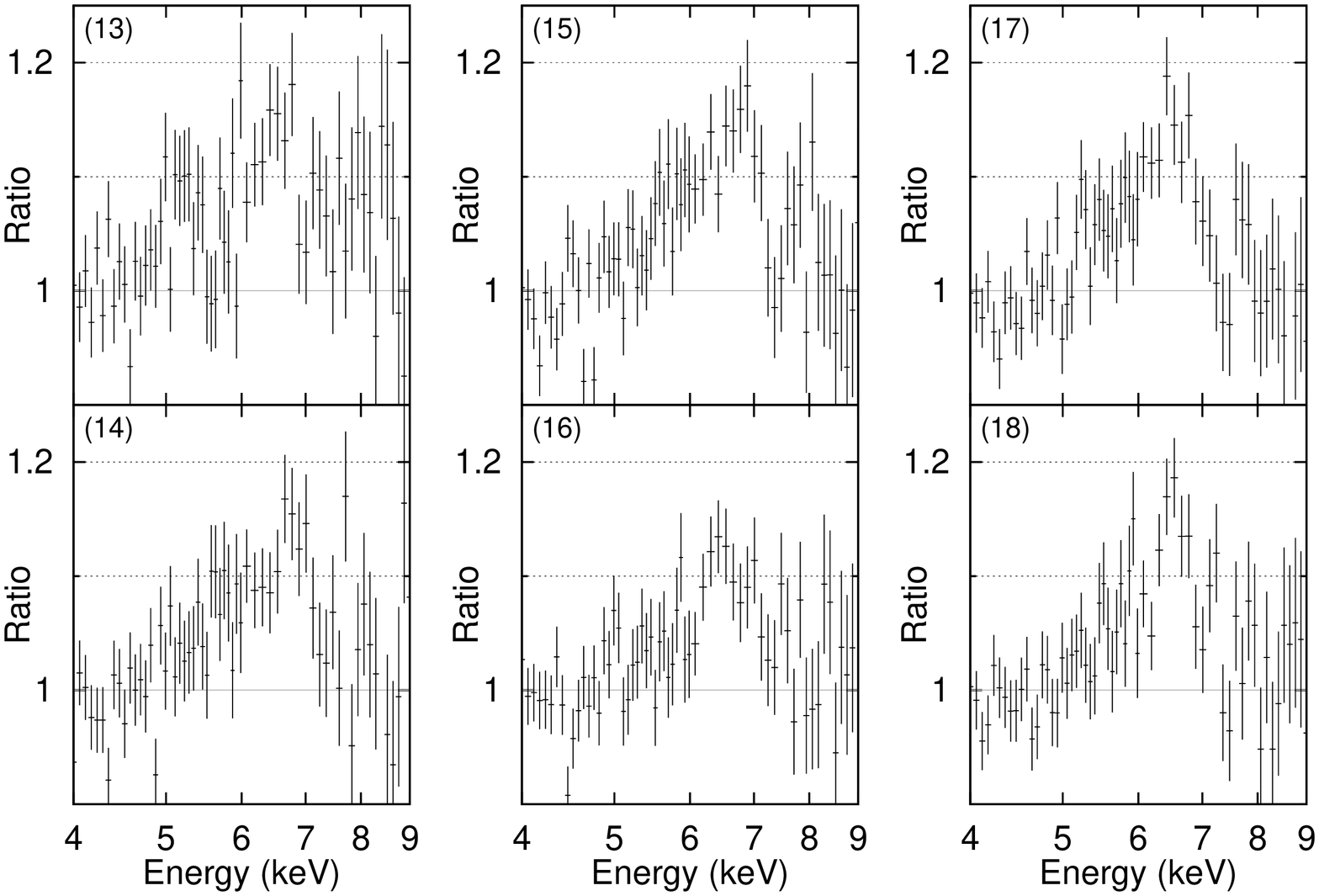}}
\resizebox{10cm}{!}{\includegraphics[angle=270]{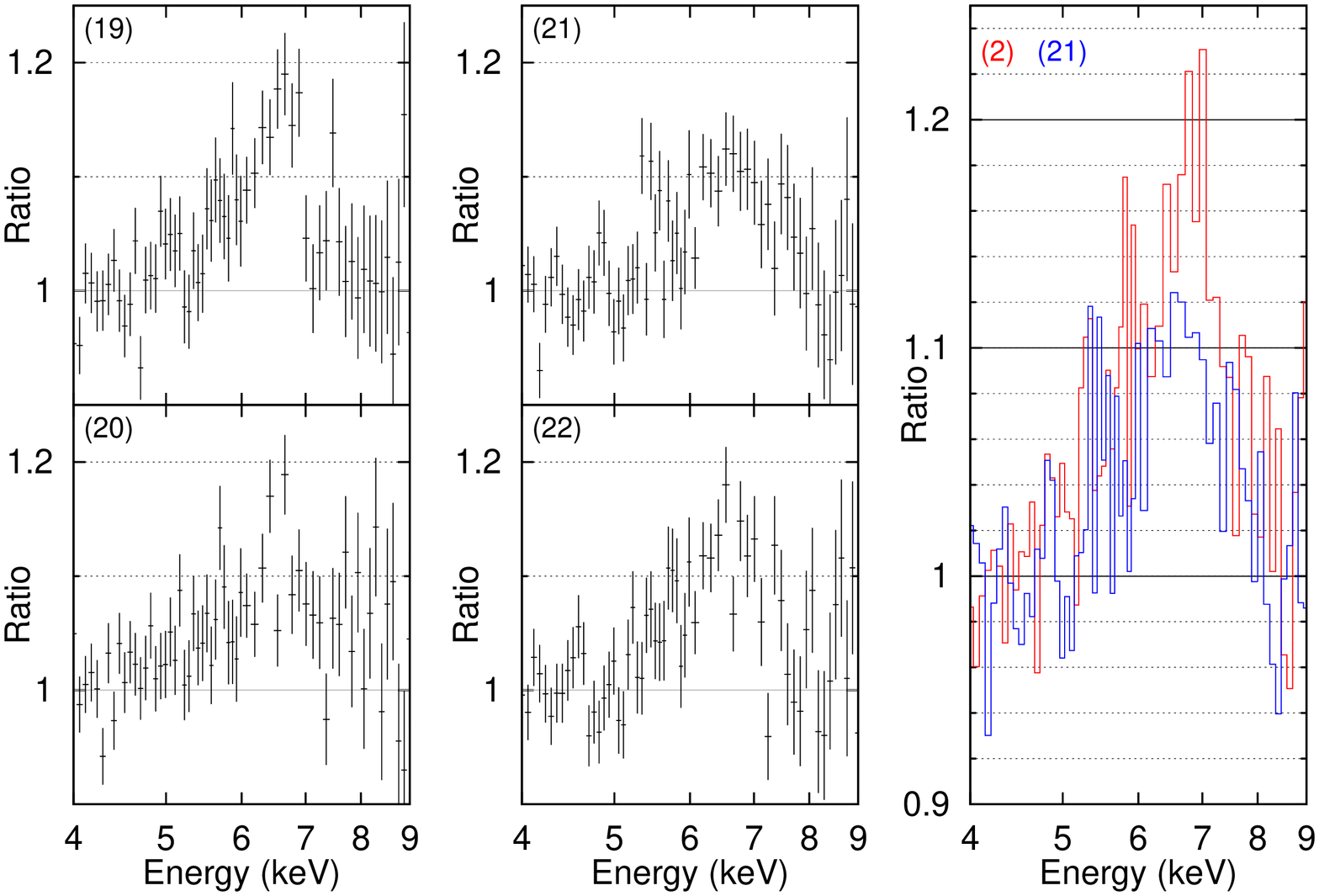}}
}
  \caption{Iron line profiles of the time-sliced spectra. The numbers in the parentheses show the orbit of Suzaku. The vertical axes show the ratio of the spectra to the power law model.
The right bottom figure shows comparison of orbit 2 (red) and orbit 21 (blue), where the error bars are not shown for clarity.
  }\label{f6-orbit1}
  \end{center}
\end{figure}
\end{landscape}
\begin{figure}[htbp]
  \begin{center}
    \includegraphics[width=100mm,angle=270]{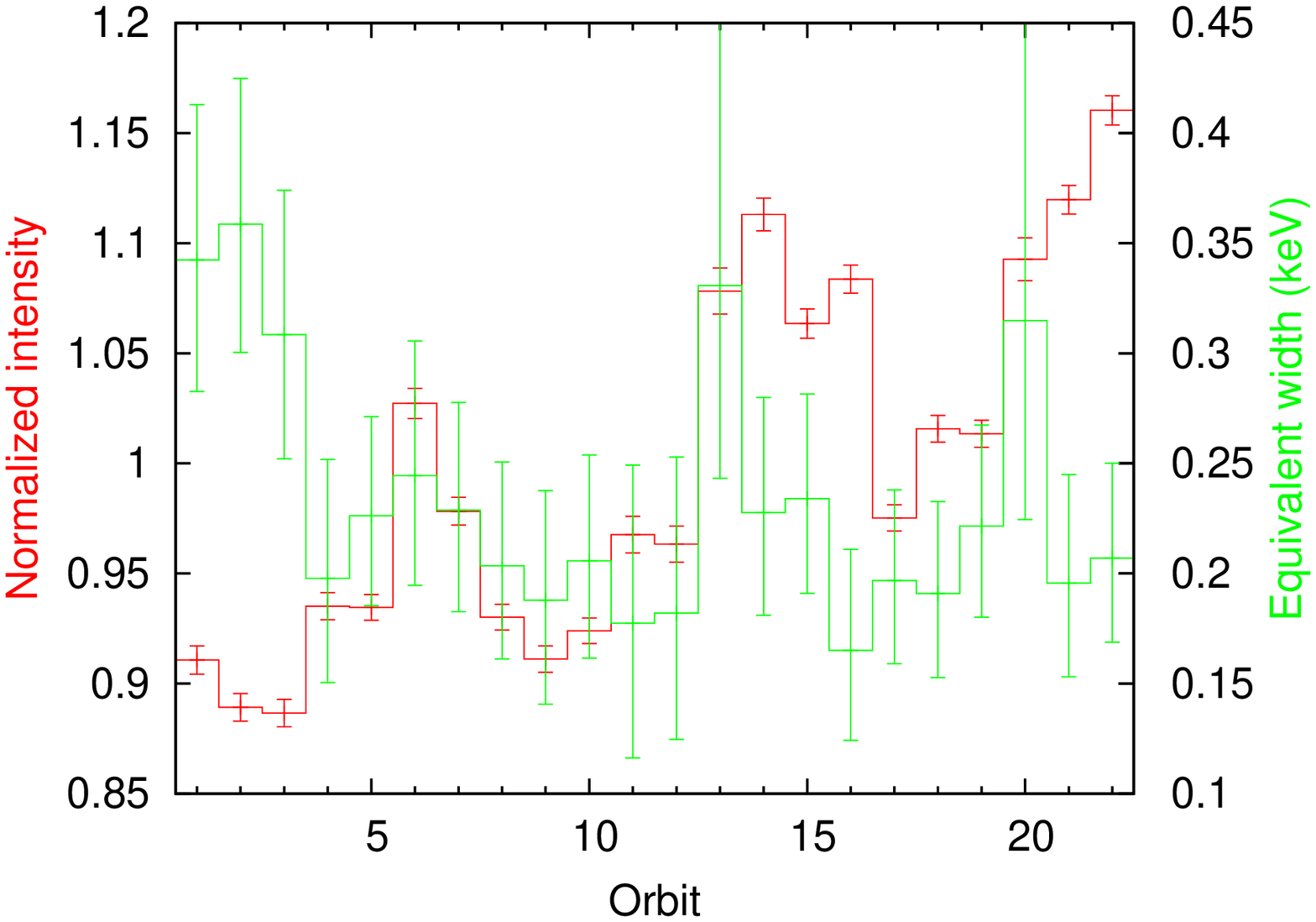}\\
    \includegraphics[width=100mm,angle=270]{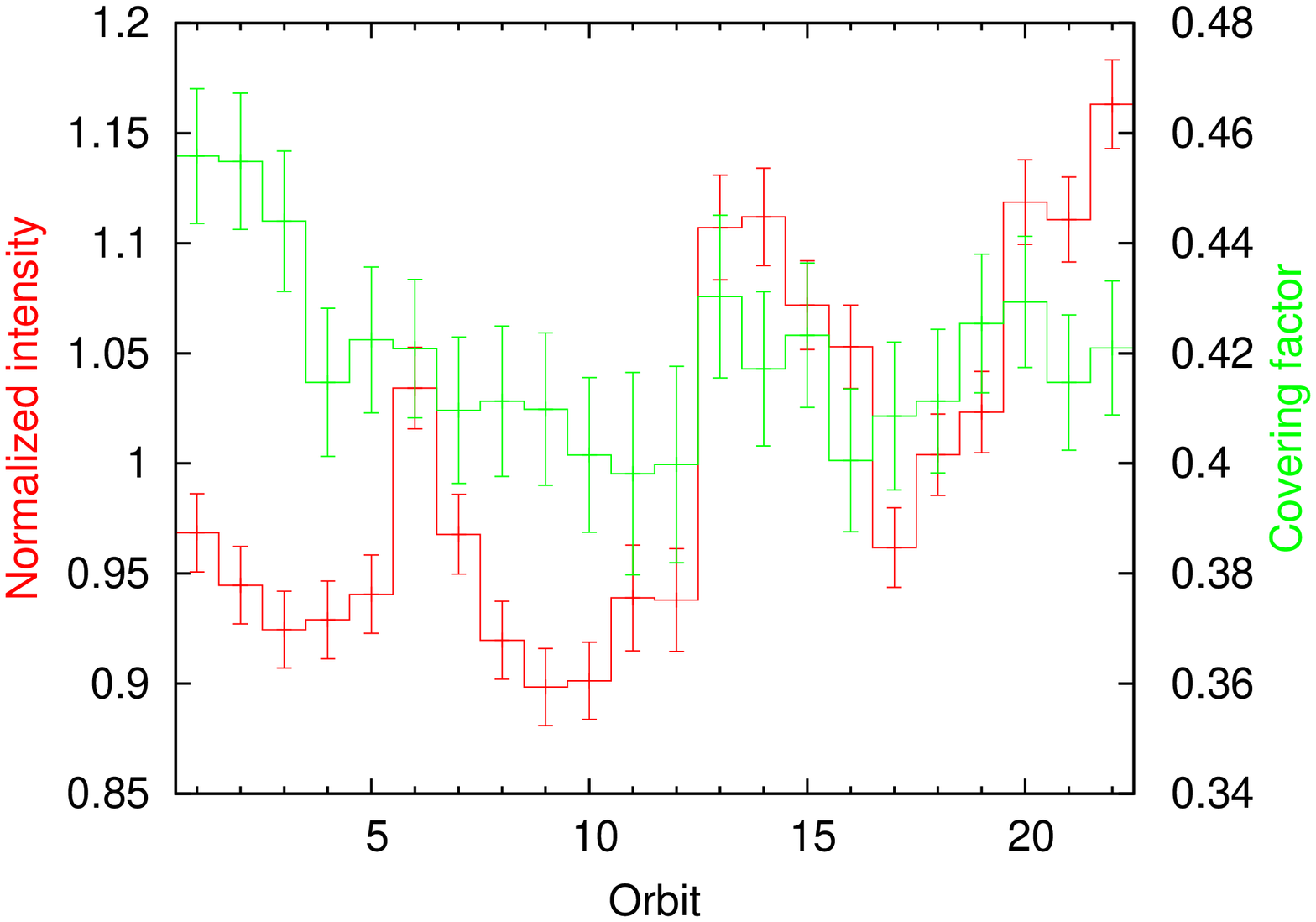}\\
  \end{center}
  \caption{(Upper panel) The time variation of the equivalent width (EW) of the iron line profiles and the normalized flux of the continuum, when the 2--12 keV is fitted with the relativistic disk reflection model. (Lower panel) The time variation of the covering factors and the normalized flux of the continuum.
  }\label{f6-EW}
\end{figure}
\begin{figure}[htbp]
  \begin{center}
    \includegraphics[width=120mm]{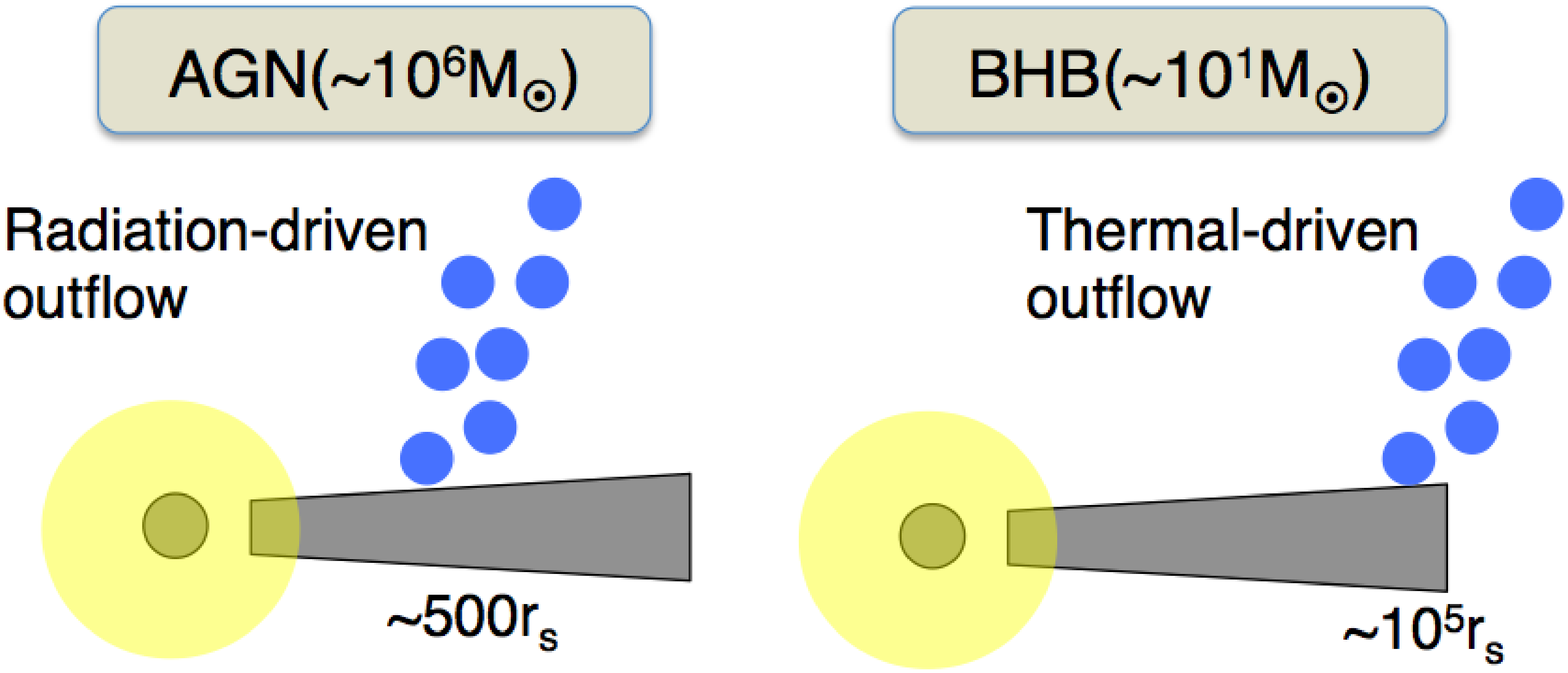}
  \end{center}
  \caption{Schematic figures of outflow in AGN and BHB
  }\label{f:outflow}
\end{figure}

\clearpage 

\bibliographystyle{aa}
\bibliography{00}
\end{document}